\documentclass[11pt]{article}
\usepackage[letterpaper, portrait, margin=1in]{geometry}
\usepackage[hypertexnames=false,colorlinks=true,linkcolor=blue,citecolor=ForestGreen]{hyperref}
\usepackage{algorithm}
\usepackage[noend]{algpseudocode}
\usepackage{url}
\usepackage{amsmath,amssymb,amsthm}
\usepackage{thmtools,thm-restate}
\usepackage[noabbrev,capitalise,nameinlink]{cleveref}
\usepackage{mathtools}
\usepackage{xspace}
\usepackage{verbatim}
\usepackage{mathrsfs} \usepackage[usenames,dvipsnames,svgnames,table]{xcolor}
\usepackage{pgf}
\usepackage[dvipsnames]{xcolor}
\usepackage{todo}
\usepackage{tabularx}
\usepackage{enumitem}
\usepackage{bm}
\usepackage{multirow}
\usepackage{diagbox}
\usepackage{nicematrix}
\usepackage{parskip}
\usepackage{adjustbox}
\usepackage{tikz}
\usepackage{tikz-3dplot}
\usepackage{transparent}
\usepackage{subcaption}

\usetikzlibrary{decorations.markings}

\tikzset{
  add slashes/.style={
    decoration={
      markings,
      mark=at position 0.5 with {
        \draw[-,line width=0.5pt]
            (-0.20,-0.15) -- (0.10,0.15)
            (-0.10,-0.15) -- (0.20,0.15);
      }
    },
    postaction={decorate}
  }
}

\usepackage{dsfont}

\newcommand*\ie{i.\kern.1em e., }
\newcommand*\eg{e.\kern.1em g., }
\newcommand*\cf{c.\kern.1em f.\ }
\newcommand*\almev{a.\kern.1em e.\ }
\newcommand*\iid{i.i.d.\ }

\topsep=1em \theoremstyle{plain}
\newtheorem{theorem}{Theorem}[section]
\newtheorem{lemma}[theorem]{Lemma}

\newtheorem{proposition}[theorem]{Proposition}
\newtheorem{claim}[theorem]{Claim}
\newtheorem{corollary}[theorem]{Corollary}

\newtheorem{observation}[theorem]{Observation}

\newtheorem{definition}[theorem]{Definition}
\newtheorem{problem}[theorem]{Problem}

\crefname{claim}{Claim}{Claims}
\crefname{fact}{Fact}{Facts}

\theoremstyle{definition}

\newtheorem{remark}[theorem]{Remark}
\newtheorem{example}[theorem]{Example}

\theoremstyle{plain}

\newcommand{\ignore}[1]{}

   \DeclareMathOperator{\poly}{poly}

\newcommand{\dist}{\mathsf{dist}}

\newcommand{\Accept}{\mathsf{Accept}}
\newcommand{\Reject}{\mathsf{Reject}}
\newcommand{\Abstain}{\mathsf{Abstain}}

\newcommand{\R}{\mathbb{R}}

\newcommand{\N}{\mathbb{N}}

\newcommand{\Exu}[2]{\underset{#1} \bE \left[ #2 \right] }

\newcommand{\Pru}[2]{\underset{ #1 } \Pr \left[ #2 \right]}
\newcommand{\Pruc}[3]{\underset{ #1 }\bbP \left[ #2 \;\; \mid \;\; #3 \right]}

\newcommand{\define}{\vcentcolon=}

\renewcommand{\epsilon}{\varepsilon}

\newcommand{\floor}[1]{\ensuremath{\left\lfloor #1 \right\rfloor}}

\newcommand{\inn}[1]{\langle #1 \rangle}

\newcommand{\ind}[1]{\mathds{1} \left[ #1 \right] }

\newcommand{\zo}{\{0,1\}}

\newcommand{\cA}{\ensuremath{\mathcal{A}}}

\newcommand{\cC}{\ensuremath{\mathcal{C}}}
\newcommand{\cD}{\ensuremath{\mathcal{D}}}
\newcommand{\cE}{\ensuremath{\mathcal{E}}}
\newcommand{\cF}{\ensuremath{\mathcal{F}}}

\newcommand{\cL}{\ensuremath{\mathcal{L}}}

\newcommand{\cN}{\ensuremath{\mathcal{N}}}

\newcommand{\cP}{\ensuremath{\mathcal{P}}}

\newcommand{\cR}{\ensuremath{\mathcal{R}}}

\newcommand{\cT}{\ensuremath{\mathcal{T}}}
\newcommand{\cU}{\ensuremath{\mathcal{U}}}

\newcommand{\cX}{\ensuremath{\mathcal{X}}}

\newcommand{\bbD}{\ensuremath{\mathbb{D}}}

\newcommand{\bbH}{\ensuremath{\mathbb{H}}}
\newcommand{\bbI}{\ensuremath{\mathbb{I}}}
\newcommand{\bbJ}{\ensuremath{\mathbb{J}}}

\newcommand{\bbP}{\ensuremath{\mathbb{P}}}
\newcommand{\bbR}{\ensuremath{\mathbb{R}}}

\newcommand{\bA}{{\boldsymbol{A}}}
\newcommand{\bB}{\boldsymbol{B}}

\newcommand{\bE}{\mathbf{E}}

\newcommand{\bR}{\boldsymbol{R}}
\newcommand{\bS}{\boldsymbol{S}}
\newcommand{\bT}{\boldsymbol{T}}

\newcommand{\boldf}{\boldsymbol{f}} 
\newcommand{\bb}{{\boldsymbol{b}}}
\newcommand{\bd}{{\boldsymbol{d}}}

\newcommand{\bg}{\boldsymbol{g}}
\newcommand{\bh}{\boldsymbol{h}}

\newcommand{\bp}{\boldsymbol{p}}

\newcommand{\bt}{\boldsymbol{t}}

\newcommand{\bw}{\boldsymbol{w}}
\newcommand{\bx}{\boldsymbol{x}}
\newcommand{\by}{\boldsymbol{y}}

\newcommand{\abs}[1]{{\left| #1 \right|}}

\newcommand{\pbra}[1]{{\left( #1 \right)}}
\newcommand{\cbra}[1]{{\left\{ #1 \right\}}}

\newcommand{\wh}[1]{\widehat{#1}}
\newcommand{\wt}[1]{\widetilde{#1}}

\newcommand{\TV}{\mathsf{TV}}

\usepackage{array}

\usepackage{authblk}

\usepackage{microtype}
\numberwithin{equation}{section}

\usepackage{tikz}
\usetikzlibrary{arrows.meta,backgrounds,calc,decorations.pathreplacing,fit,patterns.meta,positioning,shapes.geometric}

\newcommand{\Prodreal}{{\bbD}_{\mathrm{prod,real}}}
\newcommand{\Unif}{\mathrm{Unif}}
\newcommand{\conj}{\mathrm{Conj}}
\newcommand{\DL}{\mathrm{DL}}

\newcommand{\twolines}[2]{\shortstack[l]{\rule{0pt}{3.4ex}#1\\[2pt]#2}}
\newcommand{\NS}{\mathrm{NS}}
\newcommand{\bbNS}{\mathbb{NS}}

\newcommand{\Sat}{\mathrm{Sat}}

\newcommand{\block}{\mathsf{block}}

\newcommand{\odif}[1]{\mathop{}\!\mathrm{d}#1}

\title{Testable Testing: Distribution-Robust Property Testing}

\newcommand{\AffilColumbia}{*}
\newcommand{\AffilUBC}{$\dagger$}

\author[\AffilColumbia]{Xi Chen}
\author[\AffilColumbia]{Renato Ferreira Pinto Jr.}
\author[\AffilUBC]{Nathaniel Harms}
\author[\AffilColumbia]{Rocco A. Servedio}

\affil[\AffilColumbia]{Columbia University}
\affil[\AffilUBC]{University of British Columbia}

\date{}

\begin{document}

\maketitle

\begin{abstract}

We introduce a new model for property testing of Boolean functions, which we call
\emph{distribution-robust} testing. Distribution-robust testing is motivated by the hope of
designing testers whose output can be trusted even when the underlying probability distribution does
not satisfy the assumptions of the tester, while still being more efficient than distribution-free
testers that make no assumptions at all.

The purpose of this paper is to perform a broad initial exploration of this new model. In
particular, we present a variety of general algorithmic techniques, which we apply to show that
several well-studied properties admit distribution-robust testers that are significantly more
efficient than their distribution-free counterparts.
For example, we obtain constant-query distribution-robust testers
for conjunctions, decision lists, and unions of intervals, all of which are known to require significantly more queries in the distribution-free model.
These encouraging results suggest a range of
questions for future work.

\end{abstract}

\allowdisplaybreaks

\thispagestyle{empty}
\setcounter{page}{0}
\newpage
{
\setcounter{tocdepth}{2}
\tableofcontents
}
\thispagestyle{empty}
\setcounter{page}{0}
\newpage
\setcounter{page}{1}

\newpage

\pagenumbering{arabic}

\section{Introduction} \label{sec:intro}

When designing algorithms that depend on a probability distribution, it is common to make
assumptions about this distribution. Those who engage in this pathological behaviour give many
rationalizations: Maybe efficient algorithms are impossible without some assumption, maybe it is
more realistic to assume some well-behaved distribution instead of an adversarial one, or maybe 
it is just too much work to analyze an algorithm under arbitrary distributions. The problem with
algorithms designed with distributional assumptions is that they provide no theoretical guarantees
when these assumptions fail.  Even worse, as we will see, these assumptions are often infeasible to
verify: given samples from the distribution, we cannot decide whether our assumptions hold, and
therefore the output of the algorithm is not trustworthy.

These distributional assumptions are ubiquitous in property testing.  The purpose of this paper is
to introduce and study a model of property testing which lets us design algorithms that
\emph{always} provide theoretical guarantees, regardless of the input distribution, yet \emph{also}
avoid the worst-case behaviour of unrestricted input distributions. We call this model
\emph{distribution-robust testing}. Before defining the model, let us illustrate the need for these
algorithms with an example.

\begin{example}
\label{ex:intro-monotonicity}
Consider the problem of testing monotonicity, one of the most well-studied problems in property
testing. A monotonicity tester should, given query access to an input function $f \colon \zo^n \to
\zo$ and sample access to a probability distribution $\cD$ over $\zo^n$, produce an output which
(with probability at least, say, $3/4$) satisfies:
\begin{enumerate}
\item If $f$ is monotone, the algorithm outputs $\Accept$;
\item If $f$ is $\epsilon$-far from monotone, meaning
\[
    \epsilon < \dist_\cD(f,g) \define \Pru{\bm x \sim \cD}{f(\bm x) \neq g(\bm x)}
        \qquad \forall \text{ monotone functions } g ,
\]
the algorithm outputs $\Reject$.
\end{enumerate}
If we make \emph{no} assumption on the distribution $\cD$, so that $\cD$ is an arbitrary
distribution (unknown to the algorithm except through its random samples), testing monotonicity
requires $2^{\Theta(n)}$ queries \cite{HalevyKushilevitz:07}.  But, after decades of effort
\cite{GGLRS,CS13a,CST14,KMS18,BCS18,CS19,BCS20,HarmsYoshida22,BCS23,BKKM23,BCS25},
efficient testers with query cost $\wt O(\sqrt{n})$ are known
\cite{KMS18}, along with matching lower bounds \cite{CCCPS26}, under an assumption: \emph{The distribution $\cD$ is uniform\footnote{In this case, algorithms do not need sample access to $\cD$, since they can draw their own uniform samples. However, the distributional assumption is still a crucial aspect
of the model since success is defined in terms of $\dist_\cD$.} over $\zo^n$}.
Similar bounds hold
if we weaken the
assumption so that \emph{$\cD$} is a product distribution over $\R^n$ \cite{BCS25}.
\looseness=-1

Today's user of monotonicity testers finds herself in a dilemma: if she is \emph{certain} that the
input $\cD$ is uniform (or product), then she can run the tester and trust its output.  But what if
she is not certain? One option is to try to verify the assumption: she could use samples to
\emph{test} that the distribution $\cD$ is uniform; unfortunately, this is a distribution testing
problem that requires $2^{\Theta(n)}$ samples \cite{Paninski:08}. So, to trust the output of her
algorithm, she must rely on the $2^{\Theta(n)}$ upper bound that holds when no assumptions are made.
A similar dilemma holds for most other properties, because typical distributional assumptions
(usually that the distribution is uniform, Gaussian, or a product distribution)
are inefficient \cite{Paninski:08,ADK15} or impossible to
test.
\end{example}
Our distribution-robust testing model addresses this dilemma. In the parlance of learning theory,
distribution-robust testers are ones which can detect if they are being used ``out of distribution''\!, that is, outside of the class of distributions they were designed to handle. We will see that, surprisingly, it is
often possible to design algorithms that (a) remain correct when distributional assumptions hold,
(b) remain trustworthy by avoiding incorrect answers when the assumptions fail, and (c) are still
more efficient than
previously studied
testers that make no assumptions.  We present a variety of techniques
for designing distribution-robust testers, and compare them to
the existing models.

\subsection{Distribution-Robust Testing}

Here is our definition of distribution-robust testing. Throughout, we will use the notation $\bbP$
for a ``property'' (\ie a set of boolean-valued functions $\cX \to \zo$), and $\bbD$ for a class
of probability distributions over the domain $\cX$. In standard property testing, $\bbD$ would be
interpreted as an assumption on the input distribution $\cD$, \ie we assume $\cD \in \bbD$. We want
algorithms in our new model to achieve the following: If $\cD \in \bbD$ (so the ``assumption''
holds), the tester should correctly output $\Accept$ or $\Reject$. But when $\cD \notin \bbD$ (the
``assumption'' fails), the tester \emph{should not output an incorrect answer}. To accomplish this,
we need to allow a third output: $\Abstain$. Formally:

\begin{definition}[Distribution-Robust Testing]
Let $\cX$ be a domain, let $\bbP$ be a property (\ie a set) of functions $\cX \to \zo$, and let $\bbD$
be a set of probability distributions over $\cX$. As input, the algorithm is given parameters
$\epsilon, \delta > 0$, query access to a function $f \colon \cX \to \zo$ and
access to labeled samples $(\bm{x}, f(\bm{x}))_{\bm{x} \sim \cD}$
for an arbitrary probability distribution $\cD$.
Its output should, with probability at least $1-\delta$, satisfy:
\begin{enumerate}
    \item \textbf{Completeness.} If $f \in \bbP$ (the function has the property), the algorithm
            outputs either $\Accept$ or $\Abstain$;
    \item \textbf{Soundness.} If $f$ is $\epsilon$-far from $\bbP$ under $\cD$ (\ie $\dist_\cD(f,g)
        > \epsilon$ for all $g \in \bbP$), the algorithm outputs either $\Reject$ or $\Abstain$.
    \item \textbf{Assertiveness.}
       If $\cD \in \bbD$, the algorithm does not output $\Abstain$.
\end{enumerate}
We call a tester satisfying this definition a \emph{$\bbD$-robust tester} for $\bbP$, or a
distribution-robust tester for $\bbP$ with respect to $\bbD$.
\end{definition}

Distribution-robust testers enjoy standard success probability
amplification by majority vote of $O(\log(1/\delta))$ runs, each with failure
probability (say) $1/4$. When $\delta$ is not specified, we take $\delta=1/4$.

Our definition of distribution-robust testing is strongly inspired by the recent \emph{testable
learning} algorithms of Rubinfeld and Vasilyan \cite{RubinfeldVasilyan23}, which provide an elegant
solution to a similar dilemma ocurring in learning theory: time-efficient agnostic learning
algorithms for halfspaces were known when the distribution $\cD$ is assumed to be Gaussian or
uniform, but no efficient algorithms are known without assumptions on $\cD$; furthermore, it is
prohibitively expensive to test the assumptions. Testable learning allowed efficient algorithms with
no assumptions, and there is now a rapidly growing literature on testable learning
\cite{GKK:23testable,DKKLZ:23testable,GKSV:23universal,GKSV:24efficient,DKLZ:24general,STW:24ptf,KSV:24tds,CKKSV:24discrepancy,KZ:25auctions,GKSV:26noise,LQ:26queries,DIKL:26massart,BDP:26distinguishers}.

A more general framework of ``testable algorithms''\!, which are capable of efficiently checking their own
assumptions, is in our opinion extremely natural and desirable. A couple of papers
have begun
extending this framework outside of learning theory: there are distribution testing algorithms which
test their assumptions on the quality of sample labels \cite{FH24}, and triangle-counting algorithms
which use testable arboricity assumptions \cite{ERV26}. Our paper brings the framework to property
testing, which is to say, ``testable testing''\!.

To distinguish and compare our model to
existing property testing models,
let us give a
definition of property testing that captures essentially all known property testing algorithms:

\begin{definition}[Distribution-Specific and Distribution-Free Testing]
\label{def:intro-specific}
Let $\cX$ be a domain, let $\bbP$ be a property (\ie a set) of functions $\cX \to \zo$, and let $\bbD$
be a set of probability distributions over $\cX$. As input, the algorithm is given parameters
$\epsilon, \delta > 0$, query access to a function $f \colon \cX \to \zo$ and
access to labeled samples $(\bm{x}, f(\bm{x}))_{\bm{x} \sim \cD}$
for some probability distribution $\cD \in \bbD$.
Its output should, with probability at least $1-\delta$,
satisfy:
\begin{enumerate}
    \item \textbf{Completeness.} If $f \in \bbP$, the algorithm
outputs $\Accept$;
    \item \textbf{Soundness.} If $f$ is $\epsilon$-far from $\bbP$ under $\cD$, the algorithm outputs $\Reject$.
\end{enumerate}
We call an algorithm satisfying this definition a \emph{$\bbD$-specific} tester for $\bbP$, or a
tester for $\bbP$ with respect to $\bbD$.
If
$\bbD$ is the set of all probability
distributions over $\cX$, we say the tester is \emph{distribution-free}.
\end{definition}

This definition of distribution-specific property testing is not standard, so let us briefly explain
how it captures known results in the field.  Following the introduction of general distributions
$\cD$ (as opposed to the uniform distribution) to the study of property testing \cite{GGR98}, the 
majority of property testers fall into two special cases. In the first case, $\bbD = \{\cD^*\}$
is a singleton set, meaning that the input distribution $\cD$ is assumed to be a specific
well-behaved distribution $\cD^*$ (typically the uniform distribution over domain $\cX$, or the
Gaussian distribution over $\R^n$); this includes the majority of results in the field, so we shall
not cite any specific examples.

This case also includes some ``algorithm templates'' which give
testers for any fixed singleton set $\bbD = \{\cD^*\}$ where $\cD^*$ belongs to
a class $\mathfrak D$; in other words, for every distribution $\cD^*$ in some
class $\mathfrak D$, there is a $\{\cD^*\}$-specific tester that assumes the
input $\cD = \cD^*$; for example, \cite{Blaisstoc09,BBBY12} give
$\{\cD^*\}$-specific tester for every distribution $\cD^*$ among a class of
distributions.

In the second case, $\bbD$ is the set of all distributions, meaning that the algorithm makes no
assumption on $\cD$; this is called \emph{distribution-free} testing
\cite{HalevyKushilevitz:07,HalevyKushilevitz:08,GlasnerServedio:09toc,CX16,CLSSX18,Bshouty19,FlemingYoshida20,BFH21,CP22,CFP24}.

A smaller number of testers satisfy the more general distribution-specific definition by letting
$\bbD$ be a wider class of distributions; see for example \cite{BCS20,HarmsYoshida22,BCS25} which
assume product distributions over $\R^n$, or \cite{Harms19} which assumes rotation-invariant
distributions over $\R^n$.

\subsection{Results}

The goal of this paper is to define and motivate the distribution-robust model and answer the most
basic questions about it. To begin with, it holds by definition that, for any class $\bbD$ of
distributions, the query and sample complexity of testing satisfies
\[
\bbD\text{-specific}\quad \leq \quad\bbD\text{-robust}\quad \leq \quad\text{distribution-free} ,
\]
so distribution-robust testing is indeed comparable with standard models of testing, and we will
compare them. In particular, we show:
\begin{enumerate}
    \item Most importantly, there are natural and well-studied properties where distribution-robust testing is significantly more
efficient than distribution-free testing;
\item More generally, we give upper- and lower-bounds separating distribution-robust testing from
the distribution-free and distribution-specific testing models; and
\item We present a variety of general algorithmic techniques for designing distribution-robust
    testers, from which most of our results follow as corollaries.
\end{enumerate}

In this section we will go into more detail on items 1 and 2 above. In
\cref{section:intro-techniques} we will describe our algorithmic techniques, state our formal
theorems, and give brief proof overviews.  This paper will present results for the following
properties $\bbP$ of boolean functions:
\begin{itemize}
\item \emph{Monotone functions:} Functions $f \colon \R^n \to \zo$ where $x \preceq y \implies f(x) \leq
f(y)$ where $x \preceq y$ is according to the natural partial ordering on $\R^n$.
\item \emph{Conjunctions:} Functions $f \colon \zo^n \to \zo$ of the form
$f(x) = \bigwedge_{\ell_i \in S} \ell_i$ where $S$ is a subset of the literals
$x_1, \neg x_1, \dotsc, x_n, \neg x_n$.
\item \emph{1-Decision Lists:} Functions $f \colon \zo^n \to \zo$ where there is an ordered list
    $(\alpha_1, \beta_1)$, $\dotsc$, $(\alpha_k, \beta_k)$, $(*, \beta_{k+1})$ where each $\alpha_i$ is a
    literal (\ie an input variable $x_j$ or $\neg x_j$) and each $\beta_i \in \zo$, such that, on
    input $x$, the output $f(x) = \beta_i$ where $i$ is the smallest value such that $x$
    satisfies $\alpha_i$, or $i = k+1$ if none of the literals $\alpha_j$ are satisfied.
\item \emph{Unions of $k$ Intervals:} Functions $f \colon [0,1] \to \zo$ that are the indicator
    functions of sets $S = \bigcup_{i=1}^k I_i$, where each $I_i \subseteq [0,1]$ is an interval.
\end{itemize}

\subsubsection{Qualitative Results}

We begin with a qualitative summary of our results and leave the quantitative
bounds for the section below. We establish the following figure
(\cref{figure:qual}) of pairwise comparisons between different models of
testing. We distinguish between testers that are allowed to use \emph{adaptive}
queries, \emph{non-adaptive} queries (\ie the algorithm receives its samples
upfront, and then decides the full set of query points before making any of the
queries), and \emph{sample-based} testers (which only see labeled samples
and cannot choose their queries).

We would also like to emphasize two interesting separations that do not fit into
\cref{figure:qual}: There are examples where distribution-robust testing is much
more efficient than distribution-free, even when the distribution-robust tester is
\emph{non-adaptive} (\cref{section:intro-juntas})
or
\emph{sample-based} (\cref{section:intro-sample})
while the distribution-free tester is allowed to be
\emph{adaptive}.

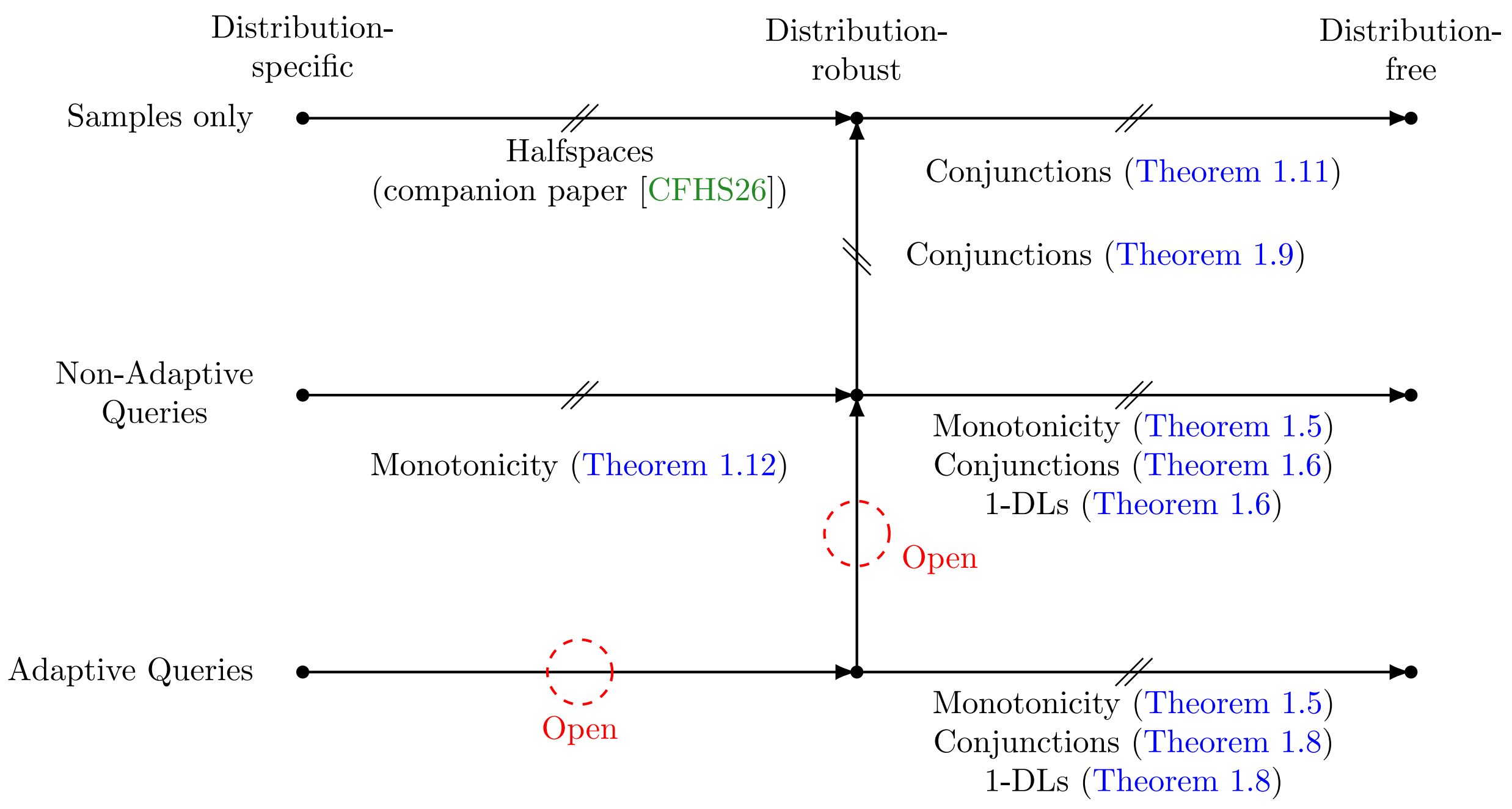
\begin{figure}[bht!]
\centering
\small
\begin{tikzpicture}[>=Latex, line width=0.8pt]

    \def\rowSep{3.0}                 \def\headerRise{0.75}            \def\sampleTextDropLeft{0.6}     \def\sampleTextDropRight{0.6}    \def\interRowTextFrac{0.50}      \def\rowTextDrop{0.78}         
    \def\leftArrowLen{6}             \def\rightArrowLen{6}          
    \pgfmathsetmacro{\ySamples}{0}
  \pgfmathsetmacro{\yNonAdaptive}{\ySamples-\rowSep}
  \pgfmathsetmacro{\yAdaptive}{\yNonAdaptive-\rowSep}

    \pgfmathsetmacro{\xSpecific}{0}
  \pgfmathsetmacro{\xRobust}{\xSpecific+\leftArrowLen}
  \pgfmathsetmacro{\xFree}{\xRobust+\rightArrowLen}

    \pgfmathsetmacro{\xSpecificRobustMid}{
    (\xSpecific+\xRobust)/2
  }
  \pgfmathsetmacro{\xRobustFreeMid}{
    (\xRobust+\xFree)/2
  }

    \pgfmathsetmacro{\xVerticalEdgeLabel}{
    \xRobust+2.7
  }

    \pgfmathsetmacro{\yAdaptiveNonAdaptiveMid}{
    (\yAdaptive+\yNonAdaptive)/2
  }
  \pgfmathsetmacro{\yHeader}{
    \ySamples+\headerRise
  }
  \pgfmathsetmacro{\ySampleTextLeft}{
    \ySamples-\sampleTextDropLeft
  }
  \pgfmathsetmacro{\ySampleTextRight}{
    \ySamples-\sampleTextDropRight
  }
  \pgfmathsetmacro{\yInterRowText}{
    \yNonAdaptive
    + \interRowTextFrac*(\ySamples-\yNonAdaptive)
  }
  \pgfmathsetmacro{\yNonAdaptiveText}{
    \yNonAdaptive-\rowTextDrop
  }
  \pgfmathsetmacro{\yAdaptiveText}{
    \yAdaptive-\rowTextDrop
  }

    \coordinate (SDS) at (\xSpecific,\ySamples);
  \coordinate (SDR) at (\xRobust,\ySamples);
  \coordinate (SDF) at (\xFree,\ySamples);

  \coordinate (NADS) at (\xSpecific,\yNonAdaptive);
  \coordinate (NADR) at (\xRobust,\yNonAdaptive);
  \coordinate (NADF) at (\xFree,\yNonAdaptive);

  \coordinate (ADS) at (\xSpecific,\yAdaptive);
  \coordinate (ADR) at (\xRobust,\yAdaptive);
  \coordinate (ADF) at (\xFree,\yAdaptive);

  \coordinate (ADSADR) at (\xSpecificRobustMid,\yAdaptive);
  \coordinate (ADRNADR) at (\xRobust,\yAdaptiveNonAdaptiveMid);

    \node[anchor=east] at (-0.4,\ySamples)
    {Samples only};
  \node[anchor=east,align=center] at (-0.4,\yNonAdaptive)
    {Non-Adaptive\\Queries};
  \node[anchor=east] at (-0.4,\yAdaptive)
    {Adaptive Queries};

    \node[align=center] at (\xSpecific,\yHeader)
    {Distribution-\\specific};
  \node[align=center] at (\xRobust,\yHeader)
    {Distribution-\\robust};
  \node[align=center] at (\xFree,\yHeader)
    {Distribution-\\free};

    \fill (SDS) circle (2pt);
  \fill (SDR) circle (2pt);
  \fill (SDF) circle (2pt);
  \fill (NADS) circle (2pt);
  \fill (NADR) circle (2pt);
  \fill (NADF) circle (2pt);
  \fill (ADS) circle (2pt);
  \fill (ADR) circle (2pt);
  \fill (ADF) circle (2pt);

    \draw[->, add slashes] (SDS) -- (SDR);
  \draw[->, add slashes] (SDR) -- (SDF);
  \draw[->, add slashes] (NADS) -- (NADR);
  \draw[->, add slashes] (NADR) -- (NADF);
  \draw[->, add slashes] (NADR) -- (SDR);
  \draw[->] (ADS) -- (ADR);
  \draw[->, add slashes] (ADR) -- (ADF);
  \draw[->] (ADR) -- (NADR);

    \node[align=center] at (\xSpecificRobustMid,\ySampleTextLeft)
    {Halfspaces \\ (companion paper \cite{CFHS26halfspaces})};

  \node[align=center] at (\xRobustFreeMid,\ySampleTextRight)
    {Conjunctions (\cref{thm:intro-conj-sample})};

  \node[align=center] at (\xVerticalEdgeLabel,\yInterRowText)
    {Conjunctions
      (\cref{thm:intro-non-adaptive-conjunction-tester})};

  \node[align=center] at (\xSpecificRobustMid,\yNonAdaptiveText)
    {Monotonicity (\cref{thm:intro-mono-lb})};

  \node[align=center] at (\xRobustFreeMid,\yNonAdaptiveText) {
        Monotonicity (\cref{thm:intro-mono-unif}) \\
        Conjunctions (\cref{thm:intro-conj-dl-na}) \\
        1-DLs (\cref{thm:intro-conj-dl-na}) };

  \node[align=center] at (\xRobustFreeMid,\yAdaptiveText) {
        Monotonicity (\cref{thm:intro-mono-unif}) \\
        Conjunctions (\cref{thm:intro-conj-dl-adaptive}) \\
        1-DLs (\cref{thm:intro-conj-dl-adaptive}) };

    \draw[color=red, dashed] (ADSADR) circle (10pt);
  \node[color=red, below, yshift=-10pt, align=center]
    at (ADSADR) {Open};

  \draw[color=red, dashed] (ADRNADR) circle (10pt);
  \node[color=red, below right, xshift=10pt, align=center]
    at (ADRNADR) {Open};

\end{tikzpicture}

\caption{Summary of separations between models. Each node represents a model in
$\{\bbD\text{-specific}, \bbD\text{-robust}, \text{distr.-free}\}
\times
\{\text{adaptive}, \text{non-adaptive}, \text{sample-based}\}$.
An arrow $A \to B$ means that the cost of a tester in model
$A$ is at most the cost in model $B$. A slashed arrow means there is an example
where the cost in model $A$ is significantly less than the cost in model $B$,
and the edge labels list the examples.
Edges that do not touch a distribution-robust endpoint are omitted.
}
\label{figure:qual}
\end{figure}

\begin{table}[hbt!]
\centering
\small
\renewcommand{\arraystretch}{1.35}
\setlength{\arrayrulewidth}{0.6pt}

\newcommand{\specificwidth}{0.185\textwidth}
\newcommand{\robustwidth}{0.215\textwidth}
\newcommand{\freewidth}{0.18\textwidth}

\begin{tabular}{@{}p{0.15\textwidth}p{0.15\textwidth}|p{\specificwidth}|p{\robustwidth}|p{\freewidth}|@{}}
\multicolumn{1}{@{}p{0.15\textwidth}}{}
& \multicolumn{1}{p{0.15\textwidth}}{}
& \multicolumn{1}{p{\specificwidth}}{\twolines{\textbf{Distribution}}{\textbf{specific}}}
& \multicolumn{1}{p{\robustwidth}}{\twolines{\textbf{Distribution}}{\textbf{robust}}}
& \multicolumn{1}{p{\freewidth}@{}}{\twolines{\textbf{Distribution}}{\textbf{free}}} \\
\cline{3-5}
\twolines{\textbf{Monotonicity}}{}
& \twolines{Non-adaptive}{$\bbD = \text{uniform}$}
& \twolines{$\widetilde{\Theta}(\sqrt{n})$}{\cite{KMS18,CCCPS26}}
& \twolines{$2^{\widetilde{O}(\sqrt{n})}$,
$2^{\Omega(n^{1/4})}$}{\cref{thm:intro-mono-unif,thm:intro-mono-lb}}
& \twolines{$2^{\Omega(n)}$}{\cite{HalevyKushilevitz:07}} \\
\cline{3-5}
& \twolines{Non-adaptive}{$\bbD = \text{product}$}
& \twolines{$n^{0.5+ o(1)}$}{\cite{BCS25}}
& \twolines{$2^{\widetilde{O}(\sqrt{n})}$}{\cref{thm:intro-mono-product}}
& \twolines{$2^{\Omega(n)}$}{\cite{HalevyKushilevitz:07}} \\
\cline{1-5}
\twolines{\textbf{Conjunctions}}{}
& Sample-based
& \twolines{$\Theta(\log n)$}{Folklore (\cref{section:sample-based})}
& \twolines{$\Theta(\log n)$}{\cref{thm:intro-conj-sample}}
& \twolines{$\widetilde \Theta(n)$}{\cite{CFP24}}\\
\cline{3-5}
& Non-adaptive
& \twolines{$\widetilde O(1/\epsilon^2)$}{\cite{DLM+:07}}
& \twolines{$O(\log n)$, $\widetilde O(1/\epsilon^2)$}{\cref{thm:intro-conj-dl-na,thm:intro-non-adaptive-conjunction-tester}}
& \twolines{$\widetilde{\Omega}(n^{1/3})$}{\cite{CX16}} \\
\cline{3-5}
& Adaptive
& \twolines{$\widetilde O(1/\epsilon)$}{\cite{Bshouty20}}
& \twolines{$\widetilde O(1/\epsilon)$}{\cref{thm:intro-conj-dl-adaptive}}
& \twolines{$\widetilde{\Theta}(n^{1/3})$}{\cite{CX16}} \\
\cline{1-5}
\twolines{\textbf{\twolines{1-Decision}{Lists}}}{}
& Non-adaptive
& \twolines{$\widetilde O(1/\epsilon^2)$}{\cite{DLM+:07}}
& \twolines{$O(\log n)$, $2^{\widetilde O(1/\epsilon)}$}{\cref{thm:intro-conj-dl-na,thm:intro-conj-dl-adaptive}}
& \twolines{$\widetilde{\Omega}(\sqrt{n})$}{\cite{CFP24}} \\
\cline{3-5}
& Adaptive
& \twolines{$\widetilde O(1/\epsilon)$}{\cite{Bshouty20}}
& \twolines{$\widetilde O(1/\epsilon)$}{\cref{thm:intro-conj-dl-adaptive}}
& \twolines{$\widetilde{O}(n^{11/12}),\ \widetilde{\Omega}(\sqrt{n})$}{\cite{CFP24}} \\
\cline{1-5}
\twolines{\textbf{k-Intervals}}{}
& Non-adaptive
& \twolines{$O(1/\epsilon^4)$}{\cite{BBBY12}}
& \twolines{$O(1/\epsilon^4)$}{\cref{thm:intro-intervals}}
& \twolines{$\Omega(\sqrt{k})$}{\cite{BBBY12}} \\
\cline{3-5}
\end{tabular}

\caption{A comparison of bounds (total sample and query complexity)
for testing various classes of boolean functions.
The $\widetilde O(\cdot)$ notation hides factors polylogarithmic
in its argument.
When $\epsilon$ does not
appear in the bound, the big-$O$ notation hides a factor $\poly(1/\epsilon)$.
All results are for
singleton $\bbD$ comprising the uniform distribution over $\zo^n$
(or $[0,1]$ for $k$-Intervals),
except when $\bbD$ is the
class of product distributions over $\R^n$ in the second row.
}
\label{table:quant}
\end{table}

\subsubsection{Quantitative Results}

To describe our results quantitatively, it is convenient to present them in a table; see
\cref{table:quant}. Note that each row of the table shows that
distribution-robust testing is more efficient than distribution-free testing. Moreover,
as many rows exhibit, distribution-robust testing often achieves the same efficiency as
the distribution-specific testing result but with a stronger qualitative guarantee.

While most of our results are positive, we also show a lower bound of
$2^{\widetilde \Omega(n^{1/4})}$ for distribution-robust testing of
monotonicity. This lower bound exemplifies another reason why the
distribution-robust model is interesting, which is that it can reveal a new sort
of brittleness or fragility in property testers. It is well known that that
there can be large gaps in complexity between algorithms designed for specific
distributions and those which must be correct on arbitrary distributions (\eg
the known $O(\sqrt n)$ vs.~$2^{\Theta(n)}$ bounds for testing monotonicity in
the uniform vs.~distribution-free models). Our lower bound shows that a similar
exponential gap appears even if the tester doesn't need to be \emph{correct} on
arbitrary distributions, but only needs to \emph{detect} that it is being
used ``out of distribution''\!, where it has no correctness guarantee.

\subsubsection{Companion Paper: Halfspaces}
\label{section:intro-companion}

Whereas the goal of the present paper is to give a general introduction to the distribution-robust
testing model, including some initial positive results, a companion paper \cite{CFHS26halfspaces}
gives a more focused and technically challenging study of a single property: halfspaces. 

A halfspace is a function $f \colon \R^n \to \zo$ of the form $f(x) = \mathbf{1}[\inn{w,x} \geq t]$ for
some vector $w \in \R^n$ and threshold $t \in \R$. There are halfspace testers  with respect to
the standard Gaussian, or the uniform distribution over $\zo^n$, that use only
$\poly(1/\epsilon)$ queries \cite{MORS:10}. If only samples are allowed, $\widetilde \Theta(\sqrt
n)$ samples are necessary and sufficient when the distribution is assumed to be the standard Gaussian \cite{BBBY12}
or merely rotation-invariant \cite{Harms19}. On the other hand, distribution-free halfspace testing
requires $\wt \Omega(n)$ samples \cite{BFH21} and essentially the same number of queries \cite{CP22}.
\looseness=-1

The companion paper \cite{CFHS26halfspaces} shows that distribution-robust halfspace testing can be
done with $\widetilde O(n^{2/3})$ queries, significantly better than $\widetilde \Omega(n)$ required
for distribution-free testing. On the other hand, it shows that
sample-based distribution-robust testers with
respect to the standard Gaussian require $\widetilde \Omega(n^{2/3})$ samples, not as efficient as
the $\widetilde \Theta(\sqrt n)$ bound for distribution-specific testing.

\subsection{Techniques and Theorem Statements}
\label{section:intro-techniques}

Let us now describe our algorithmic techniques. 
We present four general techniques, from which
most of our results follow as corollaries or consequences. The most important of these is
\emph{distribution-picky testing}: all of our algorithms (except the one sample-based algorithm)
work by a general reduction from distribution-robust testing to distribution-picky testing.
The remaining three techniques are:
\begin{enumerate}
\item Testing by function reconstruction (\cref{section:intro-reconstruction,section:simulation}),
    which is a broad generalization of the classic testing-by-learning approach of \cite{GGR98}.
    Testing-by-learning immediately gives us some initial non-adaptive testers for monotonicity,
    conjunctions, and decision lists.
\item Approximation by juntas (\cref{section:intro-juntas,section:juntas}), which allows us to
    design more efficient function reconstructors for conjunctions and decision lists, and gives
adaptive algorithms for these classes that improve on the earlier non-adaptive ones.
\item Distribution testing and refinements thereof
    (\cref{section:intro-distribution-testing,sec:DRT-via-DT-and-DST}), which
    are the starting point toward our main result for unions of intervals
    (obtained via a targeted argument).
\end{enumerate}

Our sample-based results for conjunctions use ad-hoc arguments
(\cref{section:intro-sample,section:sample-based}).

Finally, we describe our lower bounds for non-adaptive monotonicity testing in
\cref{section:intro-lb,sec:monotone-lower}.

\subsubsection{Algorithmic Technique 0: Distribution-Picky Testing}
\label{section:intro-picky}

A challenge with designing distribution-robust testers is that, unlike standard testers, we may
encounter the following puzzle: upon seeing query and sample results that we would not expect in the
case $f \in \bbP$ and $\cD \in \bbD$, it can be unclear whether we are allowed to $\Abstain$
(because $\cD \notin \bbD$) or we must $\Reject$ (because $f \notin \bbP$ but $\cD \in \bbD$).
Accordingly, the most helpful concept for distribution-robust testing that we have found so far is
to define a weaker model called \emph{distribution-picky testing}, which we show simplifies this
choice in some situations. All of our upper bounds, except the one sample-based algorithm, will use
this concept (and the upper- and lower-bounds for halfspaces in the companion paper
\cite{CFHS26halfspaces} use it too).

\begin{definition}[Distribution-Picky Testing]
\label{def:intro-picky}
Let $\cX$ be a domain, let $\bbP$ be a property (\ie a set) of functions $\cX \to \zo$, and let $\bbD$
be a set of probability distributions over $\cX$. As input, the algorithm is given parameters
$\epsilon, \delta > 0$, query access to a function $f \colon \cX \to \zo$ and
access to labeled samples $(\bm{x}, f(\bm{x}))_{\bm{x} \sim \cD}$
for some probability distribution $\cD$.
Its output should, with probability at least $1-\delta$,
satisfy:
\begin{enumerate}
    \item \textbf{Completeness.} If $f \in \bbP$ \emph{and} $\cD \in \bbD$, the algorithm
outputs $\Accept$;
    \item \textbf{Soundness.} If $f$ is $\epsilon$-far from $\bbP$ under $\cD$, the algorithm
outputs $\Reject$.
\end{enumerate}
We call an algorithm satisfying this definition a \emph{$\bbD$-picky} tester for $\bbP$, or a distribution-picky tester for $\bbP$ with respect to $\bbD$.
\end{definition}

Compared to a distribution-free tester, a distribution-picky tester is only required to
output $\Accept$ under the additional condition that $\cD \in \bbD$ (hence it is ``picky'' about the distributions it accepts).
Compared to a distribution-robust tester, it is allowed to output
$\Reject$ when $\cD \not\in \bbD$, even if $f \in \bbP$.
A distribution-robust
tester can be transformed into a distribution-picky tester by replacing every $\Abstain$ with a
$\Reject$, so lower bounds on distribution-picky testing imply lower bounds for distribution-robust
testing.

Less obvious is that, when queries are allowed, we can also turn distribution-picky testers for
singleton classes $\bbD = \{ \cD^* \}$ (as well as more general classes that admit ``sample
simulators'', see \cref{def:sample-simulator}) into distribution-robust testers. The idea is:
\begin{enumerate}
\item Run the distribution-picky tester $O(1)$ times using the input distribution $\cD$;
\item Run the distribution-picky tester $O(1)$ times using the input distribution $\cD^*$ (known to the algorithm
since it is fixed);
\item Output their majority answers if the two
experiments have similar acceptance probability, otherwise output $\Abstain$.
\end{enumerate}
The resulting tester satisfies the assertiveness condition of distribution-robust testing
because Steps~1 and~2 are identical when $\cD=\cD^*$; its soundness is
immediate from the soundness of the picky tester in Step~1; and its completeness comes from Step~2,
hinging on the fact that the condition $f \in \bbP$ is independent from the input distribution.
This reduction greatly simplifies the task of designing distribution-robust testers. Let
us move on to our techniques for designing picky testers.

\subsubsection{Algorithmic Technique I: Picky Testing by Function Reconstruction}
\label{section:intro-reconstruction}

A staple of property testing is the \emph{testing-by-learning} technique of \cite{GGR98}. The idea
is that, for any class $\bbP$ of functions $\cX \to \zo$ and any class of distributions $\bbD$
over $\cX$, if there is a \emph{proper learner} for $\bbP$ under $\bbD$, with query cost
$q(\epsilon)$, then there is a distribution-specific tester for $\bbP$ under $\bbD$ with cost
$q(\epsilon/2) + O(1/\epsilon)$. Recall that this works as follows:
\begin{enumerate}
\item Run the proper learning algorithm on input $f$ and distribution $\cD \in \bbD$ to get function
        $h \in \bbP$. If $f \in \bbP$ then this will (with high probability) satisfy $\dist_\cD(f,h)
        < \epsilon/2$.
\item Query $f$ and $h$ on $O(1/\epsilon)$ random points $\bm x \sim \cD$ to test that
$\dist_\cD(f,h) < \epsilon$.
\end{enumerate}
But, in fact, we can observe that this is stronger than a distribution-specific tester: it is
distribution-picky! The correctness of step 2
for the case when $f$ is far from $\bbP$ under $\cD$
does not require that $\cD \in \bbD$, so the tester
only passes if $\dist_\cD(f,h) < \epsilon$, regardless of whether $\cD \in \bbD$. This means that,
using our reduction above, we can immediately get distribution-robust testers for singleton classes
$\bbD = \{\cD^*\}$ that are significantly more efficient than distribution-free testers, as long as
proper learning under $\cD^*$ is more efficient than distribution-free testing.

This is the case for monotone functions, conjunctions, and decision lists under $\cD^* =
\Unif(\zo^n)$, which immediately grants us the following results. Using the proper learner of
\cite{LRV22} for monotone functions:

\begin{restatable}{theorem}{thmintromonounif}
\label{thm:intro-mono-unif}
There is a non-adaptive $\{\Unif(\zo^n)\}$-robust tester for monotone functions $\zo^n \to \zo$
with query, sample, and time cost $n^{O(\sqrt n / \epsilon)}$.
\end{restatable}

This is compared with the $2^{\Omega(n)}$ lower bound for distribution-free testing of monotone
functions \cite{HalevyKushilevitz:07}. Next, using a straightforward approach to learning conjunctions and decision lists (\Cref{fact:unif-learn-conj-DL}) we obtain:

\begin{restatable}{theorem}{thmintroconjdlna}
\label{thm:intro-conj-dl-na}
\label{thm:DR-conj-DL-via-learning}
The classes of conjunctions and 1-decision lists have
non-adaptive
distribution-robust testing-by-learning
algorithms, with respect to $\bbD = \{\Unif(\zo^n)\}$,  with sample complexity 
$O(1/\epsilon)$
and non-adaptive query complexity
$O\left( \frac{\log(n) \log(1/\epsilon)}{\epsilon}\right)$.
\end{restatable}
This is an exponential improvement over the distribution-free bounds of $\widetilde \Theta(n^{1/3})$
for conjunctions \cite{CX16} and $\widetilde \Omega(\sqrt n)$ for 1-decision lists \cite{CFP24}. But
we can also go beyond basic testing-by-learning; here are two ways:

\textbf{Function reconstruction.} The first way is by generalizing proper learning to what we call
\emph{function reconstruction} (\cref{def:function-reconstruction}). For a property $\bbP$ and input
function $f$, a function reconstructor is an algorithm which draws samples and makes queries to $f$,
and then outputs a \emph{query simulator} $G_f$. The query simulator takes input queries $x$, makes
its own queries to $f$, and then outputs a value $g(x)$. With high probability, this function $g$
should satisfy
\begin{enumerate}
    \item $\dist_\cD(g,\bbP) < \epsilon$ (\ie $g$ itself is close to the property); and
    \item If $f \in \bbP$ and $\cD \in \bbD$, then $\dist_\cD(f,g) < \epsilon$.
\end{enumerate}
This is just like a proper learner except that the output is only \emph{close} to proper, and also
it does not need to output an explicit description of $g$, only an algorithm that simulates it. This
allows us to use a variety of techniques from the literature on local correction to design more
efficient function reconstructors; we will elaborate on this when we discuss Algorithmic Technique II.

\textbf{Sample simulators.} The second way to improve on the basic testing-by-learning approach is
to allow some non-singleton classes of distributions $\bbD$. Given samples from a distribution
$\cD$, a \emph{sample simulator} produces samples from another distribution $\cD' \in \bbD$ such
that, if $\cD \in \bbD$, then $\cD = \cD'$. For example, if $\bbD$ is the class of product
distributions over $\bbR^n$, then by drawing $n$ samples $\bm x^{(1)}, \dotsc, \bm x^{(n)} \sim \cD$
from an arbitrary distribution (not necessarily product), we can combine them into one sample point
$(\bm x^{(1)}_1, \dotsc, \bm x^{(n)}_n)$ which is distributed as the product distribution with the
same marginals as $\cD$. As long as we have a sample simulator, we can do the reduction from
distribution-robust testing to distribution-picky testing.

In this way, applying a proper learning algorithm of \cite{HarmsYoshida22,Black24} for monotone
functions $\bbR^n \to \zo$ under the class $\Prodreal$ of product distributions, we can improve our
result for monotonicity:

\begin{theorem}
\label{thm:intro-mono-product}
There is a non-adaptive distribution-robust monotonicity tester over $\R^n$, with respect to the set
$\Prodreal$ of product distributions, using $2^{\widetilde O(\sqrt n / \epsilon)}$ samples and
queries.
\end{theorem}

\subsubsection{Algorithmic Technique II: Approximation by Juntas}
\label{section:intro-juntas}

While proper learning is a useful first example of function reconstruction,
there are natural cases where more efficient reconstruction is
possible. Using ideas from ``testing by implicit learning''
\cite{S10survey} and local correction, we show that the classes of conjunctions and 1-decision lists
admit function reconstructors, and hence distribution-robust testers, with
essentially optimal $\wt O(1/\epsilon)$ cost.

Specifically,
several prior works \cite{DLM+:07,AlonW12,Bshouty20} have provided methods of roughly the following
form: Let $\bbJ \subseteq \bbP$ be a subset of the function class that depend on only a small number
of variables $k$, \ie the subclass of $k$-juntas within $\bbP$. Given a function $f$ that is in
$\bbP$ (or maybe just \emph{close} to $\bbP$), output a procedure which, given a query $x$, makes
some queries to $f$, and then outputs the value $g(x)$ for some $g \in \bbJ$ that is close to $f$
under the uniform distribution. In other words, the method provides query access to a $k$-junta $g$
that approximates $f$. For example, \cite{AlonW12} construct \emph{self-correctors} (which query $f$
on a uniformly random set of points to compute $g(x)$), and \cite{Bshouty20} uses
\emph{restrictions} (identify a subset $S$ of the variables and permanently fix their values).

These ``local correction'' algorithms assume that $f$ itself is close to $\bbP$, which means that, if
$f$ is \emph{far} from $\bbP$, the simulated function $g$ is not guaranteed to have the properties
we require from the output of a function reconstructor. By combining these procedures with
\emph{testers} for the subclass $\bbJ$ (which can be much cheaper than testers for the whole class
$\bbP$), we can ensure that the simulated function $g$ is in $\bbP$ (or close to it) and satisfy the
requirements of a function reconstructor.

Using adaptive algorithms of \cite{Bshouty20}, we get adaptive testers for conjunctions and
$1$-decision lists that are more efficient than those from testing-by-learning, and in particular
are independent of $n$:

\begin{theorem}
\label{thm:intro-conj-dl-adaptive}
There are distribution-robust testers for conjunctions and 1-decision lists over $\zo^n$,
with respect to $\bbD = \{ \Unif(\zo^n) \}$, using $\widetilde O(1/\epsilon)$ 
samples and adaptive queries.
\end{theorem}

Since $q$ adaptive queries can be simulated with $2^q$ non-adaptive ones, this also gives
non-adaptive testers with query cost $\smash{2^{\wt O(1/\epsilon)}}$ for these classes, compared with $O(\log
n)$ above.  Using the self-correctors of \cite{AlonW12}, we also give a \emph{non-adaptive} tester
for conjunctions that is only quadratically worse than the adaptive one:

\begin{restatable}{theorem}{thmintrononadaptiveconj}
    \label{thm:intro-non-adaptive-conjunction-tester}
    There is a distribution-robust tester for conjunctions over
    $\zo^n$, with respect to $\bbD = \{\Unif(\zo^n)\}$, using
    $O(1/\epsilon)$ samples and $\wt O(1/\epsilon^2)$ non-adaptive queries.
\end{restatable}

\subsubsection{Algorithmic Technique III: Distribution Testing}
\label{section:intro-distribution-testing}

It is a good exercise to show that, equipped with a distribution tester for a singleton
$\bbD = \{\cD^*\}$, \ie an identity tester, we can turn a distribution-specific tester for $\bbP$
into a distribution-picky (and therefore distribution-robust one): informally, an identity
tester ensures that $\dist_\TV(\cD ,\cD^*) < \epsilon$ (else we abstain), and in
this case a $\{\cD^*\}$-specific tester can stand in for a $\{\cD\}$-specific tester.

More interestingly, one can replace a full-fledged identity tester with a weaker
variant recently systematized in \cite{BDP:26distinguishers} that
is only ``sensitive'' to
a class of \emph{distinguishers} (here, $\bbP$).
Very briefly, the weaker variant only needs to reject distributions
$\cD$ whose non-membership in $\bbD$ is witnessed by a distinguisher $h \in \bbP$
via $\abs{\bE_\cD[h] - \bE_{\cD^*}[h]} > \epsilon$, whereas standard distribution
testing must work against all Boolean-valued distinguishers $h$.
Together with results for testing distributions against $k$-interval
distinguishers \cite{DiakonikolasKN15} and testing unions of intervals under
$\Unif([0,1])$ \cite{BBBY12},
this gives a
distribution-robust tester for unions of $k$ intervals using $O(\sqrt{k})$
queries and samples.

It turns out that we can improve upon the above black-box result with a targeted
argument. We extend the structural result of \cite{BBBY12} for
distribution-specific testing, linking distance to the class of unions of
$k$ intervals with a notion of noise sensitivity, to get our main result for
this class:

\begin{restatable}{theorem}{thmintrointervals}
\label{thm:intro-intervals}
For every distribution $\cD^*$ over $[0,1]$, there is a $\{\cD^*\}$-robust tester
for unions of $k$ intervals, with non-adaptive query and sample cost $O(1/\epsilon^4)$.
\end{restatable}

\subsubsection{Sample-Based Testing of Conjunctions}
\label{section:intro-sample}

Not all of our results follow from general techniques. In particular, to separate
sample-based distribution-robust testing from distribution-free testing, we give an ad-hoc sample-based
tester for conjunctions which matches the query complexity of the testing-by-learning
\cref{thm:intro-conj-dl-na}:

\begin{theorem}
\label{thm:intro-conj-sample}
There is a sample-based distribution-robust tester for conjunctions over $\zo^n$, with respect to
$\bbD = \{ \Unif(\zo^n) \}$, using $O\left(\frac{\log(n)\log(1/\epsilon)}{\epsilon}\right)$
samples. On the other hand, any sample-based tester requires $\Omega(\log n)$ samples for
sufficiently small constant $\epsilon > 0$.
\end{theorem}

This is an exponential improvement over the distribution-free bound of $\widetilde \Theta(n^{1/3})$
\cite{CX16}, even when the distribution-free tester is allowed to use queries. Our tester is quite
simple; basically, it must only rule out the $n^{O(\log(1/\epsilon))}$ conjunctions of length at
most $\log(1/\epsilon)$, since larger conjunctions are satisfied with probability
$< \epsilon$ under $\Unif(\zo^n)$, and the tester may abstain if $\cD$
violates this condition.
\looseness=-1

The lower bound is simple and holds even for distribution-specific testing: it is known that it
requires at least $\Omega(\log n)$ samples to distinguish a random dictator function (which is a trivial
conjunction) from a random function under $\Unif(\zo^n)$.

\subsubsection{Lower Bound for Testing Monotonicity}
\label{section:intro-lb}

Our argument is inspired by the $2^{\Omega(n^{1/4})}$ lower bound for non-adaptive
\emph{tolerant} (distribution-specific) monotonicity testing from \cite{CDLNS24}.
The main tool is random Talagrand DNFs: roughly, these are used to obtain approximate partitions of the Boolean cube into many incomparable regions that index into subcubes over a randomly chosen set of ``action'' variables. Using the randomness of the choice of action variables, constructions can hide violations
of monotonicity in these subcubes so that, informally, for any non-adaptive algorithm most pairs of queried
points are either too close together or in different regions, and both cases
are nearly uninformative.

In \cite{CDLNS24}, the hard input $f$ is close to or far from monotone
(on average) in each subcube,  which gives a tolerant testing lower bound. For
us, each subcube is either perfectly monotone or far from monotone, which
naively sounds easier to distinguish. On the other hand, since our NO
distribution $\cD$ need not be uniform, we hide the violations in a small number
of subcubes which are hard to detect, yet well-represented under $\cD$,
so that the construction is far from monotone.

\begin{theorem}
\label{thm:intro-mono-lb}
Any non-adaptive distribution-robust tester for  monotone Boolean functions over $\zo^n$ with respect to $\bbD=\{\Unif(\zo^n)\}$ must make at least $2^{\Omega(n^{1/4})}$ queries or samples.
\end{theorem}

\subsection{Discussion and Open Problems}

To begin with, it is interesting that we did not yet
find any use for the techniques from testable learning, despite the conceptual similarity; one
reason is that the goal of property testing is to achieve significantly better complexity than the
associated learning algorithms;
a second reason is that testable learning is essentially only interesting in the agnostic
rather than realizable setting, whereas we show that distribution-robust testing is already
interesting in the standard (non-tolerant) setting of property testing; and a third
reason is that most testable learning results aim to
improve time complexity rather than sample  or query complexity.

As highlighted in \cref{figure:qual}, some general questions remain open about adaptive testers:

\begin{problem}
Show a separation between adaptive and non-adaptive distribution-robust testing.
\end{problem}

\begin{problem}
Show a separation between adaptive distribution-specific and distribution-robust testing.
\end{problem}

A more specific question about adaptive testing is whether one can improve on the $2^{\widetilde
O(\sqrt n)}$ bound for non-adaptive monotonicity testing by using adaptivity:

\begin{problem}
Is there an adaptive $\poly(n)$-query distribution-robust tester for monotonicity?
\end{problem}

One of the questions we were most interested in is distribution-robust halfspace
testing. The companion paper \cite{CFHS26halfspaces} separates this problem from
distribution-free testing (using non-adaptive queries) and distribution-specific
testing (using samples), but does not yet achieve tight bounds.

\begin{problem}
\label{problem:halfspaces}
What is the complexity of distribution-robust halfspace testing?
\end{problem}

We include in the companion paper \cite{CFHS26halfspaces} some more discussion of aspects
of \cref{problem:halfspaces}, including conjectured bounds informed by AI outputs.

Another concrete question concerns non-adaptive $k$-junta testing. This class
has distribution-free, and hence distribution-robust, testers with query and
sample complexity $\poly(k/\epsilon)$ \cite{CLSSX18,Bshouty19}. However, for
\emph{non-adaptive} algorithms, distribution-free testing requires
$2^{\Theta(k)}$ queries \cite{CLSSX18} while the distribution-specific query
complexity is $\tilde{\Theta}(k^{3/2})$ \cite{Blais08,CSTWX17}.

\begin{problem}
    Is there a non-adaptive distribution-robust $k$-junta tester with $\poly(k)$
    complexity?
\end{problem}

Finally, our theory of sample-based distribution-robust testing is still
incipient: \cref{thm:intro-conj-sample} gives an upper bound for conjunctions,
and the companion paper \cite{CFHS26halfspaces} gives a lower bound for
halfspaces. It remains to further chart the landscape of results and techniques
in this model.

\subsection{Notation and Conventions}
\label{sec:notation-conventions}

Throughout this paper, $\cX$ will be the domain of the Boolean functions that are
being tested. Usually we will have either $\cX=\zo^n$, $\cX = \R^n$, or $\cX
= [0,1]$. For a set $S$ we write $\Unif(S)$ to denote the uniform distribution
over $S$.

We write $\Delta(\cX)$ for the set of all probability distributions over $\cX$.
We write $\dist_\TV(\cD, \cE)$ for the total variation distance between
probability distributions $\cD$ and $\cE$. For an event $E$, we write $\neg E$
for the complement (negation) of $E$. We denote random variables by boldface
symbols such as $\bm{x}, \bm{A}$.

\section{Algorithmic Technique 0: Distribution-Picky Testing}

Recall the motivation for distribution-picky testing (\cref{def:intro-picky})
discussed in \cref{section:intro-picky}: it can be conceptually challenging to
decide, as a designer of distribution-robust testing algorithms, whether the
algorithm should output $\Reject$ or $\Abstain$ upon seeing data that differs
from what we would expect under the ``perfect'' case of a good distribution $\cD
\in \bbD$ and good function $f \in \bbP$. The distribution-picky testing model
deals with this issue by simplifying the palette of outputs back down to $\Accept$
and $\Reject$: a distribution-picky tester is only required to accept when $f
\in \bbP$ and $\cD \in \bbD$ (the ``perfect'' case), and only required to reject
when $\dist_\cD(f, \bbP) > \epsilon$.

We require the concept of a \emph{sample simulator}. 

\begin{definition}[Sample Simulator]
\label{def:sample-simulator}
Let $\bbD$ be a class of distributions. A \emph{sample simulator} for $\bbD$ with simulation cost
$k$ is an algorithm which, given sample access to an arbitrary distribution $\cD$, provides sample
access to another distribution $\cD' \in \bbD$; specifically, upon request, the algorithm draws
$k$ samples from $\cD$ and produces 1 sample from $\cD'$. We require that, if $\cD \in \bbD$, then
$\cD' = \cD$.
\end{definition}

Most often, we will use a trivial sample simulator: when $\bbD = \{\cD^*\}$ is a singleton,
we just sample from the (known) distribution $\cD^*$.

\begin{algorithm}[ht]
    \caption{Reduction to Picky Testing using Sample Simulation}     
    
    \begin{algorithmic}
        \State \textbf{Requirements}:   
        \State \qquad $\bullet$ $\cR$ is a sample simulator for $\bbD$ with simulation
            cost $k$.
        \State \qquad $\bullet$ $\cT$ is a distribution-picky tester for $\bbP$
        with respect to $\bbD$.
        \State \textbf{Input}:
        \State \qquad $\bullet$ Query access to $f : \cX \to \zo$, 
                \State \qquad $\bullet$ Access to i.i.d.~samples $(\bx,f(\bx))_{\bx \sim \cD}$,
        \State \qquad $\bullet$ Closeness and failure parameters $\epsilon,\delta \in (0,1)$.
        \Procedure{DRT-via-DPT}{}
            \State 1. Let $\cD' \in \bbD$ be the distribution simulated by $\cR$
                    given sample access to $\cD$.
            \State 2. Set $r = O(\log(1/\delta))$ with an appropriate constant.
                \State \phantom{2.} Run $r$ iterations of $\cT$ on $f$ and $\cD$, with distance parameter $\epsilon$.
                \State \phantom{2.} Let $\bm{t}_1$ be the fraction of times it outputs $\Accept$.
                \Statex
                \Comment{Uses labeled samples from $\cD,f$ and queries to $f$.}
            \State 3. Run $r$ iterations of $\cT$ on $f$ and $\cD'$,
                                with distance parameter $\epsilon$.
                \State \phantom{3.} Let $\bm{t}_2$ be the fraction of times it outputs $\Accept$.
                \Statex \Comment{Uses sample simulator $\cR$ and queries to $f$.}
            \State 4. If $|\bm{t}_1 - \bm{t}_2| > 0.01$ then output $\Abstain$.
            \State \phantom{4.} Otherwise output $\Accept$ if $\bm{t}_1 > 1/2$ and $\Reject$ if $\bm{t}_1 \leq 1/2$. 
        \EndProcedure
    \end{algorithmic}
    \label{alg:DRT-via-DPT}
\end{algorithm}

The next lemma provides our first general technique for distribution-robust testing: it says that if we have a sample simulator and a distribution-picky tester, then we get a distribution-robust tester. Most of our algorithmic results will actually be distribution-picky testers, and we will use \Cref{lem:reduction-to-picky} to obtain distribution-robust testers from them.

\begin{lemma}[Reduction to Picky Testing]
\label{lem:reduction-to-picky}
Let $\bbP$ be any property of functions $\cX \to \zo$ and $\bbD$ any class of distributions over
$\cX$. Let $\cR$ be a sample simulator with simulation cost $k$ and let $\cT$ be a
distribution-picky tester for $\bbP$ with respect to $\bbD$ with sample cost $s(\epsilon,\delta)$ and query cost
$q(\epsilon, \delta)$. Then \cref{alg:DRT-via-DPT} is a $\bbD$-robust tester for $\bbP$ with
sample cost
$O\pbra{(k+1) \cdot s\pbra{\epsilon, \frac{1}{4}} \log\pbra{\frac{1}{\delta}}}$ and query cost
$O\pbra{\pbra{q\pbra{\epsilon, \frac{1}{4}} + s\pbra{\epsilon, \frac{1}{4}}} \cdot \log\pbra{\frac{1}{\delta}}}$.
If $\cT$ is non-adaptive, then so is \cref{alg:DRT-via-DPT}.
\end{lemma}
\begin{proof}
The sample and query complexity are immediate from inspection of the algorithm
(in particular, each
labeled
sample from $\cD'$
can be obtained using
$k$ samples from $\cD$
via
the simulator $\cR$, plus one query to $f$),
and so is the non-adaptivity claim.
Let us now establish
correctness.

Let $p_1$ be the probability that $\cT$ outputs $\Accept$ on $f$ and $\cD$, and let $p_2$ be the
probability that $\cT$ outputs $\Accept$ on $f$ and $\cD'$. Let $\bm t_1, \bm t_2$ be the random
variables which are the observed fraction of times these subroutines output $\Accept$ in Steps~2 and~3, respectively. By choosing $r = O(\log(1/\delta))$ appropriately, the Chernoff bound ensures that
$|p_1 - \bm t_1| < 0.005$ and $|p_2 - \bm t_2| < 0.005$ with probability at least $1-\delta$. Let us
assume this event occurs.

\textbf{Completeness.} Suppose $f \in \bbP$. Since the simulated distribution $\cD' \in \bbD$, the
distribution-picky tester $\cT$ running on $\cD'$ will output $\Accept$ with probability $p_2 \geq
3/4$, so $\bm t_2 > 3/4 - 0.05$.  If $\bm t_1 \leq 1/2$ then $|\bm t_1
- \bm t_2| > 0.2$ and the algorithm will output $\Abstain$, so it is not possible for the algorithm
to output $\Reject$.

\textbf{Soundness.} Suppose $\dist_\cD(f,\bbP) > \epsilon$. Then the distribution-picky tester $\cT$
running on $\cD$ will output $\Accept$ with probability $p_1 \leq 1/4$, so $\bm t_1 < 1/4
+ 0.05$, so it is not possible for the algorithm to output $\Accept$.

\textbf{Assertiveness.} Suppose $\cD \in \bbD$. By the properties of $\cR$, we have $\cD = \cD'$.
Therefore $p_1 = p_2$ since the two runs of the tester $\cT$ are running on identical inputs. Then
$|\bm{t}_1 - \bm{t}_2| < 0.01$, so the algorithm will not output $\Abstain$.
\end{proof}

\section{Algorithmic Technique I: Function Reconstruction}
\label{section:simulation}

Thanks to \cref{alg:DRT-via-DPT}, designing distribution-robust testers now reduces to designing
distribution-picky testers. 
In this section we give a general technique for designing distribution-picky testers which generalizes the classic testing-by-learning reduction of \cite{GGR98} (it is not hard to see that the basic \cite{GGR98} testing-by-learning technique gives distribution-picky testers).
We will see in this section that
testing-by-learning gives good upper bounds on distribution-robust testers that can be much more
efficient than distribution-free testing. Looking ahead, this generalization can also produce testers that are
much more efficient than testing-by-learning; we will see this later, in
\cref{sec:junta-approx}.

We will need the notion of a \emph{query simulator}, which is an algorithm that uses queries to $f$ to simulate queries to another function $g$:

\begin{definition}[Query Simulator]
\label{def:query-simulator}
For functions $f$ and $g$, a \emph{query simulator for $g$ based on $f$}, with simulation
cost $k$, is an algorithm which, given query $x$, makes at most $k$ queries to $f$ and then outputs
$g(x)$.
\end{definition}

We will also need the notion of a \emph{function reconstructor}. This is an algorithm which does some preprocessing to produce a query
simulator. For intuition, one can imagine a proper learning algorithm: A proper learning algorithm
for $\bbP$ (\cref{def:proper-learning}) uses samples and queries to output a function $g \in
\bbP$, and then queries to $g$ can be simulated without using any further queries to $f$. A function
reconstructor generalizes proper learning by allowing the simulator to continue querying $f$ to
process new queries, and also allowing the function $g$ to lie slightly outside $\bbP$.

\begin{definition}[Function Reconstructor] \label{def:function-reconstruction}
Let $\bbP$ be a class of functions $\cX \to \zo$ and let $\bbD$ be a class of distributions over
$\cX$. A \emph{function reconstructor} for $\bbP$ with respect to
$\bbD$, with sample cost $s(\epsilon,\delta)$,
query cost $q(\epsilon,\delta)$, and simulation cost $k$, is an algorithm whose input is:
\begin{itemize}
\item Access to i.i.d.~samples $(\bx,f(\bx))$ with $\bx \sim \cD$ where $\cD$ is an unknown
arbitrary distribution over $\cX$ and $f \colon \cX \to \zo$ is an arbitrary function;
\item Query access to the unknown function $f \colon \cX \to \zo$;
\item Closeness and failure parameters $\epsilon, \delta \in (0,1)$.
\end{itemize}
The algorithm draws at most $s(\epsilon,\delta)$ samples from $\cD$, makes at most
$q(\epsilon,\delta)$ queries to $f$, and then outputs a query simulator for a function $\bg$ based
on $f$, with simulation cost $k$, such that, with probability at least $1-\delta$, the following
properties hold:
\begin{enumerate}
    \item $\bg$ satisfies $\dist_\cD(\bg,\bbP) < \epsilon$; and
    \item if $f \in \bbP$ and $\cD \in \bbD$, then $\dist_\cD(f,\bg) < \epsilon$.
\end{enumerate}
\end{definition}

The following algorithm and lemma show that if we have a function reconstructor for a property $\bbP$ then we can get a distribution-picky tester for that property:

\begin{algorithm}[ht]
    \caption{Picky Testing using Function Reconstruction     }     
    
    \begin{algorithmic}
        \State \textbf{Requirements}:   
        \State \qquad $\bullet$ $\cF$ is a function reconstructor for $\bbP$ under $\bbD$.
        \State \textbf{Input}:
        \State \qquad $\bullet$ Query access to $f : \cX \to \zo$, 
        \State \qquad $\bullet$ Access to i.i.d.~samples $(\bx,f(\bx))_{\bx \sim \cD}$,
                        \State \qquad $\bullet$ Closeness and failure parameters $\epsilon,\delta \in (0,1)$.
        \Procedure{DRT-via-Function-Reconstruction}{}
            \State 1. Run the function reconstructor $\cF$ on $f$ and $\cD$ with
                    accuracy $\epsilon/8$ and confidence $7/8$.
                \State \Comment{Uses labeled samples from $\cD,f$ and queries to $f$.}
            \State 2. Let $\bg$ be the function whose simulator based on $f$ is produced by
                    the previous step.
            \State 3. Query $f(\bm x)$ and $\bg(\bm x)$ on $r = O(1/\epsilon)$ random points $\bm x \sim \cD$ drawn from $\cD$; let $\bt$ be the fraction of these queries with $f(\bm x) \neq \bg(\bm x)$.
                \State \Comment{Uses samples from $\cD$ and queries to $f$.}
            \State 4. If $\bt \leq \epsilon/2$, output $\Accept$, otherwise $\Reject$.
        \EndProcedure
    \end{algorithmic}
    \label{alg:DPT-via-Query-Simulator}
\end{algorithm}

\begin{lemma}[Testing by function reconstruction]
\label{lemma:testing-by-query-simulator}
Let $\bbP$ be any property of functions $\cX \to \zo$ and $\bbD$ any class of distributions.
If $\cF$ is a function reconstructor for $\bbP$ under $\bbD$ with sample cost $s(\epsilon,\delta)$, query
cost $q(\epsilon,\delta)$,
and simulation cost $k$,
then \cref{alg:DPT-via-Query-Simulator}
is a $\bbD$-picky tester for $\bbP$ with sample cost $s(\epsilon/8, 1/8) + O(1/\epsilon)$ and
query cost $q(\epsilon/8, 1/8) + O((k+1)/\epsilon)$.
\end{lemma}
\begin{proof}
The sample and query cost of the algorithm 
follow directly from
inspection of the algorithm, so let us turn our attention to correctness.

Let $\bd \define \dist_\cD(f,\bg)$.
By the Chernoff bound, if we use a sufficiently large constant in
the setting $r = O(1/\epsilon)$, then
the following event $E_1$ occurs with probability at least $7/8$:
if $\bd < \epsilon/8$ then $\bt < \epsilon/4$, and if $\bd > 7\epsilon/8$ then
$\bt > 3\epsilon/4$.
Additionally, let $E_2$ be the event that the output $\bg$ of the function reconstructor $\cF$
satisfies the two properties described in \cref{def:function-reconstruction}, which
occurs with probability at least $7/8$. We show below that the output of \cref{alg:DPT-via-Query-Simulator} is correct conditional on event $E_1 \cap E_2$, which occurs with probability
at least $3/4$ by a union bound.

\textbf{Completeness.} Suppose $f \in \bbP$ and $\cD \in \bbD$. By
event $E_2$,
we have $\bd = \dist_\cD(f,\bg) < \epsilon/8$, and thus by event $E_1$ we have
$\bm t < \epsilon/4$, so the algorithm correctly outputs $\Accept$.

\textbf{Soundness.} Suppose $\dist_\cD(f,\bbP) > \epsilon$.
Since $\dist_{\cD}(g,\bbP) < \epsilon/8$ by event $E_2$,
we must have $\bd = \dist_\cD(f,\bg) > 7\epsilon/8$. Then $\bm t > 3\epsilon/4$
by event $E_1$, so the algorithm correctly outputs
$\Reject$.
\end{proof}

We will use this general algorithm template to give both adaptive and non-adaptive
distribution-picky testers. Recall the definition of non-adaptivity:
\begin{definition}[Non-Adaptivity]
\label{def:adaptivity}
When a randomized algorithm is given query access to $f \colon \cX \to \zo$ and access to samples
$(\bm x, f(\bm x))_{\bm x \sim \cD}$, we say it is \emph{non-adaptive} if its queries and samples
can be made as follows:
\begin{enumerate}
    \item Draw a batch of labelled samples $(\bm x, f(\bm x))$.
    \item Based on these samples, choose (possibly randomly) a set $Q \subset \cX$ points.
    \item Query all points $x \in Q$ and then compute the output.
\end{enumerate}
\end{definition}

\begin{remark}[Non-Adaptive Picky Testing via Function Reconstruction]
    \label{rem:reconstruction-adaptivity}
    There is a subtlety in judging the adaptivity or non-adaptivity of
    \cref{alg:DPT-via-Query-Simulator} in general, since non-adaptivity of the function
reconstructor and the query simulator do not compose. In particular, we can have
    \begin{enumerate}
        \item A function reconstructor $\cF$ which is non-adaptive and outputs a query simulator $S$, such that
        \item The query simulator $S$ is non-adaptive (meaning that on each query $x$ it makes
non-adaptive queries to $f$ and outputs $g(x)$), and
        \item $S$ gets queried on a non-adaptively chosen set of query points $Q$,
    \end{enumerate}
        while still \cref{alg:DPT-via-Query-Simulator} being adaptive as a whole. This is because
        the simulator $S$ may depend on the query values made by $\cF$. We will make non-adaptivity
        arguments as needed.
\end{remark}

In the rest of this section we present some instantiations of \Cref{lemma:testing-by-query-simulator}. \Cref{sec:testing-by-learning} briefly notes  how proper learning yields distribution-robust-testing using this approach, and \Cref{sec:applic-monotone,sec:applic-conj-DL} give concrete applications to the classes of monotone functions, conjunctions, and decision lists.

\subsection{Testing by Learning} \label{sec:testing-by-learning}

\begin{definition}[Proper Learning]
\label{def:proper-learning}
Let $\bbP$ be any class of functions $\cX \to \zo$ and $\bbD$ any class of distributions over
$\cX$. A proper learner for $\bbP$ under $\bbD$ with sample complexity
$s(\epsilon,\delta)$ and query complexity $q(\epsilon, \delta)$, is an algorithm $\cL$ which is
given as input:
\begin{itemize}
\item  Access
        to i.i.d.~samples $(\bx,f(\bx))$ with $\bx \sim \cD$ where $\cD$ is an unknown arbitrary distribution over $\cX$ and $f \colon \cX \to \zo$ is an arbitrary function;
                \item Query access to the function $f \colon \cX \to \zo$;
    \item Closeness and failure parameters $\epsilon, \delta \in (0,1)$.
\end{itemize}
The algorithm makes at most $q(\epsilon,\delta)$ queries to $f$ and draws at most
$s(\epsilon,\delta)$ samples from $\cD$.  It outputs a function $\bg \in \bbP$ with the guarantee
that, if $f \in \bbP$ and $\cD \in \bbD$, then with probability at least $1-\delta$, the output $\bg$
satisfies $\dist_\cD(f,\bg) < \epsilon$.
\end{definition}

A proper learning algorithm for $\bbP$ under $\bbD$ may be transformed trivially into a function
reconstructor
with simulation cost 0
by simply outputting a query simulator for the learned function $\bg$. So we have the following:

\begin{lemma}[Distribution-Robust Testing by Learning]
\label{lem:testing-by-learning}
Let $\bbP$ be any class of functions $\cX \to \zo$,
let $\cD^*$ be any distribution over $\cX$,
and let $\bbD = \{ \cD^* \}$.
Suppose there is a proper learning algorithm for $\bbP$ under $\bbD$ with query
cost $q(\epsilon, \delta)$
and sample cost $s(\epsilon, \delta)$.
Then there is a $\bbD$-robust tester for $\bbP$ with sample cost
$O\pbra{s\pbra{\frac{\epsilon}{8}, \frac{1}{8}} + \frac{1}{\epsilon}}$
and query cost
$O\pbra{s\pbra{\frac{\epsilon}{8}, \frac{1}{8}}
+ q\pbra{\frac{\epsilon}{8}, \frac{1}{8}}
+ \frac{1}{\epsilon}}$.
Moreover, if the proper learning algorithm is non-adaptive then so is the
robust tester.
\end{lemma}
\begin{proof}
    \cref{lemma:testing-by-query-simulator} gives a $\bbD$-picky tester for
    $\bbP$, from which \cref{lem:reduction-to-picky} gives the desired
    $\bbD$-robust tester with the announced query and sample costs. If the
    learning algorithm is non-adaptive, then so is the
    $\bbD$-picky tester by using the fact that the query
    simulator makes no additional queries, and hence the $\bbD$-robust
    tester is non-adaptive by \cref{lem:reduction-to-picky}.
\end{proof}

\subsection{Application: Monotone Functions}
\label{sec:applic-monotone}

For our first application, we use testing-by-learning to get a distribution-robust tester for
monotonicity that is significantly more efficient than distribution-free testing.

Let us first show off the simplest application of our general algorithm: We can obtain a
distribution-picky tester for the singleton class $\bbD = \{ \Unif( \zo^n ) \}$ using the basic
testing-by-learning reduction of \cite{GGR98}. All we need is the following theorem:

\begin{theorem}[\cite{LRV22}]
\label{thm:proper-learning-monotone-cube}
There is a sample-based proper learner for the class of monotone functions $\zo^n \to \zo$
under $\bbD = \{ \Unif(\zo^n) \}$, with sample cost $n^{O(\sqrt n / \epsilon)}$, and time cost
polynomial in the number of samples. 
\end{theorem}

By applying \cref{lem:testing-by-learning}, we immediately obtain:

\begin{corollary}
There is a $\{\Unif(\zo^n)\}$-robust tester for monotone functions $\zo^n \to \zo$
with query, sample, and time cost $n^{O(\sqrt n / \epsilon)}$.
\end{corollary}

We can use the more general power of our algorithm, using sample simulators, to upgrade this result
to the domain $\bbR^n$, for the class $\Prodreal$ of product distributions over $\R^n$.

\begin{proposition}
For domain $\bbR^n$, there is a sample simulator for $\Prodreal$ with simulation cost $n$.
\end{proposition}
\begin{proof}
The simulator performs the following, given sample access to an arbitrary distribution $\cD$ over
$\bbR^n$. When a sample is requested:
\begin{enumerate}
\item Draw $n$ independent samples $\bm x^{(1)}, \dotsc, \bm x^{(n)} \sim \bbD$;
\item Output the value $\bm y \define (\bm x^{(1)}_1, \dotsc, \bm x^{(n)}_n)$, where $\bm x_i$ is
the $i^{th}$ coordinate of $\bm x \in \bbR^n$.
\end{enumerate}
The sample $\bm y$ is distributed as $\cD' \in \Prodreal$ whose marginals on each
coordinate are equal to the marginals of $\cD$. When $\cD$ is a product distribution, $\cD' = \cD$
as desired.
\end{proof}

Now it suffices to plug in a proper learner for monotone function $\bbR^n \to \zo$ under
$\Prodreal$. This follows by combining the results of \cite{HarmsYoshida22,Black24}. Since this
result is not stated explicitly in the literature (and requires some inspection of the proofs), we
give a proof sketch in \cref{section:appendix-monotone}.

\begin{restatable}{theorem}{thmmonotoneproduct}
\label{thm:proper-learning-monotone-product}
There is a sample-based proper learner for the class of monotone functions $\bbR^n \to \zo$
under $\Prodreal$, with sample cost $2^{\widetilde O(\sqrt n / \epsilon)}$.
\end{restatable}

We remark that the method of \cite{LRV22} for obtaining \emph{time} bounds that are polynomial in
the sample size do not immediately apply; we leave this as an open problem.

\subsection{Application: Conjunctions and Decision Lists} \label{sec:applic-conj-DL}

As another simple application of \Cref{lem:testing-by-learning}, we give distribution-robust
algorithms for testing conjunctions and 1-decision-lists under the uniform distribution on
$\zo^n$, i.e.~when the set of good distributions is the singleton $\bbD=\{\Unif(\zo^n)\}.$ We
recall the following well known facts about learnability of conjunctions and decision lists under
the uniform distribution (for completeness we sketch the simple proof in \Cref{ap:conjDL}):

\begin{restatable}{fact}{factuniflearnconjdl}
\label{fact:unif-learn-conj-DL}
Let $\cX = \zo^n$, let $\bbD=\{\Unif(\zo^n)\}$, and let $\bbP$ be either the class
$\bbP_{\conj}$ of all Boolean conjunctions over $\cX$ or the class $\bbP_{\DL}$ of all
1-decision-lists over $\cX.$ There is a proper nonadaptive query learner for $\bbP$ under
$\Unif(\zo^n)$ with query complexity $q(\epsilon,\delta) = O\pbra{{\frac {(\log n) \cdot
(\log(1/\epsilon)) + \log(1/\delta)}{\epsilon}}}$.
\end{restatable}

Combining \Cref{lem:testing-by-learning} and \Cref{fact:unif-learn-conj-DL},
we immediately obtain \cref{thm:intro-conj-dl-na}, restated below for convenience.

\thmintroconjdlna*

We remark that \Cref{thm:intro-conj-dl-na} has much better efficiency than any algorithms for distribution-free testing of conjunctions or 1-decision-lists,
even if adaptive queries are allowed,
as witnessed by the following polynomial lower bounds, 
(see \cite{CX16}, Theorem~1.2 and \cite{CFP24}, Theorem~1.2):
    There is an absolute constant $\epsilon>0$ such that any distribution-free $\epsilon$-testing algorithm for conjunctions (1-decision-lists, respectively) over $\zo^n$ must make $\tilde{\Omega}(n^{1/3})$ queries ($\tilde{\Omega}(n^{1/2})$ queries, respectively).
\Cref{thm:DR-conj-DL-via-learning} and these distribution-free lower bounds together establish
the separations depicted in \cref{figure:qual} between the distribution-robust
and distribution-free models with queries.
(As mentioned earlier, in \Cref{sec:junta-approx} we will give
a stronger separation for adaptive algorithms
by providing \emph{constant-query} distribution-robust testers for conjunctions and 1-decision-lists.)

\section{Algorithmic Technique II: Approximation by Juntas}
\label{sec:junta-approx}
\label{section:juntas}

A number of results for distribution-specific property testing with respect to
the uniform distribution \cite{DLM+:07,Bshouty20} have been obtained for
properties that have \emph{junta subclass approximators},  \ie which are such
that there is a subclass $\bbJ \subseteq \bbP$, consisting of $k$-juntas for
some suitable value of $k$, such that every $f \in \bbP$ is close to some junta
$g \in \bbJ$. In this section we give a general technique for obtaining
distribution-robust testing algorithms for such properties, and 
we apply this method to obtain
$\tilde{O}(1/\epsilon)$-query testers (with no dependence on the dimension
parameter $n$) for conjunctions and decision lists, improving on the
non-adaptive
distribution-robust testers for these classes that were given in
\Cref{thm:DR-conj-DL-via-learning}
at the cost of using adaptive queries.

\subsection{Adaptive Testers using Query Simulators}

We will use junta approximation techniques to prove the following:
\begin{theorem}
[Adaptive testing of conjunctions and 1-decision-lists]
\label{thm:more-efficient-DR-conj-DL}
Let $\bbP$ be either the class $\bbP_{\conj}$ of all Boolean conjunctions over $\cX=\zo^n$ or the class $\bbP_{\DL}$ of all 1-decision-lists over $\cX.$
There is an adaptive distribution-robust tester for $\bbP$ with respect to the uniform distribution over $\cX$ with sample complexity $\widetilde O(1/\epsilon)$ and query complexity $\widetilde O(1/\epsilon)$.
\end{theorem}

By \cref{lemma:testing-by-query-simulator}, it suffices to design function reconstructors for the
classes of conjunctions and decision lists. We accomplish this using results of \cite{Bshouty20}.
From that paper, we can obtain two procedures, which are similar for conjunctions and decision
lists:
\begin{enumerate}
\item An adaptive query algorithm which, given query access to $f \colon \zo^n \to \zo$, identifies a
small set $S \subseteq [n]$ of variables and a string $w \in \zo^n$, and defines a function
\[
    g(x) \define f(x_S \circ w_{\overline S}) ,
\]
where $x_S \circ w_{\overline S}$ denotes the string obtained by replacing the values of $x$ on the
coordinates $\overline S = [n] \setminus S$ with the values of $w$. This gives a query simulator
which, on query $x$, can compute the value $g(x)$ using 1 query to $f$. This procedure satisfies:
\medskip

    \begin{claim}
    \label{claim:bsh-simulator}
        There is an algorithm which, given query access to an arbitrary function $f \colon \zo^n
\to \zo$, makes at most $\widetilde O(1/\epsilon)$ queries and outputs a query simulator for a
function $g$. If $f$ is a conjunction, the following holds: $g$ is a conjunction of length at most
$O(\log(1/\epsilon))$, and with probability at least $1-\delta$, $\dist(f,g) < \epsilon$ under
$\Unif(\zo^n)$.
The same holds for decision lists.
    \end{claim}

\item A distribution-free tester for the subclass of functions $g$ obtained by the restrictions of
functions $f \in \bbP$ given above; in each case, these restricted functions are functions in $\bbP$
which depend on a small number of variables.

\medskip

    \begin{claim}[\cite{Bshouty20}, Theorem 33]
    \label{claim:bsh-tester}
        For every $K$,
        there is a distribution-free tester for the class of conjunctions of length at most $K$, with sample and query cost $\widetilde O(K/\epsilon)$.
    \end{claim}
    \begin{claim}[\cite{Bshouty20}, Theorem 39]
  For every $K$, there is a distribution-free tester for the
        class of 1-decision lists of length at most $K$, with sample and query
        cost $\widetilde O(K^2/\epsilon)$.
    \end{claim}
\end{enumerate}
For both conjunctions and decision lists, the first procedure is adaptive because it uses binary
search to identify coordinates $S$; any following queries to the function $g$ are therefore also
adaptive because they use queries to $f$ that depend on $S$.

These two procedures can be combined to get a function reconstructor (the formal claims of these two
procedures are stated in the proofs below, and more detail is given in
\cref{section:bshouty-appendix}, since \cite{Bshouty20} does not state some of these claims explicitly).
The following lemma suffices to establish
\cref{thm:more-efficient-DR-conj-DL} using \cref{lemma:testing-by-query-simulator}.

\begin{lemma}
\label{lemma:function-reconstruction-conj-dl}
There is a function reconstructor for conjunctions over $\bbD = \{\Unif(\zo^n)\}$ with sample and
query cost $\widetilde O(1/\epsilon)$, and simulation cost $1$. The same holds for decision lists.
\end{lemma}
\begin{proof}
We start with conjunctions.

    The function reconstructor performs the following,
    with $K = O(\log(1/\epsilon))$ being the length
    of the conjunction produced in the first claim:
    \begin{enumerate}
        \item Run the algorithm from the first claim with confidence parameter $\delta/2$ to obtain a simulator for a function $g$;
        \item Run the tester from the second claim with value $K$ with confidence parameter $\delta/2$, on function $g$ under distribution $\cD$. If it outputs $\Accept$ then output the
        simulator for $g$,
        otherwise output the simulator for the constant 0
        function.
    \end{enumerate}
    Suppose neither step fails; this occurs with probability at least $1-\delta$. If step 2 outputs $\Accept$, then
    the algorithm outputs the simulator for $g$, which
    must satisfy $\dist_\cD(g,\bbP) < \epsilon$ because the tester passed. If step 2 outputs $\Reject$, the algorithm outputs a conjunction. 
    
    If $f$ is a conjunction and $\cD = \Unif(\zo^n)$,
    then $g$ is a conjunction on $K$ variables, and $\dist_{\cD}(f,g) < \epsilon$ by the guarantee on the
    first algorithm. The tester in step 2 will $\Accept$,
    so the algorithm outputs $g$ as desired.

    For decision lists, we use the same argument, with the same claims as above, except with the
    word ``conjunction'' replaced with ``decision list''\!.
\end{proof}

\subsection{Non-Adaptive Tester for Conjunctions}
\label{sec:non-adaptive-conjunctions}

\cref{thm:more-efficient-DR-conj-DL} already implies non-adaptive testers for
conjunctions and 1-decision lists with sample complexity $\wt O(1/\epsilon)$ and
query complexity $2^{\wt O(1/\epsilon)}$, \ie independent of the domain size,
since any adaptive tester may be made non-adaptive by preemptively querying all
points that appear in its decision tree. In this subsection, we improve this
non-adaptive bound for conjunctions:

\thmintrononadaptiveconj*

We will again apply \cref{lemma:testing-by-query-simulator} (with the reduction to
distribution-picky testing in \cref{lem:reduction-to-picky}) by creating a function reconstructor.
One remark before we begin:

\begin{remark}
\label{remark:randomized-query-simulator}
Our function reconstructor will produce a \emph{randomized} query simulator, meaning that on each
query $x$, the simulator makes $k$ \emph{random} queries to $f$ and then, with probability at least
$2/3$, outputs the correct values $g(x)$. However, our \cref{def:query-simulator} and analysis of \cref{alg:DPT-via-Query-Simulator} assumed that the simulator always produces the correct answer. We can make the randomized query simulator satisfy this property by amplifying its error probability to $\delta \ll 1/q$ where $q$ is the number of queries to the simulator made by \cref{alg:DPT-via-Query-Simulator}, and then applying a union bound over all $q$ queries to bound the probability that any of them produce an incorrect answer. (This is done in the standard way of running the simulator $O(\log(1/\delta))$ independent times on each input $x$ and taking the majority). This incurs an extra multiplicative $O(\log q)$ query cost to the algorithm; in our application this is $O(\log(1/\epsilon))$.
\end{remark}

Our function reconstructor will use two tools. The first is a \emph{self-corrector}:

\begin{definition}[Self-Corrector, \cite{AlonW12}]
Let $\bbP$ be a class of functions $\cX \to \zo$. A \emph{self-corrector} $\cC$ for $\bbP$, with
error rate $\eta$, is an algorithm that is given query access to a function $f$. On any query $x$,
it produces an output $\bm b(x)$ which is random and independent of previous query values.  It
satisfies the property that, if $f$ is $\eta$-close to a unique function $g \in \bbP$ under
$\Unif(\cX)$, then on each query $x$, $\bm b(x) = g(x)$ with probability at least $2/3$.
\end{definition}

\begin{theorem}[\cite{AlonW12}]
For all $k$, there is a self-corrector for the class of $k$-juntas with any error rate $\eta <
2^{-k-3}$ and query cost $O(2^k)$, which  queries $f$ on a randomly chosen set of points depending only on the input point $x$.
\end{theorem}

The next is a distribution-specific tester for the class $\bbP_{\leq k}$
of conjunctions of length at most $k$:

\begin{theorem}[\cite{DLM+:07}, Appendix~C]
There is a distribution-specific non-adaptive tester for $\bbP_{\leq k}$ under $\Unif(\zo^n)$
with query cost $\poly(k) \cdot \widetilde O(1/\epsilon^2)$.
\end{theorem}

\begin{algorithm}[ht]
    \caption{Non-adaptive conjunction reconstructor}
    \begin{algorithmic}
        \State \textbf{Requirements}: For all $k$,
        \State \qquad $\bullet$ $\cT_k$ is a non-adaptive $\bbP_{\leq k}$ tester under $\Unif(\zo^n)$.
        \State \qquad $\bullet$ $\cC_k$ is a non-adaptive $k$-junta self-corrector under $\Unif(\zo^n)$,
            for any error rate $\eta < 2^{-k-3}$.
        \State \textbf{Input}:
        \State \qquad $\bullet$ Query access to $f : \zo \to \zo$;
        \State \qquad $\bullet$ Access to i.i.d.~samples $(\bx,f(\bx))_{\bx \sim \cD}$;
        \State \qquad $\bullet$ Closeness  parameter $\epsilon \in (0,1)$.
        \Procedure{Conjunction-Reconstructor}{}
            \State 1. Set $\wh{\bm{p}} \gets $ empirical estimate of $\Pru{\bm{x} \sim
            \cD}{f(\bm{x}) = 1}$ over $\Theta(1/\epsilon)$ draws.
            \Statex \Comment{Uses labeled samples from $\cD, f$.}
            \State 2. If ($\wh{\bm{p}} \le \epsilon/2$), output a query simulator for the
                constant 0 function (conjunction $x_1 \wedge \neg x_1$).
            \State 3. Let $k = \lfloor \log(4/\epsilon) \rfloor$, run $\cT_k$ on $f$
                with closeness parameter $\epsilon/64$.
            \State \phantom{3.} If the tester rejects, output a query simulator for the
                constant 0 function.
                \State \Comment{Uses queries to $f$.}
            \State 4. Output the $k$-junta self-corrector $\cC$ with input $f$ and error rate
                $\epsilon/64 < 2^{-k-3}$.
                        \EndProcedure
    \end{algorithmic}
    \label{alg:non-adaptive-conjunction-reconstructor}
\end{algorithm}

\begin{proof}[Proof of 
\cref{thm:intro-non-adaptive-conjunction-tester}]
The theorem follows from \cref{lemma:non-adaptive-conjunction-reconstructor} below,
combined with 
\cref{remark:randomized-query-simulator} to derandomize the query simulator; 
\cref{alg:DPT-via-Query-Simulator}
    will query the simulator $O(1/\epsilon)$ times, so it suffices to repeat each
query to the simulator $O(\log(1/\epsilon))$ times. Since the query simulator in
\cref{lemma:non-adaptive-conjunction-reconstructor} makes queries to the input $f$ that depend only
on the query $x$ to the simulator (and not on the queries made by
\cref{alg:non-adaptive-conjunction-reconstructor}),
the resulting algorithm is non-adaptive.\footnote{
    In more detail, the choice of \cref{alg:non-adaptive-conjunction-reconstructor} to output a
    query simulator for the constant $0$ function or the self-corrector $\cC$ depends on the binary
    output of the tester $\cT_k$, but the tester may non-adaptively carry out both branches and
    decide at the end, after seeing the results of its queries, which output to produce.}
\end{proof}

\begin{lemma}
\label{lemma:non-adaptive-conjunction-reconstructor}
\cref{alg:non-adaptive-conjunction-reconstructor} is a
non-adaptive
function reconstructor for the class of
conjunctions under $\bbD = \{ \Unif(\zo^n) \}$, with sample cost $O(1/\epsilon)$, query cost
$\widetilde O(1/\epsilon^2)$, and simulation cost $\widetilde O(1/\epsilon)$. It outputs a
(randomized) query simulator which, on any query $x$, queries the input $f$ on a random set of
points depending only on $x$.
\end{lemma}
\begin{proof}
The simulation cost is either 0 (if the constant 0 function is the output) or the query cost of 
$\cC$, which is $O(2^k) = O(1/\epsilon)$. The algorithm is non-adaptive because the tester
$\cT_k$ is non-adaptive, and the reconstructor does not make any other queries.

Let $p \define \Pru{\bm x \sim \cD}{f(\bm x) = 1}$. By the Chernoff bound, we have with probability
at least $9/10$ the event $p \leq \epsilon/4 \implies \wh{\bm p} \leq \epsilon/2$ and $p > \epsilon
\implies \wh{\bm p} > \epsilon/2$. Assume this event occurs, and that the tester $\cT_k$ succeeds.

Let $g$ be the function whose query simulator is produced. We first show that $\dist_\cD(g,
\bbP_\conj) < \epsilon$. If $g$ is the constant 0 function then this is true, so assume $g$ is the
function computed by the self-corrector $\cC$. Then $p > \epsilon/4$ and the tester $\cT_k$ accepts,
so $f$ is $\epsilon/64$-close
to some conjunction $h \in \bbP_{\leq k}$ under $\Unif(\zo^n)$.
    It follows that $h$ is the unique $k$-junta that is $\epsilon/64$-close to $f$ under
    $\Unif(\zo^n)$, since otherwise, by the triangle inequality, there would be two distinct
    $k$-juntas that are at most $2 \cdot \epsilon/64 < 2^{-k}$ apart, a contradiction.
So the self-corrector computes $g = h \in \bbP_\conj$.

Now suppose $f \in \bbP_\conj$ and $\cD = \Unif(\zo^n)$. First consider the case $\wh{\bm p} \leq
\epsilon/2$ so $g$ is the constant 0 function. Then $p < \epsilon$ so $\dist_\cD(f,g) = p <
\epsilon$ as desired.

Next, suppose $\wh{\bm p} > \epsilon/2$, so $p > \epsilon/4$. Since $f$ is a conjunction and $\cD =
\Unif(\zo^n)$, we have $\epsilon/4 < p \leq 2^{-L}$ where $L$ is the number of variables in the
conjunction defining $f$. Therefore $L \leq \lfloor \log(4/\epsilon) \rfloor = k$, so $\cT_k$ will
accept. As before, $f$ is the unique $k$-junta that is $\epsilon/64$-close to itself,
so the self-corrector computes $g = f$.
\end{proof}

\section{Algorithmic Technique III: Distribution Testing}
\label{sec:DRT-via-DT-and-DST}

It is natural to try to use distribution testers to design
distribution-robust testers. One of the first ideas one might think of is the
following. An \emph{identity tester} tests whether an unknown probability
distribution $\cD$ is equal to (or far from) a fixed reference distribution
$\cD^*$. Informally, it draws samples from $\cD$ and its output should (with
high probability) satisfy:
\begin{itemize}
    \item If $\cD = \cD^*$, output $\Accept$;
    \item If $\dist_\TV(\cD, \cD^*) > \epsilon$, output $\Reject$.
\end{itemize}
As mentioned in the introduction, it is a good exercise to show how to combine
a tester for $\cD^*$ with a $\cD^*$-specific tester for a property $\bbP$ into
a $\{ \cD^* \}$-picky tester. Essentially, the procedure is:
\begin{enumerate}
    \item Run the distribution tester on input distribution $\cD$;
        if it rejects, output $\Reject$.
    \item Run the $\bbD$-specific tester for $\bbP$ on input $f$
        (and distribution $\cD^*$). If it rejects, output $\Reject$,
        otherwise $\Accept$.
\end{enumerate}
The reason this works is that, if $\dist_\TV(\cD, \cD^*) < \epsilon$
and $\dist_{\cD^*}(f,h) < \epsilon$, then $\dist_{\cD}(f,h) < 2\epsilon$, so when the tester accepts we can be confident that
$\dist_{\cD}(f,\bbP) < 2\epsilon$.

The problem with this strategy is that identity testing is quite expensive,
requiring $\Omega(|\cX|^{1/2})$ samples when $\cD^*$ is the uniform distribution over a finite domain $\cX$ \cite{Paninski:08}. In this section we will give three results:
\begin{enumerate}
    \item A general method for improving upon the above identity testing method, using the notion of \emph{fooling distance} from \cite{BDP:26distinguishers} (\Cref{sec:fooling-distance}).
    
    \item An example application of this technique, where we get a distribution-robust tester with cost $O(\sqrt k / \epsilon^2)$ for the class $\bbI_k$ of unions of $k$ intervals over $[0,1]$, with respect to any fixed singleton class $\bbD = \{\cD^*\}$ (\cref{sec:union-of-intervals}).
    \item An ad-hoc proof improving this result for $\bbI_k$
        to query and sample cost $O(1/\epsilon^4)$ (\cref{sec:ad-hoc}).
\end{enumerate}

\subsection{General Method: Fooling Distance} \label{sec:fooling-distance}

What makes the identity-testing method work is the implication
\[
    (\dist_\TV(\cD, \cD^*) < \epsilon \wedge (\dist_{\cD^*}(f, h)
    < \epsilon) \implies \dist_\cD(f, h) < 2\epsilon 
\]
for any two functions $f, h$, which holds because the guarantee $\dist_\TV(\cD,
\cD*) < \epsilon$ bounds the difference in probabilities between $\cD$ and
$\cD^*$ of the event $\bm x \in \{ x \in \cX \mid f(x) \neq h(x) \}$. Ensuring the guarantee $\dist_\TV(\cD, \cD^*) < \epsilon$ is the job of the identity tester, and this is expensive.

We can get more efficient algorithms by observing that, for input $f$, we only
care about the family of events of the form
\[
    \bm x \in \{ x \in \cX \mid f(x) \neq g(x) \} \text{ for } g \in \bbP,
\]
rather than arbitrary $g$. If we write 
\[
    f \oplus \bbP \define \{ f \oplus g \mid g \in \bbP \} ,
\]
then we want to replace the expensive identity tester with a tester that outputs $\Reject$
when
\[
    \epsilon < \dist_{f \oplus \bbP}(\cD, \cD^*) \define \sup_{h \in f \oplus \bbP} 
    \left| \Exu{\cD}{ h } - \Exu{\cD^*}{ h } \right| .
\]
where we have here used the definition of \emph{fooling distance} between two distributions (c.f.~\cite{BDP:26distinguishers}): For a family $\bbH$ of functions $\cX \to \zo$,
the $\bbH$-fooling distance between distributions $\cD$ and $\cE$ is
\[
    \dist_\bbH(\cD, \cE) \define \sup_{h \in \bbH} \left| \Exu{\cD}{ h } - \Exu{\cE}{ h } \right| .
\]
We would then like to replace the identity tester with a tester that accepts when $\cD = \cD^*$ and rejects when $\dist_{f \oplus \bbP}(\cD, \cD^*) > \epsilon$. Formally, we would
like a tester satisfying the following definition for $\bbH = f \oplus \bbP$:

\begin{definition}
An \emph{identity tester} for $\cD^*$ \emph{against $\bbH$} with sample cost $s(\epsilon,\delta)$ is an algorithm that draws $s(\epsilon,\delta)$ samples
from input distribution $\cD$ and whose output satisfies the following, with probability at least $1-\delta$:
\begin{enumerate}
    \item If $\cD = \cD^*$ then the algorithm outputs $\Accept$;
    \item If $\dist_\bbH(\cD, \cD^*) > \epsilon$ then the algorithm outputs $\Reject$.
\end{enumerate}
\end{definition}
Na\"ively attempting to apply this definition with $\bbH = f \oplus \bbP$ is awkward because
we would need to design a tester whose definition depends on the unknown input $f$. We would
like a single algorithm that works against all $f \oplus \bbP$ simultaneously. This is
given by the following lemma of
\cite{BDP:26distinguishers}:

\begin{lemma}[Lemma~4.7 of \cite{BDP:26distinguishers}]
    \label{lemma:error-distinguisher-tester}
    Let $\cX$ be a domain, $\cD^* \in \Delta(\cX)$ be a distribution, and $\bbP \subseteq \zo^\cX$
    be a property. Suppose algorithm $\cT$ is an identity tester for $\cD^*$ against $\bbP$ with
    sample complexity $s(\epsilon, \delta)$. Then there is an algorithm $\cT'$ which, given sample
    access to input distribution $\cD \in \Delta(\cX)$ and query access to input function $f : \cX
    \to \zo$, uses at most $O\pbra{s\pbra{\frac{\epsilon}{7}, \frac{\delta}{4}} +
    \frac{\log(1/\delta)}{\epsilon^2}}$ samples and queries and has the following
    performance guarantee:
    \begin{enumerate}
        \item Completeness: If $\cD = \cD^*$, then $\cT'$ outputs $\Accept$ with probability at
            least $1-\delta$.
        \item Soundness: If $\dist_{f \oplus \bbP}(\cD, \cD^*) > \epsilon$, then $\cT'$ outputs
            $\Reject$ with probability at least $1-\delta$.
    \end{enumerate}
\end{lemma}

    At a high level, the algorithm of \cite{BDP:26distinguishers} works by creating two ``synthetic
    distributions'' $\mu$ and $\nu$, each of which combines the conditional distributions of $\cD$
    and $\cD^*$ on $f^{-1}(0)$ and $f^{-1}(1)$; samples from $\mu$ and $\nu$ can be obtained using
    samples from $\cD$ and $\cD^*$ along with queries to $f$. The algorithm performs the test $\cT$
    on $\mu$ and $\nu$, and accepts if and only if both tests accept. When $\cD = \cD^*$, we have
    $\mu = \nu = \cD^*$ and the tests accept, so $\cT'$ accepts. On the other hand, when $\dist_{f
    \oplus \bbP}(\cD, \cD^*) > \epsilon$, it can be shown that $\mu$ or $\nu$ is far from $\cD^*$
    under $\dist_\bbP$, so that at least one execution of $\cT$ rejects and hence so does $\cT'$.

Moreover, inspecting \cite{BDP:26distinguishers} shows that the algorithm $\cT'$ obtained above is
non-adaptive, because it only queries $f$ at points sampled from $\cD$ or $\cD^*$, with no
dependence between the queries. We can now present a formal generalization of the identity testing algorithm.

\begin{algorithm}[ht]
    \caption{Distribution-picky tester via fooling distance}
    \begin{algorithmic}
        \State \textbf{Requirements}:
        \State \qquad $\bullet$ $\cT_{\cD^*}$ is an identity tester for $\cD^*$ against $\bbP$.
        \State \qquad $\bullet$ $\cT'_{\cD*}$ is the algorithm constructed from $\cT_{\cD^*}$ in
            \cref{lemma:error-distinguisher-tester}.
        \State \qquad $\bullet$ $\cT_{\bbP}$ is a $\{\cD^*\}$-specific tester for $\bbP$.
        \State \textbf{Input}:
        \State \qquad $\bullet$ Query access to $f : \cX \to \zo$, 
        \State \qquad $\bullet$ Access to i.i.d.~samples
        $(\bx,f(\bx))_{\bx \sim \cD}$,
        \State \qquad $\bullet$ Closeness parameter $\epsilon \in (0,1)$.
        \Procedure{PickyTest-via-FoolingDistance}{}
            \State 1. Run the tester $\cT'_{\cD^*}$ using samples from $\cD$ with distance parameter $\epsilon/2$ and 
            \Statex \hspace{\algorithmicindent} failure probability $1/8$. If it outputs $\Reject$, output $\Reject$.
            \Statex \Comment{Uses samples from $\cD$ and queries to $f$}.
            \State 2.
                        Run the $\{\cD^*\}$-specific 
            tester $\cT_{\bbP}$ with distance parameter $\epsilon/2$ and
            \Statex \hspace{\algorithmicindent} failure probability $1/8$  on input $f$.  
            \Statex \hspace{\algorithmicindent} If $\cT_{\bbP}$ outputs $\Reject$, then output $\Reject$.
                \Comment{Uses queries to $f$.}
            \State 3. Output $\Accept.$
                    \EndProcedure
    \end{algorithmic}
    \label{alg:DPT-by-fooling-distance}
\end{algorithm}

\begin{lemma}[Picky testing via fooling distance]
    \label{lem:DPT-by-fooling-distance}
    Let $\cX$ be a domain, $\bbP \subset \zo^\cX$ be a property, and $\cD^*$
    be a distribution. Suppose that algorithm $\cT_{\cD^*}$ is an identity tester for $\cD^*$
    against $\bbP$
    with sample complexity $s(\epsilon, \delta)$ and that algorithm $\cT_{\bbP}$ is a $\{\cD^*\}$-specific tester for $\bbP$ with query\footnote{Recall
    that, since the
    family $\bbD$ for $\cT_{\bbP}$ consists of a single distribution, there is no need for $\cT_{\bbP}$ to draw samples.} complexity $q(\epsilon,\delta)$.   
    Then \Cref{alg:DPT-by-fooling-distance}
        is a distribution-picky tester
    for $\bbP$ with respect to $\bbD = \{\cD^*\}$
    with sample complexity
    $m = O\pbra{s\pbra{\frac{\epsilon}{14}, \frac{1}{32}} + \frac{1}{\epsilon^2}}$
    and query complexity $m + q\pbra{\frac{\epsilon}{2}, \frac{1}{8}}$.
    Moreover, that tester is non-adaptive if $\cT_{\bbP}$ is.
\end{lemma}
\begin{proof}
    The sample complexity and query complexity of
    \Cref{alg:DPT-by-fooling-distance} are evident from inspection,
    as is the non-adaptivity claim,
    so it remains to verify correctness.

    \textbf{Completeness.} Suppose $f \in \bbP$ and the input distribution is $\cD=\cD^*$. By the
    completeness guarantees of the tester $\cT'_{\cD^*}$ from
    \cref{lemma:error-distinguisher-tester} and the distribution-specific tester $\cT_{\bbP}$,
    each tester outputs $\Accept$ with probability at least $7/8$, so, by the union bound, the
    algorithm fails with probability at most $1/4$.

    \textbf{Soundness.} Suppose $\dist_\cD(f,\bbP)>\epsilon$. If
    $\dist_{f\oplus \bbP}(\cD,\cD^*)>\epsilon/2$, then the soundness guarantee of $\cT'_{\cD^*}$
    from \cref{lemma:error-distinguisher-tester} implies that it will $\Reject$ with probability at
    least $7/8$, and hence the whole algorithm rejects with probability at least $7/8$.

    Now suppose $\dist_{f \oplus \bbP}(\cD,\cD^*) \le \epsilon/2$. Then for all $g \in \bbP$,
    \[
        \abs{\dist_\cD(f, g) - \dist_{\cD^*}(f, g)}
        = \abs{\Exu{\cD}{f \oplus g} - \Exu{\cD^*}{f \oplus g}}
        \le \sup_{h \in f \oplus \bbP} \abs{\Exu{\cD}{h} - \Exu{\cD^*}{h}}
        = \dist_{f \oplus \bbP}(\cD, \cD^*) \leq \epsilon/2 ,
    \]
    which implies
    \[
        \dist_{\cD^*}(f,g) \geq \dist_\cD(f,g) - \frac{\epsilon}{2} > \frac{\epsilon}{2} .
    \]
    Since this holds for all $g \in \bbP$, the tester $\cT_{\bbP}$ will $\Reject$ with probability
    at least $7/8$ as desired.
\end{proof}

\subsection{Application: Unions of Intervals} 
\label{sec:union-of-intervals}

As an example application, we use the fooling distance method to get the following result on
distribution-robust testing of unions of $k$ intervals.  Let $\bbI_k$ denote the class of unions of
at most $k$ intervals over $[0,1]$, that is, $f : [0,1] \to \zo$ is in $\bbI_k$ if and only if there
exist intervals $\pbra{[\ell_i, r_i] : i \in [k']}$ for some $k' \le k$ such that $f(x) = 1$ if and
only if $x \in [\ell_i, r_i]$ for some $i \in [k']$.

\begin{theorem}
    \label{thm:unions-intervals-general}
    Let $\cD^*$ be any fixed distribution over $[0,1]$. There is a non-adaptive $\{\cD^*\}$-robust
    tester for $\bbI_k$ with sample complexity
    $O\pbra{\frac{\sqrt{k}}{\epsilon^2}}$ and query complexity $O\pbra{\frac{\sqrt{k}}{\epsilon^2} +
    \frac{1}{\epsilon^4}}$.
\end{theorem}

To prove this theorem, we require two results from the literature.  The first, from \cite{BBBY12},
gives a distribution-specific tester for $\bbI_k$ with respect to any fixed distribution $\cD^*$
over $[0,1]$ using $O(1/\epsilon^4)$ queries.

\begin{theorem}[\cite{BBBY12}]
    Let $\cD^*$ be a fixed distribution over $[0,1]$. There is a non-adaptive distribution-specific
    tester for $\bbI_k$ with respect to $\bbD = \{\cD^*\}$ with query complexity $q(\epsilon,
    \delta) = O\pbra{\frac{\log(1/\delta)}{\epsilon^4}}$.
\end{theorem}

\begin{remark}
    The result of \cite{BBBY12} holds in the more challenging model of \emph{active} property
    testing, where the algorithm must draw samples from $\cD^*$ and can only query $f$ at points
    that appeared in the sample. In that context, the number of samples drawn from $\cD^*$ is also
    relevant, and \cite{BBBY12} showed that $O(\sqrt{k}/\epsilon^5)$ samples suffice. In the present
    setting,
    the distribution $\cD^*$ is fixed and known to the algorithm, so
    the distribution-specific tester can draw its own samples and only requires query access
    to $f$.
\end{remark}

The second result, from \cite{DiakonikolasKN15}, gives a distribution tester for any fixed
distribution $\cD^*$ against family $\bbI_k$ (in the language of that paper, this is the task of
identity testing against \emph{$\cA_k$-distance}) using $O(\sqrt{k}/\epsilon^2)$ samples.

\begin{theorem}[\cite{DiakonikolasKN15}]
    Let $\cD^*$ be a fixed distribution over $[0,1]$. There is a distribution tester for $\cD^*$
    against $\bbI_k$ with sample complexity $s(\epsilon, \delta) = O\pbra{\frac{\sqrt{k}
    \log(1/\delta)}{\epsilon^2}}$.
\end{theorem}

We get \cref{thm:unions-intervals-general} by applying these two theorems with
\cref{lem:DPT-by-fooling-distance} to get a $\{\cD^*\}$-picky tester for every fixed distribution
$\cD^*$ over $[0,1]$, and then applying \cref{lem:reduction-to-picky} to get the $\{\cD^*\}$-robust
tester.

\cref{thm:unions-intervals-general} is a positive result in the sense that no distribution-free tester for
$\bbI_k$ with query complexity $o(k)$ is known. On the other hand, the best known query complexity
lower bound for distribution-free testing $\bbI_k$ is $\Omega(\sqrt{k})$ \cite{BBBY12}, so
\cref{thm:unions-intervals-general} does not establish that $\bbI_k$ witnesses a separation between the two
models. Next, we show a much stronger positive result for distribution-robustly testing $\bbI_k$ by
directly extending the techniques of \cite{BBBY12}.

\subsection{Ad-Hoc Improvement for Unions of Intervals}
\label{sec:ad-hoc}

We now give a better tester for unions of $k$ intervals, albeit with a specialized technique that
is specific to this class, unlike the general method above.

\thmintrointervals*

Our tester essentially upgrades the distribution-specific tester of \cite{BBBY12} to be
distribution-picky.

\begin{remark}
\label{remark:intervals-reduction-to-unif}
For simplicity of exposition, we make the following simplifications.

First, we will prove the theorem only for the case $\cD^* = \Unif([0,1])$. By
an argument of \cite{BBBY12}, the $\{\cD^*\}$-specific testing in the general case reduces to the
uniform case, and the same argument works for distribution-picky testing.

Second, to eliminate boundary edge cases, it is helpful to assume that $f = 0$ near the endpoints of
the domain, say, on $[0, 1/4) \cup (3/4, 1]$. This is without loss of generality for
distribution-picky testing with respect to $\cU$ for the following reason. On input function $f$ and
input distribution $\cD$, a tester may simulate 1)~query access to function $f' : [0,1] \to \zo$
given by $f'(x) \define f\pbra{2x-\frac{1}{2}}$ for $x \in [1/4, 3/4]$, and $f'(x) \define 0$
otherwise; and 2)~sample access to distribution $\cD'$ given as follows: to produce a sample from
$\cD'$, with probability $1/2$ sample $\bm{x} \sim \cD$ and output $\frac{1}{4} + \frac{\bm{x}}{2}$,
and with probability $1/2$ output a uniformly random sample from $[0, 1/4) \cup (3/4, 1]$. If $f \in
\bbI_k$, then $f' \in \bbI_k$. Also, $\dist_{\cD'}(f', \bbI_k) = \frac{1}{2} \dist_\cD(f, \bbI_k)$.
Moreover, if $\cD = \cU$ then $\cD' = \cU$ as well. Therefore, on inputs $f$ and $\cD$, it suffices
for the distribution-picky tester to simulate queries from $f'$ and samples from $\cD'$ with
distance parameter $\epsilon/2$. Therefore, going forward we assume without loss of generality that
$f = 0$ on $[0, 1/4) \cup (3/4, 1]$.
\end{remark}

\subsubsection{Noise Sensitivity}

The key observation of \cite{BBBY12} is that the \emph{noise sensitivity} of $f$ characterizes its
distance to the class $\bbI_k$ of unions of $k$-intervals. For parameter $\rho > 0$, the \emph{local
$\rho$-noise sensitivity} of $f$ at point $x \in [0,1]$ is
\[
    \NS_\rho(f, x) \define \Pru{\bm{y} \sim_\rho x}{f(x) \ne f(\bm{y})} \,,
\]
where we write $\bm{y} \sim_\rho x$ to denote that $\bm{y}$ is uniformly distributed in the interval
$(x-\rho, x+\rho)$, and set $f(x) \define 0$ at $x \in \R \setminus [0,1]$ by
convention\footnote{This is slightly different from the convention in \cite{BBBY12}, which restricts
random variables to $[0,1]$, but the difference is not important and this choice is more
convenient for our purposes.}.  The \emph{$\rho$-noise sensitivity} of $f$ with respect to
distribution $\cE$ (over $[0,1]$) is
\[
    \bbNS_\rho^\cE(f) \define \Exu{\bm{x} \sim \cE}{\NS_\rho(f, \bm{x})} \,.
\]
(\cite{BBBY12} only define this for the uniform distribution $\cE = \cU$, but we require the general
case.)

Noise sensitivity characterizes the distance to $\bbI_k$ as follows.
Fix $\rho \define \frac{\epsilon^2}{64 k}$, as in
\cref{alg:DPtester-unions-intervals}. The next claim is from \cite{BBBY12}.

\begin{lemma}[Proposition~3.3 of \cite{BBBY12}]
    \label{lemma:unions-intervals-completeness-structural}
    Suppose $f \in \bbI_k$. Then $\bbNS_\rho^\cU(f) \le k \rho$.
\end{lemma}

The final claim we require is a generalization of a proof from \cite{BBBY12}. By following
the proof of Lemma~3.4 of \cite{BBBY12}, which we do in \cref{ap:omitted-proofs}, we get:

\begin{restatable}{lemma}{lemmaunionsintervalssoundnessstructural}
    \label{lemma:unions-intervals-soundness-structural}
    Suppose $\max\cbra{\bbNS_\rho^\cU(f), \bbNS_\rho^\cD(f)} \le k \rho \pbra{1 +
    \frac{\epsilon}{4}}$. Then $\dist_\cD(f, \bbI_k) \le \epsilon$.
\end{restatable}

\subsubsection{Algorithm and Analysis}

\cref{alg:DPtester-unions-intervals} extends the algorithm of \cite{BBBY12} by separately estimating
both $\bbNS_\rho^\cU$ and $\bbNS_\rho^\cD$ and tracking how the structural results of \cite{BBBY12}
depend on these two quantities. 

\begin{algorithm}[t!]
    \caption{Distribution-picky tester for unions of $k$ intervals with respect to $\Unif([0,1])$}
    \begin{algorithmic}
        \State \textbf{Input}: \State \quad $\bullet$ Query access to $f : [0,1] \to \zo$, 
        \State \quad $\bullet$ access to \iid samples $(\bm{x},
        f(\bm{x}))_{\bm{x} \sim \cD}$,
        \State \quad $\bullet$ parameters $k \in \N$ and $\epsilon \in (0,1)$.
        \Procedure{PickyTester-Intervals}{$f, k, \epsilon$}
            \State 1. Set $\rho \define \frac{\epsilon^2}{64 k}$ and $r \define
            \Theta\pbra{\frac{1}{\epsilon^4}}$.
            \State 2. Sample \iid pairs
                $(\bm{x}_i \sim \cU, \bm{y}_i \sim_\rho \bm{x}_i)_{i \in [r]}$ and
                $(\bm{x}'_i \sim \cD, \bm{y}'_i \sim_\rho \bm{x}'_i)_{i \in [r]}$.
            \Statex \Comment{Uses samples from $\cD$.}
            \State 3. Let
                $\bm{Z}_\cU \define \frac{1}{r} \sum_{i=1}^r \ind{f(\bm{x}_i) \ne f(\bm{y}_i)}$
                and
                $\bm{Z}_\cD \define \frac{1}{r} \sum_{i=1}^r \ind{f(\bm{x}'_i) \ne f(\bm{y}'_i)}$
            \Statex \Comment{Uses queries to $f$.}
            \State 4. If $\max\cbra{\bm{Z}_\cU, \bm{Z}_\cD} \le k \rho
            \pbra{1+\frac{\epsilon}{8}}$, output $\Accept$. Otherwise, output
            $\Reject$.
        \EndProcedure
    \end{algorithmic}
    \label{alg:DPtester-unions-intervals}
\end{algorithm}

The following theorem implies \cref{thm:intro-intervals} via
\cref{lem:reduction-to-picky} and \cref{remark:intervals-reduction-to-unif}.

\begin{theorem}
    \cref{alg:DPtester-unions-intervals} is a non-adaptive distribution-picky tester for $\bbI_k$
    with respect to $\bbD = \{\Unif([0,1])\}$ with query complexity $O(1/\epsilon^4)$ and sample complexity
    $O(1/\epsilon^4)$.
\end{theorem}
\begin{proof}
    The query and sample cost claims are immediate, as is non-adaptivity, so we now show
    correctness.

    \paragraph*{Completeness.} Suppose $f \in \bbI_k$ and $\cD = \cU$.
    By \cref{lemma:unions-intervals-completeness-structural},
    $\bbNS_\rho^\cU(f) \le k \rho$. It follows by a Chernoff bound that
    $\bm{Z}_\cU \le k \rho \pbra{1 + \frac{\epsilon}{8}}$ with probability at
    least $7/8$, as long as the constant in the choice of $r =
    \Theta(1/\epsilon^4)$ is sufficiently large. Since $\cD = \cU$, we similarly
    have that $\bm{Z}_\cD \le k \rho \pbra{1 + \frac{\epsilon}{8}}$ with
    probability at least $7/8$. By a union bound, both events hold
    simultaneously with probability at least $3/4$, and in this case the tester
    outputs $\Accept$.

    \paragraph*{Soundness.} Suppose $\dist_\cD(f, \bbI_k) > \epsilon$.
    By \cref{lemma:unions-intervals-soundness-structural},
    $\max\cbra{\bbNS_\rho^\cU(f), \bbNS_\rho^\cD(f)} > k \rho \pbra{1 +
    \frac{\epsilon}{4}}$. Then, as above, a Chernoff bound implies that, with
    probability at least $3/4$, either $\bm{Z}_\cU > k \rho \pbra{1 +
    \frac{\epsilon}{8}}$ or $\bm{Z}_\cD > k \rho \pbra{1 +
    \frac{\epsilon}{8}}$, and hence the tester outputs $\Reject$.
\end{proof}

\section{Sample-Based Testing}
\label{section:sample-based}

The previous sections established the separations between query-based
distribution-specific, distribution-robust, and distribution-free testing that
were summarized in \Cref{figure:qual} and \Cref{table:quant}. 
In this section we establish the claimed separations for \emph{sample-based} testers; recall that these are algorithms that
do not make any queries but instead only use random labeled samples $(\bx,f(\bx))_{\bx \sim \cD}$.

\paragraph*{Prelude: Separating query- and sample-based distribution-robust testing.}
Before turning to separations between different sample-based testing models,  we first observe that for some natural properties, sample-based distribution-robust testers can sometimes inherently be less efficient than query-based distribution-robust testers. This is an easy consequence of the results we established earlier for distribution-robust testing of conjunctions using queries and the following folklore lower bound for sample-based conjunction testing:

\begin{observation}[Query-based distribution-robust testers can be more efficient than sample-based distribution-robust testers] \label{obs:query-vs-sample-DR}
Any sample-based distribution-robust tester for the class of conjunctions with respect to the uniform distribution must use at least $\Omega(\log n)$ samples even if the distance parameter $\epsilon$ is an absolute constant such as 0.01.  
\end{observation}

\Cref{obs:query-vs-sample-DR} is an easy consequence of the well known fact that any algorithm that only receives, say, ${\frac 1 2} \log n$ independent samples of the form $(\bx,\boldf(\bx))_{\bx \sim \Unif(\zo^n)}$ cannot distinguish between the two possibilities that (i) $\boldf$ is a uniform random Boolean function (in which case $\boldf$ is far from every conjunction with extremely high probability), versus (ii) $\boldf$ is a uniform random dictator drawn from $\{x_1,\dots,x_n\}$ (in which case $\boldf$ is a conjunction).
\Cref{thm:intro-non-adaptive-conjunction-tester} and
\Cref{obs:query-vs-sample-DR} together show that conjunctions are
distribution-robustly testable with constant complexity if both
queries and
samples are allowed (even without adaptivity), but if queries are disallowed then at least a logarithmic
number of samples is required.

\paragraph*{Separating distribution-specific and distribution-robust sample-based
testing.}
As explained in the introduction, this separation
is based on a result from a companion paper \cite{anonymous:DRHT} on distribution-robust testing of halfspaces over $\R^n$ with respect to the standard Gaussian distribution
$\bbD=\{\cN(0,I_n)\}$.
As was described earlier,
the main lower bound result of \cite{anonymous:DRHT} states that every sample-based distribution-robust halfspace tester with respect to the standard Gaussian $\cN(0, I_n)$ requires at least
$\wt \Omega(n^{2/3})$
samples.
In contrast,
sample-based distribution-specific halfspace testers with respect to
$\cN(0, I_n)$
are known that use only $\tilde{O}(n^{1/2}) \cdot
\poly(1/\epsilon)$ samples \cite{BBBY12,Harms19}.

\paragraph*{Separating distribution-robust and distribution-free sample-based testing.}
As stated earlier, \cite{CX16} showed that any distribution-free tester for
conjunctions must have complexity $\tilde{\Omega}(n^{1/3})$ even if both
adaptive queries and samples are allowed. In contrast, conjunctions
can be distribution-robustly tested with respect to $\Unif(\zo^n)$ using
exponentially fewer samples even if no queries are allowed:

\begin{algorithm}[ht]
    \caption{Distribution-robust sample-based tester for conjunctions with respect to $\Unif(\zo^n)$}
    \begin{algorithmic}
        \State \textbf{Input}: \State \quad $\bullet$ Sample access to i.i.d.~labeled examples $(\bx,f(\bx))_{\bx \sim \cD}$ for some $\cD \in
            \Delta(\cX)$,
        \State \quad $\bullet$ Closeness parameter $\epsilon \in (0,1)$.
        \Procedure{SampleBasedDistributionRobustConjTester}{$f, \cD, \epsilon$}
	   \State 1.  Let $g_1,\dots,g_K$ be a list of all conjunctions of Boolean literals over $\zo^n$ of length 
\Statex \hspace{\algorithmicindent} at most $L$, where $L := \lceil \log(16/\epsilon)\rceil $.
            \State 2. Use a batch of $m$ (specified below) i.i.d.~samples $(\bx,f(\bx))_{\bx \sim \cD}$ to obtain, for each $i \in [K]$, 
            \Statex \hspace{\algorithmicindent} an empirical estimate $\widehat{\bd}_i$ of $\dist_{\cD}(f,g_i)$. If some $i \in [K]$ has its estimate at most $3\epsilon/8$, 
            \Statex \hspace{\algorithmicindent} then output $\Accept$.
               \Comment{Uses samples from $\cD$.}
            \State 3. Use a batch of $m$ i.i.d.~samples $(\bx,f(\bx))_{\bx \sim \cD}$ to obtain, for each $i \in [K]$ such that 
            \Statex \hspace{\algorithmicindent} the conjunction $g_i$ has length exactly $L$, an empirical estimate $\wh{\bp}_i$ of $\Pr_{\bx \sim \cD}[g_i(\bx)=1].$
            \Statex \hspace{\algorithmicindent}
            If one of these $i$'s has $\wh{\bp}_i > 3 \epsilon /32$ then output $\Abstain$.
                           \Comment{Uses samples from $\cD$.}
            \State 4. Output $\Reject.$ 
        \EndProcedure
    \end{algorithmic}
    \label{alg:sample-based-conj}
\end{algorithm}

The following theorem, along with \cref{obs:query-vs-sample-DR}, establishes
\cref{thm:intro-conj-sample}.

\begin{theorem}
[Sample-based distribution-robust conjunction testing with logarithmically many samples]
\label{thm:DRsampleconj}
\Cref{alg:sample-based-conj} is a sample-based distribution-robust tester for the class of conjunctions over $\cX = \zo^n$ with respect to $\bbD=\{\Unif(\zo^n)\}$ that uses $O\pbra{\frac{\log(n)\cdot \log(1/\epsilon)}{\epsilon}}$ samples.
\end{theorem}

\begin{proof}
Let $\bbP$ denote the class of all conjunctions over $\zo^n$. 
The list length $K$ from Step~1 has size
\[
        K = \sum_{\ell=0}^L \binom{2n}{\ell} \le n^{O(\log(1/\epsilon))},
\]
and hence $\log K=O(\log(n) \cdot \log(1/\epsilon))$.

In Step~2, let $d_i \define \dist_{\cD}(f,g_i)$ be the true distance between $f$ and $g_i$ under $\cD$.
Taking 
\[
    m \define C\log(K)/\epsilon
\]
for a sufficiently large universal constant $C$, by a multiplicative Chernoff bound and union bound we have that with probability at least $7/8$, for all $i \in [K]$ it holds that  if $d_i < \epsilon/4$ then $\wh{\bd}_i\le 3\epsilon/8$ and if $d_i > \epsilon/2$ then $\wh{\bd}_i > 3\epsilon/8.$

In Step~3, let  $p_i \define \Pru{\bx\sim\cD}{g_i(\bx)=1}$ be the true probability that $g_i$ is satisfied by a random draw from $\cD$ for each length-$L$ conjunction $g_i$. 
Another multiplicative Chernoff bound and union bound imply that with probability at least $7/8$, Step~3 outputs $\Abstain$ whenever some length-$L$ conjunction has $p_i>\epsilon/8$, and it does not not output $\Abstain$ when every length-$L$ conjunction has $p_i\le \epsilon/16$.

The total number of samples used by the algorithm is $2m$, which is as claimed in \Cref{thm:DRsampleconj}. Let $E_2$ be the successful Step~2 event and $E_3$ the successful Step~3 event, so both $E_2$ and $E_3$ hold with probability at least $3/4$.  It remains to establish correctness conditioned on $E_2\cap E_3$.  

    \textbf{Completeness.}  Suppose $f\in\bbP$, i.e.~$f$ is a conjunction.  If $f$ has at most $L$ literals then $f=g_i$ for some $i\in[K]$, so $d_i=0<\epsilon/4$ and the tester accepts in Step~2.  Otherwise, let $f^*$ be obtained from $f$ by deleting literals until exactly $L$ literals remain.  Then $f^*=g_i$ for some length-$L$ conjunction in the list.  If $\dist_\cD(f,f^*)<\epsilon/4$, then because of $E_2$ the Step~2 test for this $g_i$ causes the tester to accept.  If $\dist_\cD(f,f^*)\ge \epsilon/4$, then every satisfying assignment of $f$ is also a satisfying assignment of $f^*$, and the points on which $f$ and $f^*$ differ are all points with $f^*(x)=1$ and $f(x)\ne 1$.  Therefore
\[
    \Pru{\bx\sim\cD}{f^*(\bx)=1}
    \ge \dist_\cD(f,f^*)
    \ge \epsilon/4,
\]
so if the tester reaches Step~3, then because of $E_3$ the tester will abstain.  Hence conditioned on $E_2\cap E_3$ the tester never rejects a conjunction.

    \textbf{Soundness.} Suppose $\dist_\cD(f,\bbP)>\epsilon$.  Then for every $i\in[K]$ we have
$d_i=\dist_\cD(f,g_i)>\epsilon>\epsilon/2$, so because of $E_2$ the tester does not accept in Step~2, hence it either abstains or rejects as required.

    \textbf{Assertiveness.}  Suppose $\cD=\Unif(\zo^n)$.  Every conjunction of length exactly $L$ is satisfied under $\cD$ with probability $2^{-L}\le \epsilon/16$, so because of $E_3$ the tester does not abstain in Step~3 regardless of $f$.  This concludes the proof.
\end{proof}

\def\calU{\mathcal{U}}
\def\calD{\mathcal{D}}
\def\calT{\mathcal{T}}
\def\calS{\mathcal{S}}
\def\calA{\mathcal{A}}
\def\calE{\mathcal{E}}

\def\bc{\boldsymbol{c}}

\def\bcalD{\boldsymbol{\mathcal{D}}}

\def\bT{{\boldsymbol{T}}}
\def\bA{{\boldsymbol{A}}}

\def\bgamma{\boldsymbol{\gamma}}

\def\Dyes{\mathfrak{D}_{\text{yes}}}
\def\Dno{\mathfrak{D}_{\text{no}}}
\def\ff{\boldsymbol{f}}
\def\bg{\boldsymbol{g}}
\def\Mono{\textsc{Mono}}
\def\ALG{\textsc{Alg}}
\def\bb{\boldsymbol{b}}

\def\calF{\mathcal{F}}
\def\calG{\mathcal{G}}

\def\fT{\frak{T}}
\def\fS{\frak{S}}

\newcommand{\eps}{\epsilon}

\section{Lower Bound for Testing Monotonicity}
\label{sec:monotone-lower}

While our main interest is in establishing \emph{positive} results which show that distribution-robust testing can be significantly more efficient than distribution-free testing. another natural question that suggests itself is to establish \emph{lower} bounds showing that distribution-robust testing can be more challenging than standard-model testing.  Such results are of interest because they serve as reassurance that the distribution-robust testing model is not simply the standard model ``in disguise,'' i.e.~they rule out the possibility that every property that is testable in the standard model is also testable in the distribution-robust model with the same complexity.

As mentioned earlier, in an unpublished manuscript \cite{anonymous:DRHT} it has been shown that for sample-based testing, the class of halfspaces over $\R^n$ witnesses a separation between the distribution-robust and standard models.  In more detail, \cite{anonymous:DRHT} shows that any sample-based distribution-robust testing algorithm for the class of halfspaces under the standard Normal distribution $N(0,I_n)$ must use $\tilde{\Omega}(n^{2/3})$ samples even when the distance parameter $\epsilon$ is $\Theta(1)$, which is larger than the $\tilde{O}(n^{1/2})$ upper bound on sample-based halfspaces testing with respect to the $N(0,I_n)$ distribution (again for distance parameter $\epsilon = \Theta(1)$) that has been shown in \cite{BBBY12,Harms19}.  

Here we show a different and larger separation, between the \emph{non-adaptive} standard model and the non-adaptive distribution-robust testing model.  This separation is for the class of monotone functions over $\zo^n$.  We recall that the celebrated work of Khot, Minzer and Safra \cite{KMS18} gave a non-adaptive algorithm for testing monotone functions with respect to $\Unif(\zo^{n})$ that has query complexity $\tilde{O}(\sqrt{n})$ for constant closeness parameter $\epsilon$.  In contrast, the following result
(which is a sharpening of \cref{thm:intro-mono-lb})
gives a mildly exponential lower bound for non-adaptive distribution-robust testers:

\begin{theorem}
[Mildly exponential lower bound for non-adaptive distribution-robust monotonicity testing over $\zo^n$]
\label{thm:mono-lb}
There are absolute constants $\epsilon_0,c>0$ such that the following holds.  Let
$n$ be sufficiently large and let
\[
  \frac{1}{n^{1/6}}\leq \epsilon\leq \epsilon_0. 
    \]
Every non-adaptive distribution-robust tester for monotonicity over $\cX = \zo^n$ with closeness parameter $\epsilon$ with
respect to the uniform distribution must either draw at least
\[
  2^{c \sqrt{n}/\epsilon}
\]
labeled samples or make at least
\[
  2^{c n^{1/4}/\sqrt{\epsilon}
    }
\]
non-adaptive queries.  
\end{theorem}

\subsection{Notation and Parameters}

It will be convenient to carry out the proof over $\{0,1\}^n$. Let $\calU$ denote the
uniform distribution over $\{0,1\}^n$.  For $x\in\{0,1\}^n$, let $|x|$ denote
its Hamming weight.  

A Boolean function $f:\{0,1\}^n\rightarrow \{0,1\}$ is said to be
  monotone if $f(x)\le f(y)$ for every pair of $x$ and $y$ satisfying $x\preceq y\in \{0,1\}^n$.
Given a distribution $\calD$ over $\{0,1\}^n$ and $f:\{0,1\}^n\rightarrow \{0,1\}$,
  we write $\dist_\calD(f,\Mono)$ to denote the distance of $f$ to monotonicity with
  respect to $\calD$:
$$
\dist_\calD(f,\Mono)=\min_{g} \Big\{\Pr_{\bx\sim \calD} \big[f(\bx)\ne g(\bx)\big]\Big\},
$$
where the minimum is taken over all monotone functions $g:\{0,1\}^n\rightarrow \{0,1\}$.

Choose absolute constants $c_r,c_\Delta,c_h>0$ and then choose $\epsilon_0>0$
sufficiently small.  Let $r$ be the largest odd integer at most
$c_r\sqrt n/\epsilon$, and set
\begin{equation}
  \Delta \define \left\lfloor\frac{c_\Delta n}{r}\right\rfloor,
  \qquad
  h \define \left\lfloor c_h\sqrt r\right\rfloor,
  \qquad
  \rho \define \frac{\Delta}{\sqrt n}.
  \label{eq:drpt-parameters}
\end{equation}
The constants are fixed so that, for all sufficiently large $n$,
\begin{equation}
  r=\Theta\!\left(\frac{\sqrt n}{\epsilon}\right),
  \qquad
  \Delta = \Theta(\eps \sqrt{n}),
  \qquad
  20h\leq\Delta\leq\frac{\sqrt n}{20},
  \qquad \Theta(\eps) =
  \rho\geq C_{\mathrm{far}}\epsilon,
  \label{eq:drpt-scale}
\end{equation}
where $C_{\mathrm{far}}>0$ is a sufficiently large absolute constant.  It is easy to check that these
conditions are compatible: first take $c_r$ sufficiently small relative to
$c_\Delta$ and $c_h$, and then take $\epsilon_0$ sufficiently small.  Notice in
particular that $r=o(n)$.

We will use the following standard fact about matchings in the hypercube (see, for example,
\cite[Claim~12]{CDLNS24}).

\begin{lemma}\label{lem:action-matching}
For every odd positive integer $r$, there is a perfect matching $M$ of
$\{0,1\}^r$ such that every matched pair $(u^-,u^+)$ satisfies
\[
  u^-\prec u^+ 
  \qquad\text{and}\qquad
  |u^-|+|u^+|=r.
\]
\end{lemma}

\begin{proof}
For each integer $0\leq k<r/2$, consider the bipartite comparability graph
between the $k$-th and $(r-k)$-th Hamming layers.  This graph is regular on both
sides, and hence it has a perfect matching by Hall's theorem.  Taking the union
of these matchings over all $k<r/2$ proves the lemma.
\end{proof}

For each $A\subset[n]$ of size $r$, we write $M_A$ to denote the perfect matching of $\{0,1\}^n$ obtained by applying
\Cref{lem:action-matching} on the coordinates in $A$ independently for every
fixed setting of the coordinates in the complement $\overline A$.  Thus every edge
$(x,y)\in M_A$ satisfies
\begin{equation}
  x\prec y,
  \qquad
  x_{\overline A}=y_{\overline A},
  \qquad
  |x_A|+|y_A|=r.
  \label{eq:drpt-matching}
\end{equation}

\subsection{Overview of the Proof of \Cref{thm:mono-lb}}

In \Cref{sec:construction}, we will construct a pair of probability distributions $\Dyes$ and $\Dno$ over pairs $(f,\calD)$ where $f$ is a function from $\{0,1\}^n$ to $\{0,1\}$ and $\cD$ is a probability distribution over $\{0,1\}^n$. The yes
distribution always has $\calD=\calU$ and always produces a monotone function.
With constant probability, the no distribution produces a function that is
$\Omega(\rho)$-far from monotone under its accompanying distribution; by
\eqref{eq:drpt-scale}, this is more than $\epsilon$.  The main indistinguishability lemma
shows that no deterministic non-adaptive algorithm using at most $s$ labeled
samples and $q$ queries can distinguish the two distributions with constant
advantage.

More precisely, after defining the distributions in \Cref{sec:construction} we will prove the following three lemmas.

\begin{lemma}\label{lem:lem1}
A pair $(\ff,\bcalD)\sim\Dyes$ always satisfies $\bcalD=\calU$ and
$\ff\in\Mono$.
\end{lemma}

Given this we will write $(\ff,\calU)$ to denote a draw from $\Dyes$ from now on.

\begin{lemma}\label{lem:lem2}
There are absolute constants $c,p_0>0$ such that a pair
$(\bg,\bcalD)\sim\Dno$ satisfies
\[
  \dist_{\bcalD}(\bg,\Mono)\geq c\rho>\epsilon
\]
with probability at least $p_0$.
\end{lemma}

For the last lemma, let $s$ and $q$ be the following parameters:
\begin{equation}
s=\exp\big(c_0 r\big)\quad \text{and}\quad q=\exp\big(c_0 \sqrt{r}\big),
\label{eq:drpt-s-q}
\end{equation}
where $c_0$ is a sufficiently small positive constant to be specified later.

We consider any 
  deterministic, non-adaptive algorithm $\ALG$ (recall \cref{def:adaptivity}) working on a given
pair $(f,\calD)$ (drawn either from
  $\Dyes$ or $\Dno$)
in the following fashion:
  \begin{flushleft}\begin{itemize}
  \item Phase 1: $\ALG$ receives 
    $s$ independent labeled samples drawn from $\calD$ (i.e., 
    $s$ pairs\\ $(\bx^1,f(\bx^1)),\ldots,(\bx^s,f(\bx^s))$ with $\bx^1,\ldots,\bx^s$ drawn independently from $\calD$;
  \item Phase 2: Based on labeled samples received in Phase 1, $\ALG$ (deterministically)\\ makes a single batch of 
    $q$ queries on $f$;
  \item Phase 3: Based on information received in Phases 1 and 2, $\ALG$ accepts or rejects.
  \end{itemize}\end{flushleft}

The last lemma, which is the most challenging one among the three, is as follows:

\begin{lemma}\label{lem:lem3}
Every such deterministic non-adaptive algorithm $\ALG$ satisfies
\[
\left|
\Pr_{(\ff,\calU)\sim\Dyes}[\ALG\text{ accepts}]
-
\Pr_{(\bg,\bcalD)\sim\Dno}[\ALG\text{ accepts}]
\right|=o_n(1),
\]
where the probabilities include the randomness of the labeled samples.
\end{lemma}

\Cref{thm:mono-lb} then follows from these three lemmas by standard arguments.
In particular, notice that by our constructions of $\Dyes$ and $\Dno$,
the lemmas yield a lower bound for distribution-picky, and hence also
distribution-robust monotonicity testing with respect to $\cU$.

\subsection{The Construction}\label{sec:construction}

Choose a constant $c_1$ satisfying $c_0 \ll c_1\ll1$ and define
\begin{equation}
  N \define 2^r,
  \qquad
  L \define \left\lfloor\exp(c_1r)\right\rfloor.
  \label{eq:drpt-budgets}
\end{equation}
In particular we choose $c_1>2c_0$, so $L\gg s^2$, while $c_1$ is still small enough for all
of the exponential union bounds below.

\paragraph*{Talagrand terms and the multiplexer.}
We use $T=(T_1,\ldots,T_N)$ to denote a sequence of $N$ terms, where each $T_\ell$
is of the form $T_\ell: [r] \to [n]$ and should be interpreted as a term of length at most 
$r$. We will abuse the notation and write $T_\ell$ to denote a Boolean
function $T_\ell:\{0,1\}^n\rightarrow \{0,1\}$, where
\[
    T_\ell(x) \define \bigwedge_{j=1}^r x_{T_\ell(j)}.
\]

Each $T$ defines the following \emph{multiplexer} map $\Gamma_T$, which maps
each $x\in \{0,1\}^n$ either to a special symbol in $\{0^*,1^*\}$, or to a subset of
$[N]$ of size at most $2$ (including the empty set): Given $x\in \{0,1\}^n$,
\begin{enumerate}
  \item if $|x|>n/2+\Delta$, set $\Gamma_T(x)=1^*$;
  \item if $|x|<n/2-\Delta$, set $\Gamma_T(x)=0^*$;
  \item if $|x|\in[n/2-\Delta,n/2+\Delta]$ (hereafter,
  the \emph{middle layers}), let 
  $\Sat_T(x) \define \{\ell:T_\ell(x)=1\}$ and set
  \[
    \Gamma_T(x)=
    \begin{cases}
      \emptyset,&\Sat_T(x)=\emptyset,\\
      \{\ell\},&\Sat_T(x)=\{\ell\},\\
      \{\ell_1,\ell_2\},&|\Sat_T(x)|\geq2,
    \end{cases}
  \]
  where $\ell_1<\ell_2$ are the two smallest indices in $\Sat_T(x)$ in the last
  case.
\end{enumerate}
We write $\calT$ to denote the uniform distribution over all such
sequences $T$, where each term $T_\ell$ is independently drawn by sampling
its $r$ coordinates independently and uniformly from $[n]$.

\paragraph*{Action variables and support set.}
Let $A$ denote a size-$r$  subset of $[n]$; the variables in $A$ are called
\emph{action} variables. We write $\calA$ to denote the uniform distribution
over all size-$r$ subsets of $[n]$.

Once $A$ is fixed, the \emph{support set}
$S$ will be a set of $2L$ points in $\{0,1\}^n$ that consists of $L$ pairs of points from $M_A$.
We write $\calS_A$ to denote the uniform distribution over all such size-$(2L)$ subsets of $\{0,1\}^n$, given $A$.

\paragraph*{Term labels.}
We also associate with each term $T_\ell$ a binary label $b_\ell \in \zo$, and
write $B = (b_1, \dotsc, b_N)$, which will be drawn uniformly below.

\subsubsection{Construction of \texorpdfstring{$\Dyes$}{Dyes}}

To draw $(\ff,\calU)\sim \Dyes$, we first draw random objects $\bT,\bA,\bS$ and $\bB$ discussed above as follows:
\begin{enumerate}
\item $\bT\sim \calT$, which is used to obtain the multiplexer map $\Gamma_{\bT}$;
\item $\bA\sim \calA$ contains the $r$ action variables;
\item 
$\bS\sim \calS_{\bA}$ is a size-$(2L)$ subset of $\{0,1\}^n$; and
\item $\bB=(\bb_1,\ldots,\bb_N)$ is drawn uniformly from $\{0,1\}^N$.
\end{enumerate}
Note that the definition of the function $\ff$ below will actually not use $\bA$ and $\bS$; they are
  useful for the proof of \cref{lem:lem3}
  (which proceeds by constructing a coupling).
It should also be noted that the dependence among $\bT,\bA,\bS$ and $\bB$ is only between $\bA$ and $\bS$.
The distribution $\bcalD$ is set to $\calU$, the uniform distribution over $\{0,1\}^n$.

The function $\ff$ is defined as follows using $\bT$ and $\bB$: Given any $x\in \{0,1\}^n$,
\begin{enumerate}
\item If $\Gamma_{\bT}(x)=1^*$ or $|\Gamma_{\bT}(x)|=2$, set $\ff(x)=1$;
\item If $\Gamma_{\bT}(x)=0^*$ or $\Gamma_{\bT}(x)=\emptyset$, set $\ff(x)=0$;
\item Otherwise, we have $\Gamma_{\bT}(x)=\{\ell\}$ for some $\ell\in [N]$, and set $\ff(x)=\bb_\ell$.
\end{enumerate}

\subsubsection{Construction of \texorpdfstring{$\Dno$}{Dno}}

To draw $(\bg,\bcalD)\sim \Dno$, we first draw the same random objects $\bT,\bA,\bS$ and $\bB$ as in $\Dyes$, and set $\bcalD$ to be the uniform distribution over the $2L$ points in
the support $\bS$. 

\begin{figure}[H]
\centering
\begin{tikzpicture}[
  x=1cm,y=1.32cm,
  font=\footnotesize,
  >={Latex[length=2.0mm,width=1.35mm]},
  cube/.style={draw=black!72,line width=.85pt},
  zoom/.style={->,draw=black!78,line width=.85pt},
  support/.style={circle,draw=black,fill=black,minimum size=2.4mm,inner sep=0pt},
  witness/.style={circle,draw=black,double,double distance=.55pt,fill=black,
                  minimum size=3.8mm,inner sep=0pt}
]

\coordinate (cube-top) at (4.00,11.20);
\coordinate (cube-right) at (7.75,6.20);
\coordinate (cube-bottom) at (4.00,1.20);
\coordinate (cube-left) at (.25,6.20);

\fill[blue!18] (cube-top) -- (6.25,8.20) -- (1.75,8.20) -- cycle;
\fill[orange!14] (1.75,4.20) -- (6.25,4.20) -- (cube-bottom) -- cycle;
\fill[black!2]
  (1.75,4.20) -- (6.25,4.20) -- (cube-right) --
  (6.25,8.20) -- (1.75,8.20) -- (cube-left) -- cycle;

\begin{scope}
  \clip (1.75,4.20) -- (6.25,4.20) -- (cube-right) --
        (6.25,8.20) -- (1.75,8.20) -- (cube-left) -- cycle;
  \path[fill=yellow!12,draw=black!35,line width=.55pt]
    (1.88,4.42) -- (4.12,4.42) -- (5.55,8.02) -- (3.40,8.02) -- cycle;
  \path[fill=yellow!12,draw=black!35,line width=.55pt]
    (3.08,4.42) -- (5.42,4.42) -- (4.68,8.02) -- (2.42,8.02) -- cycle;
    \begin{scope}
    \clip (1.88,4.42) -- (4.12,4.42) -- (5.55,8.02) -- (3.40,8.02) -- cycle;
    \fill[blue!22]
      (3.08,4.42) -- (5.42,4.42) -- (4.68,8.02) -- (2.42,8.02) -- cycle;
  \end{scope}
    \fill[pattern={Lines[angle=45,distance=3.2pt,line width=.28pt]},
        pattern color=black!38]
    (4.35,4.55) -- (5.18,4.55) -- (5.03,5.55) -- (4.69,5.55) -- cycle;
  \draw[draw=black!78,line width=1.0pt]
    (4.35,4.55) -- (5.18,4.55) -- (5.03,5.55) -- (4.69,5.55) -- cycle;
\end{scope}

\draw[cube] (cube-top) -- (cube-right) -- (cube-bottom) -- (cube-left) -- cycle;
\draw[draw=black!48,dashed,line width=.65pt] (1.75,8.20) -- (6.25,8.20);
\draw[draw=black!48,dashed,line width=.65pt] (1.75,4.20) -- (6.25,4.20);

\node[anchor=east] at (1.58,8.20) {$n/2+\Delta$};
\node[anchor=east] at (1.58,4.20) {$n/2-\Delta$};
\node[align=center] at (4.00,9.52) {$\Gamma_{\bT}=1^*$\\[-1pt]\bfseries $\bg=1$};
\node[align=center] at (4.00,2.82) {$\Gamma_{\bT}=0^*$\\[-1pt]\bfseries $\bg=0$};

\node[align=center] at (1.65,5.48)
  {no term\\[-1pt]\bfseries $\bg=0$};
\node[align=center] at (3.78,5.18)
  {$\ge2$ terms\\[-1pt]\bfseries $\bg=1$};
\node[fill=white,inner sep=1pt] at (4.86,5.67) {$C_\ell$};
\node[font=\bfseries,fill=white,inner sep=1.5pt] (cell-label) at (4.78,5.05)
  {item (3)};

\node[font=\small] at (4.00,.70) {$\{0,1\}^n$};

\coordinate (branch) at (8.20,6.75);
\draw[zoom] (5.08,5.28) -- (branch);
\draw[zoom] (branch) --
  node[pos=.38,sloped,above,fill=white,inner sep=.7pt,font=\scriptsize]
    {$\ell$ not qualifying}
  (9.28,8.72);
\draw[zoom] (branch) --
  node[pos=.52,sloped,above,fill=white,inner sep=1.2pt]
    {$\ell$ qualifying}
  (10.40,3.37);

\coordinate (nq-top) at (11.35,10.62);
\coordinate (nq-right) at (13.45,8.72);
\coordinate (nq-bottom) at (11.35,6.82);
\coordinate (nq-left) at (9.25,8.72);
\fill[black!5] (nq-top) -- (nq-right) -- (nq-bottom) -- (nq-left) -- cycle;
\draw[cube,draw=black!45] (nq-top) -- (nq-right) -- (nq-bottom) -- (nq-left) -- cycle;
\fill[yellow!12]
  (11.35,10.02) -- (12.75,8.72) -- (11.35,7.42) -- (9.95,8.72) -- cycle;
\draw[draw=black!65,dashed,line width=.65pt]
  (11.35,10.02) -- (12.75,8.72) -- (11.35,7.42) -- (9.95,8.72) -- cycle;
\node[text=black!72] at (11.35,9.52) {$C_\ell$};
\node[font=\bfseries\large] at (11.35,8.70) {$\bb_\ell$};
\node[text=black!68] at (12.55,10.00) {other $\Gamma$};
\node at (11.35,6.48) {$\{0,1\}^{\bA}$};

\coordinate (q-top) at (13.05,5.92);
\coordinate (q-right) at (15.70,3.37);
\coordinate (q-bottom) at (13.05,.82);
\coordinate (q-left) at (10.40,3.37);
\fill[black!5] (q-top) -- (q-right) -- (q-bottom) -- (q-left) -- cycle;
\draw[cube,draw=black!45] (q-top) -- (q-right) -- (q-bottom) -- (q-left) -- cycle;

\begin{scope}
  \clip (13.05,5.52) -- (15.18,3.37) -- (13.05,1.22) -- (10.92,3.37) -- cycle;
  \fill[blue!18] (10.70,3.37) rectangle (15.40,5.75);
  \fill[orange!14] (10.70,1.00) rectangle (15.40,3.37);
  \fill[pattern={Lines[angle=45,distance=3.2pt,line width=.27pt]},
        pattern color=black!30] (10.70,2.99) rectangle (15.40,3.75);
\end{scope}
\draw[draw=black!65,dashed,line width=.65pt]
  (13.05,5.52) -- (15.18,3.37) -- (13.05,1.22) -- (10.92,3.37) -- cycle;
\draw[draw=black!75,line width=.75pt] (10.92,3.37) -- (15.18,3.37);
\draw[draw=black!42,dashed] (11.35,3.75) -- (14.75,3.75);
\draw[draw=black!42,dashed] (11.35,2.99) -- (14.75,2.99);

\node[font=\bfseries\large] at (12.43,4.66) {$\bb_\ell$};
\node[font=\bfseries\large] at (12.40,2.45) {$1-\bb_\ell$};
\node[anchor=west,fill=white,inner sep=1pt] at (14.62,3.75) {$r/2+h$};
\node[anchor=west,fill=white,inner sep=1pt] at (14.62,3.37) {$r/2$};
\node[anchor=west,fill=white,inner sep=1pt] at (14.62,2.99) {$r/2-h$};

\node[witness] (yp) at (13.65,4.55) {};
\node[witness] (ym) at (13.65,2.18) {};
\draw[->,draw=black!82,line width=.72pt,line join=round]
  (ym.north) -- (13.83,2.70) -- (13.49,3.06) -- (13.83,3.42) --
  (13.49,3.78) -- (13.83,4.14) -- (yp.south);
\node[anchor=west,fill=blue!18,inner sep=1pt] at (13.85,4.53) {$y^+$};
\node[anchor=west,fill=orange!14,inner sep=1pt] at (13.85,2.16) {$y^-$};
\node[support] at (13.10,4.92) {};
\node[support] at (12.85,1.78) {};

\node[font=\scriptsize,text=black!72,fill=white,inner sep=1pt] at (12.25,3.16)
  {no $\bS\cap C_\ell$ here};
\node at (13.05,.47) {$\{0,1\}^{\bA}$};
\node[text=black!68] at (14.66,5.28) {other $\Gamma$};

\node[support] at (7.95,2.10) {};
\node[anchor=west] at (8.12,2.10) {$y\in\bS\cap C_\ell$};
\node[align=left] at (8.82,.78)
  {$\bb_\ell=0:$ \textbf{violation}\\[-1pt]
   $\bb_\ell=1:$ ordered};

\end{tikzpicture}
\caption{Schematic diagram depicting the essential features of a function
$\bg$ from $\Dno$.  The large diamond
is a depiction of the whole $\{0,1\}^n$ cube; its middle slab shows regions where no term is satisfied, where $\geq 2$ terms are satisfied, and the highlighted region $C_\ell$ where a single term $T_\ell$ is satisfied.  The two arrows lead to the
possible cases for the highlighted index.  Each right-hand rhombus is a copy of
$\{0,1\}^{\bA}$ obtained by fixing the coordinates outside $\bA$; only the
colored inner region lies in $C_\ell$.  In the qualifying case, all points of
$\bS\cap C_\ell$ avoid the striped $h$-band and some matched pair
$y^-\prec y^+$ crosses the center.  In the nonqualifying case,
$\bg=\bb_\ell$ throughout the colored region.  The displayed pair is a
violation exactly when $\bb_\ell=0$.}
\label{fig:dno-construction}
\end{figure}
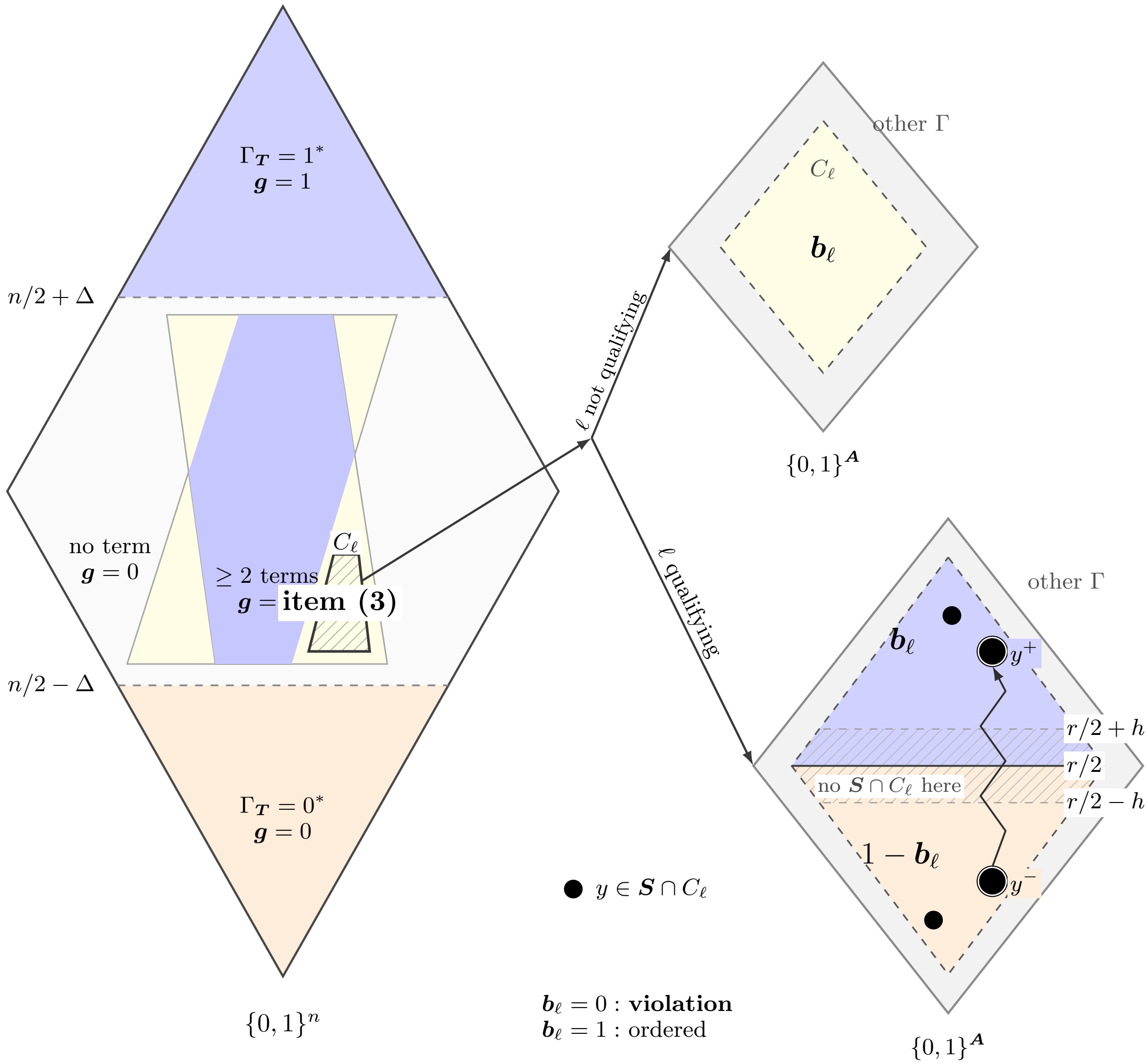

Next we define the function $\bg$:
Given any $x\in \{0,1\}^n$, 
\begin{enumerate}
\item If $\Gamma_\bT(x)=1^*$ or $|\Gamma_\bT(x)|=2$, set $\bg(x)=1$;
\item If $\Gamma_\bT(x)=0^*$ or $\Gamma_\bT(x)=\emptyset$, set $\bg(x)=0$.
\item Otherwise, we are left with the case when $\Gamma_\bT(x)=\{\ell\}$ for some $\ell\in [N]$.
For this case, we say index $\ell$ is \emph{qualifying} (with respect to $\bT$,
$\bS$ and $\bA$) if (i)~there exists at least one matched pair $(y^-,y^+)$ in $\bS$ satisfying
$\Gamma_\bT(y^-)=\Gamma_\bT(y^+)=\{\ell\}$ and 
(ii)~every point $y\in \bS$ with $\Gamma_\bT(y)=\{\ell\}$ satisfies
\[
    \abs{|y_\bA|-\frac{r}{2}}\ge h \,.
\]
Then we have the following two cases: (recall that $r$ is an odd integer)
\begin{flushleft}\begin{enumerate}
\item If $\ell$ is qualifying, set 
\begin{align*}
  \bg(x)=\begin{cases}\bb_\ell & \text{if $|x_\bA|> r/2$}\\
  1-\bb_\ell & \text{if $|x_\bA|<r/2$}\end{cases}
\end{align*}
\item Otherwise, set $\bg(x)=\bb_\ell$.
\end{enumerate}\end{flushleft}
\end{enumerate}
See \cref{fig:dno-construction} for a depiction of this construction.

\clearpage

\subsection{The YES Functions are Monotone}

\begin{proof}[Proof of \Cref{lem:lem1}]

Given $T$ and $B$, let $f$ denote the function built in $\Dyes$. We show below that $f$ is monotone.
Fixing any $x\preceq y$, it suffices to show that $f(x)=1$ implies $f(y)=1$.
Consider the following cases:
\begin{flushleft}\begin{enumerate}
\item $f(x)=1$ because $\Gamma_T(x)=1^*$:
In this case, $y$ must also be above the middle layers and thus, $\Gamma_T(y)=1^*$ and $f(y)=1$.
\item $f(x)=1$ because $|\Gamma_T(x)|=2$:
In this case, $y$ must either be above the middle layers, or be in the middle layers and satisfy at least two
  terms. As a result, we also have $f(y)=1$.
\item $f(x)=1$ because $\Gamma_T(x)=\{\ell\}$ for some $\ell\in [N]$ and $b_\ell=1$:
In this case, $y$ must either be above the middle layers, or be in the middle layers and satisfy at least two terms,
  or $\Gamma_T(y)=\{\ell\}$; all these three cases lead to $f(y)=1$.
\end{enumerate}\end{flushleft}
This finishes the proof that $f$ is monotone.
\end{proof}

\subsection{The NO Functions are Far from Monotone}

For a fixed $A$, let $\mathcal{P}_A$ be the collection of oriented edges
$(x^-,x^+)\in M_A$ satisfying
\begin{equation}
  \left||x^-_{\overline A}|-\frac{n-r}{2}\right|\leq\frac\Delta4
  \quad\text{and}\quad
    h \leq \frac{r}{2} - |x^-_A| \leq 2h.
  \label{eq:drpt-candidate-edges}
\end{equation}
Both endpoints of every edge in $\mathcal{P}_A$ lie in the middle layers, by
\eqref{eq:drpt-scale}.
Moreover, we have
\begin{equation}
  |\mathcal{P}_A|=\Theta(\rho 2^n) \,,
  \label{eq:drpt-candidate-count}
\end{equation}
since for a uniform $\bm{x} \sim \zo^n$, the first condition
in  \eqref{eq:drpt-candidate-edges}
holds with probability $\Theta(\Delta/\sqrt{n}) = \Theta(\rho)$, the second
condition holds with constant probability, and the two events are independent.

For a fixed $T$, call an edge $(x^-,x^+)\in\mathcal{P}_A$ \emph{good} if there is an index $\ell$ such
that
\begin{equation}
  \Gamma_T(x^-)=\Gamma_T(x^+)=\{\ell\}
  \label{eq:drpt-good-label}
\end{equation}
and the term $T_\ell$ contains at least $r/10$ distinct coordinates.  Given $T$ and $A$, let
$M^*(T,A)$
be the set of good edges.

\begin{claim}\label{claim:many-good-edges}
There are absolute constants $c_*,p_*>0$ such that, for every fixed $A$,
\begin{equation}
  \Pru{\bT}{|M^*(\bT,A)|\geq c_*\rho2^n} \geq p_*.
  \label{eq:drpt-many-good}
\end{equation}
\end{claim}

\begin{proof}
For each point $x \in \zo^n$, write
$p_x \define \Pr[\bm{T}_\ell(x)=1]$ (note that this is the same for all $\ell$). Then for any $x$ in the middle layers, \ie
$\abs{x} \in [n/2-\Delta, n/2+\Delta]$, \eqref{eq:drpt-parameters} gives
\begin{equation} \label{eq:nice}
  p_x
  = \pbra{\frac{|x|}{n}}^r
  = 2^{-r} (1 \pm O(\Delta/n))^r
  = 2^{-r}\exp(\pm O(r\Delta/n))
  = \Theta(2^{-r}) \,.
  \end{equation}
Fix $(x^-,x^+)\in\mathcal{P}_A$, which are in the middle layers as noted earlier.
Since $x^- \prec x^+$,
\eqref{eq:drpt-good-label} holds iff
there is exactly one term
$\bT_\ell$
satisfied by $x^+$ and that
term is already satisfied by $x^-$,
which occurs with
probability
\begin{align*}
    Np_{x^-}(1-p_{x^+})^{N-1}
  = \Theta(1) \,. \tag{by~\eqref{eq:nice} applied to both $x^+$ and $x^-$}
\end{align*}
Conditional on such a unique $\bT_\ell$ existing,
its $r$ coordinates are drawn independently and uniformly from
the $\abs{x^-} = \Theta(n)$ one-coordinates of $x^-$. Hence
edge $(x^-, x^+) \in \cP_A$ is good with probability
\begin{align*}
    &\Pru{\bT}{\text{\eqref{eq:drpt-good-label} holds}}
        \cdot \Pruc{}{\bT_\ell \text{ contains at least $r/10$ distinct coordinates}}{
        \bT_\ell(x^-) = 1} \\
    &\quad = \Theta(1) \cdot (1 - o_n(1))
    = \Theta(1) \,.
\end{align*}
Thus $\Exu{\bT}{\abs{M^*(\bT,A)}} \ge \Omega(1) \cdot \abs{\cP_A}$.
Since $\abs{M^*(\bT,A)} \le \abs{\cP_A}$, the claim follows by Markov's inequality
along with \eqref{eq:drpt-candidate-count}.
\end{proof}

\begin{proof}[Proof of \Cref{lem:lem2}]
Condition on outcomes $A,T$ of $\bA,\bT$ for which $\abs{M^*(T,A)} = \Omega(\rho 2^n)$,
which occurs with probability $\Omega(1)$ by \cref{claim:many-good-edges}.
Since $\bS$ consists of $L$
uniformly chosen edges of $M_A$, a hypergeometric tail bound shows that, with
probability $1-o_n(1)$ over $\bS$, at least $c \rho L$ of the edges selected for $\bS$ belong to
$M^*$.

We next show that, with probability $1-o_n(1)$, every
good edge in $\bS$
has a qualifying index 
and that their indices are all distinct.
By the definitions of $\cP_A$ and $M^*$,
we already have that each good edge $(\bx^-, \bx^+)$ in $\bS$ has
$\Gamma_T(\bx^-) = \Gamma_T(\bx^+) = \{\ell\}$ for
$T_\ell$ containing at least $r/10$ distinct coordinates, and moreover
$\abs{\abs{\bx^-_A}-r/2}, \abs{\abs{\bx^+_A}-r/2} \ge h$ (by \eqref{eq:drpt-candidate-edges}, recalling that $|\bx^+_A|+|\bx^-_A|=r$ by \Cref{lem:action-matching}).
Therefore it suffices to show that, with probability $1-o_n(1)$, for every good edge
$(\bx^-, \bx^+)$ in $\bS$
and different (possibly non-good) edge $(\by^-, \by^+)$ in $\bS$, neither $\by^-$
nor $\by^+$ satisfies the $T_\ell$ corresponding to $(\bx^-, \bx^+)$.
A uniformly random edge
of $M_A$ has an endpoint satisfying such $T_\ell$ with probability
at most
$2^{1-r/10}$,
and hence a union bound over the at most $L^2$ pairs of edges in $\bS$ gives failure
probability
\[
    O(L^2 2^{-r/10}) = o_n(1)
\]
by the choice of $c_1$. Thus the claim holds.

For every
good edge $(\bx^-,\bx^+)$ in $\bS$ with
distinct qualifying index $\ell$ as above,
the definition of $\bg$ gives
\[
  \bg(\bx^-)=1-\bb_\ell,
  \qquad
  \bg(\bx^+)=\bb_\ell.
\]
These
$\bb_\ell$'s are independent, so a Chernoff bound shows that,
with probability $1 - o_n(1)$,
$\Omega(\rho L)$ of these disjoint comparable edges satisfy
$\bg(\bx^-)=1$ and $\bg(\bx^+)=0$.  Any monotone function disagrees with $\bg$
on at least one endpoint of every such edge.  Since $\bcalD$ is uniform on the
$2L$ points of $\bS$, it follows that
\[
  \dist_{\bcalD}(\bg,\Mono)=\Omega(\rho)>\epsilon \,. \qedhere
\]
\end{proof}

\subsection{Indistinguishability}

\def\hh{\boldsymbol{h}}

Fix a deterministic, non-adaptive algorithm $\ALG$ with $s$ samples and $q$ queries, and let
\begin{align*}
\alpha:=\Pr_{(\ff,\calU)\sim \Dyes}\big[\ALG\ \text{accepts}\ (\ff,\calU)\big]\quad\text{and}\quad
\beta :=\Pr_{(\bg,\bcalD)\sim \Dno} \big[\ALG\ \text{accepts}\ (\bg,\calD)\big].
\end{align*}
We would like to show that $|\alpha-\beta|\le o_n(1)$.
To ease the proof, we introduce a new algorithm $\ALG'$ derived from $\ALG$ as follows.
Letting $(\hh,\bcalD)$ be a pair drawn from $\Dyes$~or $\Dno$, with
  $\bT,\bA,\bS$ and $\bB$ drawn during the process, $\ALG'$ runs on $(\hh,\bcalD)$ as follows:
\begin{flushleft}\begin{enumerate}
\item Phase 0: Recall that $\bS$ contains $L$ pairs (and thus, $2L$ points) in the perfect matching $M_\bA$.
At the beginning, $\ALG'$ receives $L$ points, one from each pair in $\bS$ drawn uniformly at random.
Let $\bR\subset \bS$ be this random set of size $L$.
$\ALG'$ receives $\bR$ and their values under $\hh$.

\item Phase 1: $\ALG'$ draws a sequence of $s$ samples $\bx^1,\ldots,\bx^s$ independently and 
  uniformly from $\bR$ and sends $(\bx^1,\hh(\bx^1)),\ldots,(\bx^s,\hh(\bx^s))$ to $\ALG$ as
  samples the latter receives in its Phase 1.

\item Phase 2: $\ALG'$ makes the same batch of $q$ queries on $\hh$ as $\ALG$ makes in Phase 2.

\item Phase 3: $\ALG'$ returns the same answer (accept or reject) as $\ALG$ does in Phase 3.
\end{enumerate}\end{flushleft}
With $\ALG'$ defined from $\ALG$ as above, we let 
\begin{align}\label{eq:defalphabeta}
\alpha':=\Pr_{(\ff,\calU)\sim \Dyes}\big[\ALG'\ \text{accepts}\ (\ff,\calU)\big]\quad\text{and}\quad
\beta' :=\Pr_{(\bg,\bcalD)\sim \Dno} \big[\ALG'\ \text{accepts}\ (\bg,\bcalD)\big].
\end{align}
Then \cref{lem:lem3} follows from the following three claims:

\begin{claim}\label{claim:simple1}
$|\alpha-\alpha'|\le o_n(1)$.
\end{claim}

\begin{claim}\label{claim:simple2}
$|\beta-\beta'|\le o_n(1)$.
\end{claim}

\begin{claim}\label{claim:main}
$|\alpha'-\beta'|\le o_n(1)$.
\end{claim}

\cref{claim:simple1,claim:simple2} are easy to prove:

\begin{proof}[Proof of \cref{claim:simple1}]
It suffices to show that the samples received by the simulation
of $\ALG$ in $\ALG'$ are indistinguishable from the uniformly distributed samples
received by the original $\ALG$.
The probability that $s$ samples from the uniform
distribution on $\bR$ use the same point of $\bR$ twice is $O(s^2/L)=o_n(1)$.
The distribution of $s$ samples drawn from $\bR$ without replacement is
the same as the distribution of $s$ \iid points drawn uniformly
from $\zo^n$ conditional on no two points lying on the same edge of $M_A$;
the latter
conditioning fails with probability at most $O(s^2/2^n)=o_n(1)$.
\end{proof}

\begin{proof}[Proof of \cref{claim:simple2}]
Compare sampling from the $2L$ endpoints in $\bS$ with
sampling from one uniformly random endpoint of each edge.  The two experiments
can be coupled perfectly unless the same edge is sampled twice,
which has probability $O(s^2/L)=o_n(1)$.
\end{proof}

We now prove \cref{claim:main} for a more general class of deterministic algorithms $\ALG^*$.
Letting $(\bh,\bcalD)$ be a pair drawn from either $\Dyes$ or $\Dno$, with $\bT,\bA,\bR,\bB$ being
  the random objects\footnote{Note that we will skip $\bS$ in the list from now on because it is uniquely determined by $\bA$ and $\bR$.} drawn during the process (recall that they are drawn from the same distribution
  in both cases), $\ALG^*$ runs as follows:
\begin{flushleft}\begin{enumerate}
\item Phase 1: At the beginning, 
$\ALG^*$ receives $\bR$. Moreover, for each $x\in \bR$, $\ALG^*$ also receives
  $\bgamma_x=\Gamma_\bT(x)$ and $\bt_x=\bh(x)$.
  We write $\bgamma=(\bgamma_x:x\in \bR)$ and $\bt=(\bt_x:x\in \bR)$.

\item Phase 2: $\ALG^*$ makes a nonadaptive batch of $q$ queries on $\bh$ based on $\bR,\bt$ and $\bgamma$.

\item Phase 3: $\ALG^*$ accepts or rejects based on the information received in Phases 1 and 2.\end{enumerate}\end{flushleft}
We similarly define $\alpha^*$ and $\beta^*$ as the two probabilities given in \Cref{eq:defalphabeta}, with $\ALG'$ replaced by $\ALG^*$.
Given that $\ALG'$ is clearly a special case of $\ALG^*$,
to prove \Cref{claim:main}, it suffices to show that $|\alpha^*-\beta^*|=o_n(1)$ for any $\ALG^*$.

To this end, we consider the following event over the three objects $\bT,\bA$ and $\bR$:

\begin{quote}\emph{
Event $E$: No two distinct points $y, z \in \bR$ satisfy any of the following conditions:
\begin{enumerate}
  \item[(a)] $y$ and $z$ have Hamming distance smaller than $n/3$; or
  \item[(b)] $y$ and $z$ are in the middle layers and
  $\bT_\ell(y) = \bT_\ell(z) = 1$ for some $\ell \in [N]$.
\end{enumerate}
}\end{quote}

Note that event $E$ implies that no distinct $y,z \in \bR$ and $\ell
\in [N]$ satisfy $\ell \in \Gamma_{\bT}(y) \cap \Gamma_{\bT}(z)$.

\begin{claim}\label{claim:event-E}
The event $E$ has probability $1-o_n(1)$.
\end{claim}

\begin{proof}
For part~(a), fix any outcome $A$ of $\bA$. Notice that the $\bar A$
part of the strings in $\bR$ are drawn \iid uniformly from $\zo^{\bar A}$, and
two such strings are at least $n/3$ apart in Hamming distance with probability
$1 - e^{-\Omega(n)}$ by a Chernoff bound since $\abs{\bar A} = n - r = n -
o(n)$. By a union bound over the $L^2 = e^{O(r)} = e^{o(n)}$ pairs in $\bR$,
this holds simultaneously for all $L^2$ pairs with probability $1 -
o_n(1)$.

For part~(b), fix $A$ and $R$ such that (a) holds, let $\bT \sim \cT$ be random,
and fix two points $y,z$ in the middle layers
at distance at least $n/3$. Then they have at most $n/3+\Delta$ common
one-coordinates, since otherwise one of the points would have
Hamming weight larger than $\frac{n}{3}+\Delta + \frac{1}{2} \cdot \frac{n}{3}
= \frac{n}{2} + \Delta$, contradicting the middle-layer hypothesis.
Hence the probability that one of the $N=2^r$ terms in $\bT$ is
satisfied by both $y$ and $z$ is at most, by a union bound,
\[
  N\left(\frac{n/3+\Delta}{n}\right)^r
  \leq\left(\frac23+o_n(1)\right)^r.
\]
A union bound over all (at most $L^2$) pairs of distinct points $y,z \in \bR$ proves the claim when $c_1$ is sufficiently small.
\end{proof}

Given the above claim, it suffices to show that conditioning on $\bT,\bA,\bR$ satisfying event $E$, the two probabilities $\alpha^*$ and $\beta^*$ are within $o_n(1)$ of each other. For this, we next show that conditioning $\bT,\bA$ and $\bR$ (which are drawn in the same way in $\Dyes$ and $\Dno$) on $E$, the distribution of $(\bR,\bt,\bgamma)$ that $\ALG^*$ receives in Phase 1 in the case of $\Dyes$ is the same as in the case of $\Dno$:

\begin{claim}
Conditioning on event $E$, the distribution of $(\bR,\bt,\bgamma)$ is the same for $\Dyes$ and $\Dno$.
\end{claim}
\begin{proof}
Fix any $T,A,R$ such that event $E$ holds.
Given that $T,A$ and $R$ are fixed, both $\bR$ and $\bgamma$ are fixed.
Given $R$ and $\gamma$, $\bt_x$ of an $x\in R$ is distributed as follows: If $\Gamma_T(x)=0^*,1^*,\emptyset$ or a set of size $2$, then $\bt_x$ is the same fixed value in $\{0,1\}$ in both cases. Otherwise (i.e., $\Gamma_T(x)=\{\ell\}$ for some $\ell\in [N]$), given that no other point $y\in R$ has $\Gamma_T(y)=\{\ell\}$ (as given by condition (b) of event $E$), $\bt_x$ is an independent uniform random bit drawn from $\{0,1\}$.
\end{proof}

To finish the proof, we fix any triple $(R,t,\gamma)$ that $\ALG^*$ may receive under event $E$, and show that conditioning on event $E$ and receiving $(R,t,\gamma)$ in Phase 1, $\ALG^*$ accepts with probabilities within $o_n(1)$ of each other in $\Dyes$ and $\Dno$. 

For this purpose, we first note that by event $E$, set $R$ and tuple $\gamma$ must satisfy
\begin{enumerate}
\item Every two distinct points $y,z\in R$ have Hamming distance at least $n/3$; and  
\item No two distinct points $y,z\in R$ have $\ell\in \gamma_y\cap \gamma_z$ for some $\ell\in [N]$. 
\end{enumerate}
Conditioning on $E$ and $\ALG^*$ receiving $(R,t,\gamma)$,
the objects $\bT,\bA$ and $\bB$ in the two cases are distributed as follows.
First, $\bT$ and $\bA$ are independently distributed in the same way in the two cases as follows:
\begin{flushleft}\begin{enumerate}
\item $\bA$ is drawn from $\calA$ (i.e., uniform over all $r$-subsets; this is because every two points $y,z\in R$ have Hamming distance at least $n/3$ and thus, always have $y_{\overline{A}}\ne z_{\overline{A}}$ and hence do not belong to the same
edge of $M_A$, for any $A$);
\item $\bT$ is uniform over all $T$ satisfying $\Gamma_T(y)=\gamma_y$ for all $y\in R$. In particular, for any $\ell\in [N]$ such that $\gamma_y=\{\ell\}$ for some $y\in R$ (which must be unique), $\bT_\ell$ is uniform over all $T_\ell$ such that 
\begin{itemize}
\item $T_\ell(y)=1$;
\item $T_\ell(z)=0$ for all $z\in R$ with $\gamma_z=\emptyset$;
\item $T_\ell(z)=0$ for all $z\in R$ with $\gamma_z=\{\ell'\}$ and $\ell' \ne \ell$;
\item $T_\ell(z)=0$ for all $z\in R$ with $\gamma_z=\{\ell_1,\ell_2\}$ and $\ell<\max(\ell_1,\ell_2)$.
\end{itemize}
Looking ahead, we will show in \Cref{claim:simpleclaim} that this distribution of $\bT_\ell$ has very small total variation distance from just drawing $r$ random variables from $\{i\in [n]: y_i=1\}$.
\end{enumerate}\end{flushleft}

Next, with $T,A$ and $R$ fixed, $\bB=(\bb_1,\ldots,\bb_N)$ is distributed in the $\Dyes$ case as follows:
\begin{flushleft}\begin{enumerate}
    \item For any $\ell\in [N]$, if no $y\in R$ has $\gamma_y=\{\ell\}$, then $\bb_\ell$ is an independent uniform random bit; 
    \item If there exists a $y\in R$ with $\gamma_y=\{\ell\}$ (which must be unique), $\bb_\ell$ is fixed to be $t_y$.
\end{enumerate}\end{flushleft}
And $\bB$ is distributed in the $\Dno$ case as follows:
\begin{flushleft}\begin{enumerate}
    \item For any $\ell\in [N]$, if no $y\in R$ has $\gamma_y=\{\ell\}$, then $\bb_\ell$ is an independent uniform random bit;
    \item If there exists a $y\in R$ with $\gamma_y=\{\ell\}$ (which must be unique), then we have two cases: (1) if $\ell$ is qualifying (given $T,A$ and $S$, the latter of which can be inferred from $A$ and $R$), then $\bb_\ell$ is fixed to be $t_y$ if $|y_A|>r/2$, or $1-t_y$ if $|y_A|<r/2$;
    (2) if $\ell$ is not qualifying, then $\bb_\ell$ is fixed to be $t_y$. 
\end{enumerate}\end{flushleft}

Given the above analysis of the distribution of $\bT,\bA$ and $\bB$ in the two cases, it is natural to 
 consider the following coupling of $\bT,\bA,\bB$ in the two cases. $\bT$ and $\bA$ are drawn as above and are shared in the two cases;
 after $\bT$ and $\bA$ are drawn, we draw $\bb_\ell$ as an independent uniform bit for each $\ell\in [N]$ such that no $y\in R$ has $\gamma_y=\{\ell\}$, which is the same in the two cases;
 for the rest of the $\bb_\ell$'s, they are set using $t$ in the two cases as described above (in different ways, when $\ell$ is qualifying).

We record the following claim about the distribution of $\bT_\ell$ when $\gamma_y=\{\ell\}$ for some $y\in R$:
\begin{claim}\label{claim:simpleclaim}
Condition on any $(R, t, \gamma)$ under event $E$. Suppose $\gamma_y = \{\ell\}$
for some $y \in R$. Then
the distribution of $\bT_\ell$ has total variation distance at most $(4/5)^r$ from the uniform distribution over all terms in which all $r$ variables are drawn
independently from $\{i\in [n]: y_i=1\}$. 
\end{claim}
\begin{proof}
Let $R'$ be the subset of $z\in R$ described above that must have $\bT_\ell(z)=0$.
Then $\bT_\ell$ is uniform over all $T_\ell$ that satisfies $T_\ell(y)=1$ and $T_\ell(z)=0$ for all $z\in R'$.
Given that every $z\in R'$ must lie in the middle layers and has Hamming distance at least $n/3$ from $y$, we have
\[
    \big|\{i:y_i=1\land z_i=0\}\big|\ge n/6-O(\Delta) \,.
\]
As a result, recalling that $\Delta = o(n)$, if we draw $\bT_\ell$ as $r$ uniform and independent draws from $\{i \in [n]:y_i=1\}$, the probability of $\bT_\ell(z)=1$ is at most $(3/4)^r$
for sufficiently large $n$.
The claim follows from a union bound over the at most $|R| = L$ points in $R'$ and by choosing the constant $c_1$ in $L$ to be sufficiently small.
\end{proof}

Given $(R,t,\gamma)$, let $Q$ be the set of $q$ queries that $\ALG^*$ makes in Phase $2$. 
To finish the proof, it suffices to show that for $\bT,\bA,\bB$ drawn in the two cases under the coupling described earlier, we have with probability at least $1-o_n(1)$, $\ff(x)=\bg(x)$ for all $x\in Q$.
To this end, we first define the following sufficient condition for this to happen:

\newcommand{\Good}{\mathsf{Good}}

\begin{equation}
\begin{split}
\Good(T,A):\quad
&\text{there do not exist }y\in R,\ z\in Q,\ \ell\in[N]\text{ such that}\\
&\gamma_y=\Gamma_T(z)=\{\ell\}
\quad\text{and}\quad
\big||y_A|-|z_A|\big|\geq h.
\end{split}
\label{eq:drpt-good-condition}
\end{equation}

\begin{claim}\label{claim:sufficientcond}
Suppose outcomes $T, A$ of $\bT, \bA$ satisfy $\Good(T, A)$. Then under the
coupling above, we have $\ff(x) = \bg(x)$ for all $x \in Q$.
\end{claim}
\begin{proof}
Fix $T$ and $A$ such that $\Good(T, A)$ holds, and consider any $x\in Q$. 
Clearly we have $\ff(x)=\bg(x)$ trivially unless $\Gamma_T(x)=\{\ell\}$ for some $\ell\in [N]$. If no $y\in R$ has $\gamma_y=\{\ell\}$, then $\bb_\ell$ is set to be the same in both cases by the coupling.
Given that $\ell$ cannot be qualifying (since no $y\in R$ has $\gamma_y=\{\ell\}$), we have in both cases that $\ff(x)$ and $\bg(x)$ are set to $\bb_\ell$.

If there exists a $y\in R$ with $\gamma_y=\{\ell\}$ (which is unique), then we always have $\ff(x)=t_y$.
For $\bg(x)$, we first note that  condition $\Good(T,A)$ implies~that $||y_A|-|x_A||<h$. If $\ell$ is not qualifying given $T,A$ and $R$, then we have $\bg(x)=t_y=\ff(x)$.
On the other hand, if $\ell$ is qualifying, then we must have $||y_A|-r/2|\ge h$;
note
that no other point $z\in R$ can even have $\gamma_y=\{\ell\}$.
Combining the two inequalities we have $\bg(x)=t_y$ as well by the definition of the coupling. This finishes the proof of the claim.
\end{proof}

Finally we show that $\Good(\bT, \bA)$ holds with high probability:

\begin{claim}
Conditional on every fixed $(R,t,\gamma)$ satisfying $E$,
$\Pr[\Good(\bT,\bA)]=1-o_n(1)$.
\end{claim}
\begin{proof}
Fix a $z\in Q$. We assume without loss of generality that $z$ is in the middle layers.
First, given that every two points in $R$ have Hamming distance at least $n/3$, there can be at most one point $y^*\in R$ with Hamming distance less than $n/6$ from $z$.

We first bound the probability of the condition failing 
because of some
$y\ne y^*$ in $R$ and $z$.
Assume without loss of generality that $\gamma_y=\{\ell\}$ for some $\ell\in [N]$.
Given that they have Hamming distance at least $n/6$ and are both in middle layers, with similar arguments to the ones used earlier, we have 
$$\big|\{i: y_i=1\land z_i=0\}\big|\ge n/12-O(\Delta).$$ 
Using \Cref{claim:simpleclaim}, the probability of $\bT_\ell(z)=1$ is at most
$\exp(-\Omega(r))$. A union bound over at most $Lq = \exp(O(r))$ pairs yields
contribution $o_n(1)$, for constants $c_0, c_1$ small enough.

Now consider the special $y^*$ (skip this case if such a $y^*$ does not exist).
Again assume without loss of generality that $\gamma_{y^*}=\{\ell\}$ for some $\ell\in [N]$.
Let $c = c_h/2$ and
consider the following two cases:
\begin{enumerate}
\item If the Hamming distance between $y$ and $z$ is less than $cn/\sqrt{r}$,
then since $\bA \sim \cA$ is uniform over $r$-subsets, the distribution of $\abs{y_{\bA}-z_{\bA}}$
is hypergeometric with expectation at most $c\sqrt{r}$, and hence
the probability that $||y_\bA|-|z_\bA||\ge h = \floor{c_h \sqrt{r}}$ is at most 
$\exp(-\Omega(\sqrt{r}))$;
\item If the Hamming distance between $y$ and $z$ is at least
$cn/\sqrt{r}$,
then similar to the previous arguments we have $|\{i:y_i=1\land z_i=0\}|\ge cn/(2\sqrt{r})-O(\Delta)$,
which is $\Omega(n/\sqrt{r})$,
since $\Delta = \Theta(n/r) = o(n/\sqrt{r})$.
Another application of \Cref{claim:simpleclaim} gives
\[
  \Pr[\bT_\ell(z)=1]\leq
  \left(1-\Omega(1/\sqrt r)\right)^r+\exp(-\Omega(r))
  =\exp(-\Omega(\sqrt r)) \,.
\]
\end{enumerate}
A union bound over the $q$ points $z \in Q$, each with at most one associated
$y^*$, yields contribution $o_n(1)$ as long as $c_0$ is sufficiently
small.
\end{proof}

This concludes the proof of \cref{claim:main}, and hence of \cref{lem:lem3} and
\cref{thm:mono-lb}.

\bibliographystyle{alpha}

\bibliography{allrefs}

\appendix

\newpage

\section{Omitted Proofs} \label{ap:conjDL}
\label{ap:omitted-proofs}

\subsection{Learning Decision Lists and Conjunctions}

\factuniflearnconjdl*
\begin{proof}[Proof sketch]
We consider the class of 1-decision-lists below (the approach also works for conjunctions).
\Cref{fact:unif-learn-conj-DL} follows from a standard Occam's Razor argument using the fact that
under the uniform distribution any 1-decision-list $f$ is $\epsilon/2$-close to some 1-decision-list
$f'$ that has length at most $k := \log(2/\epsilon)$ (simply truncate all nodes of $f$ after the
first $k$ nodes). 

In a bit more detail, the algorithm draws $q(\epsilon,\delta)$ uniform random labeled examples and
outputs the 1-decision-list of length at most $k$ that has the best agreement with these examples. A
standard multiplicative Chernoff bound gives that with probability at least $1-\delta/2$ the
1-decision-list $f'$ will disagree with at most a $3\epsilon/4$ fraction of the $q(\epsilon,\delta)$
examples, and a multiplicative Chernoff bound and union bound together give that with probability at
least $1-\delta/2$, none of the (at most $(O(n))^k$ many) 1-decision-lists of length at most $k$
that have true error rate greater than $\epsilon$ will disagree with less than a $3\epsilon/4$
fraction of the $q(\epsilon,\delta)$ examples.  So with probability at least $1-\delta$ the
1-decision-list that is output will have error at most $\epsilon$, as required.
\end{proof}

\subsection{Noise Sensitivity and Distance to Intervals}

\lemmaunionsintervalssoundnessstructural*
\begin{proof}
    We closely follow the proof of the uniform distribution case from \cite[Lemma~3.4]{BBBY12}.
    The proof proceeds in three steps. First, we construct a ``smoothed'' function $f_\rho : \R
    \to \zo$ by taking local averages of $f$ and a ``self-corrected'' function $g : \R \to \zo$ by
    applying a threshold to $f_\rho$, and show that $f$ is $\frac{\epsilon}{2}$-close to $g$ under
    $\cD$. Second, we show that $g$ is a union of $k \pbra{1 + \frac{\epsilon}{2}}$ intervals.
    Third, we show that every union of $k \pbra{1 + \frac{\epsilon}{2}}$ intervals is
    $\frac{\epsilon}{2}$-close to $\bbI_k$ under $\cD$.

    \paragraph*{Step 1.} Define $f_\rho : \R \to \R$ by $f_\rho(x) \define \Exu{\bm{y}
    \sim_\rho x}{f(\bm{y})}$. Let $\tau \define \frac{2}{\epsilon} k \rho
    \pbra{1+\frac{\epsilon}{4}} = \frac{\epsilon}{32} \pbra{1+\frac{\epsilon}{4}}$, so that
    $\bbNS_\rho^\cD(f) \le \frac{\epsilon}{2} \cdot \tau$ by hypothesis. Define $g^* : \R \to \{0,
    1, *\}$ by
    \[
        g^*(x) \define \begin{cases}
            1 & \text{if } f_\rho(x) \ge 1-\tau \\
            0 & \text{if } f_\rho(x) \le \tau \\
            * & \text{otherwise.}
        \end{cases}
    \]
    Then, define $g : \R \to \zo$ by $g(x) \define g^*(y)$ for $y = \sup \cbra{z \le x : g^*(z) \ne
    *}$. Note that, for each $x \in [0,1]$, if $\NS_\rho(f, x) \le \tau$ then $\abs{f_\rho(x) -
    f(x)} \le \tau$ and hence $g(x) = g^*(x) = f(x)$. Therefore, by the contrapositive and Markov's
    inequality,
    \[
        \dist_\cD(f, g)
        = \Pru{\bm{x} \sim \cD}{f(\bm{x}) \ne g(\bm{x})}
        \le \Pru{\bm{x} \sim \cD}{\NS_\rho(f, \bm{x}) > \tau}
        \le \frac{\bbNS_\rho^\cD(f)}{\tau}
        \le \frac{\epsilon}{2} \,.
    \]

    \paragraph*{Step 2.} We claim that $g$ is a union of $m \le k \pbra{1+\frac{\epsilon}{2}}$
    intervals. First, it is easy to check that $f_\rho$ is Lipschitz continuous with Lipschitz
    constant $\frac{1}{2\rho}$. By continuity, each left boundary point of an interval\footnote{Here
    we use the assumption that the support of $f$ is contained in $[1/4, 3/4]$
    to avoid boundary points near $0$ or $1$.} of $g^{-1}(1)$  occurs at some point $x \in [0,1]$
    contained in a minimal interval $[y, z] \subseteq [0,1]$ such that $f_\rho(y) \le \tau$,
    $f_\rho(z) \ge 1-\tau$, and $\tau \le f_\rho \le 1-\tau$ within $[y,z]$. By the intermediate value
    theorem, there exists $x^* \in [y, z]$ for which $f_\rho(x^*) = \frac{1}{2}$, and by the
    Lipschitz constant of $f_\rho$, we have that $z - x^* \ge 2\rho \pbra{\frac{1}{2}-\tau}$ and
    $x^* - y \ge 2\rho \pbra{\frac{1}{2}-\tau}$.
    Observing that $\NS_\rho(f, t) \ge \min\cbra{f_\rho(t), 1-f_\rho(t)}$,
    and letting $\ell \define 2\rho(\frac{1}{2}-\tau)$, we obtain
    \begin{equation}
        \label{eq:NS}
        \begin{aligned}
            \int_y^z \NS_\rho(f, t) \odif t
            &= \int_y^{x^*} \NS_\rho(f, t) \odif t + \int_{x^*}^z \NS_\rho(f, t) \odif t \\
            &\ge \int_0^\ell \min\cbra{f_\rho(x^*+s), 1-f_\rho(x^*+s)} \odif s
                + \int_0^\ell \min\cbra{f_\rho(x^*-s), 1-f_\rho(x^*-s)} \odif s \\
            &\ge 2\int_0^\ell \pbra{\frac{1}{2} - \frac{s}{2\rho}} \odif s
            = 2 \cdot \pbra{\frac{\ell}{2} - \frac{\ell^2}{4\rho}}
            = \ell \pbra{\frac{1}{2}+\tau}
            \ge 2\rho \pbra{\frac{1}{2}-\tau}^2 \,,
        \end{aligned}
    \end{equation}
    where the second inequality used that $f_\rho(x^*) = \frac{1}{2}$
    along with the Lipschitz property of $f_\rho$.
    Similarly, each right boundary point of an interval of $g^{-1}(1)$ yields a minimal interval $[y, z]
    \subseteq [0,1]$ satisfying \eqref{eq:NS}. Since the resulting $2m$ intervals corresponding to
    all left and right boundaries are all disjoint, we conclude that
    \[
        k \rho \pbra{1+\frac{\epsilon}{4}}
        \ge \bbNS_\rho^\cU(f)
        = \int_0^1 \NS_\rho(f, t) \odif t
        \ge 2m \cdot 2\rho \pbra{\frac{1}{2}-\tau}^2
        = m \rho (1 - 2\tau)^2 \,,
    \]
    and hence
    \[
        m \le \frac{k \pbra{1+\frac{\epsilon}{4}}}{(1 - 2\tau)^2}
        \le k \pbra{1+\frac{\epsilon}{2}} \,.
    \]

    \paragraph*{Step 3.} We claim that $g$ is $\frac{\epsilon}{2}$-close to some union of $k$
    intervals under $\cD$. Let $h : [0,1] \to \zo$ be obtained from $g$ by deleting its $m - k \le
    \frac{k \epsilon}{2}$ lightest intervals according to the measure assigned by $\cD$, \ie setting
    $h$ to $0$ on each of those intervals and copying the value of $g$ elsewhere. Then $h \in
    \bbI_k$, and since the total measure of all the intervals of $g$ under $\cD$ is at most $1$, we
    obtain that
    \[
        \dist_\cD(g, h) \le \frac{m - k}{m} = 1 - \frac{k}{m}
        \le 1 - \frac{k}{k \pbra{1+\frac{\epsilon}{2}}}
        \le \frac{\epsilon}{2} \,.
    \]
    We conclude by the triangle inequality that $\dist_\cD(f, \bbI_k) \le \dist_\cD(f, g) +
    \dist_\cD(g, h) \le \epsilon$.
\end{proof}

\subsection{Proper Learning of Monotone Functions under Product Distributions}
\label{section:appendix-monotone}

We used the following theorem to establish our upper bound on distribution-robust testing monotone
functions under the class $\Prodreal$ of product distributions over $\bbR^n$.

\thmmonotoneproduct*

In this section we will sketch a proof of this theorem, which does not appear in the literature, but
which is obtained by a combination of results of \cite{HarmsYoshida22,Black24}. The next theorem is
implicit in \cite{Black24}.

\begin{theorem}[\cite{Black24}]
\label{thm:hadley-improper-learner}
There exists an algorithm which, given access to samples from any absolutely continuous product
distribution $\cD$ labelled by a monotone function $f \colon \bbR^n \to \zo$, draws $2^{\widetilde
O(\sqrt n /\epsilon)}$ samples, uses time polynomial in the number of samples, and outputs the
following:
\begin{enumerate}
    \item A monotone function $\block \colon \bbR^n \to [N]^n$ where $N = \Theta(n/\epsilon)$;
    \item A function $g \colon [N]^n \to \zo$.
\end{enumerate}
With probability at least $9/10$, these satisfy the following properties:
\begin{enumerate}
    \item There exists a monotone function $h \colon [N]^n \to \zo$ such that
        $\Pru{\bm y \sim \Unif([N]^n)}{g(\bm y) \neq h(\bm y)} < \epsilon$.
    \item $\Pru{\bm x \sim \cD}{ g(\block(\bm x)) \neq f(\bm x) } < \epsilon$.
    \item $\dist_\TV(\block(\cD), \Unif([N]^n)) < \epsilon$, where $\block(\cD)$ is the distribution
        of $\block(\bm x)$ with $\bm x \sim \cD$.
\end{enumerate}
\end{theorem}

This is sufficient to prove \cref{thm:proper-learning-monotone-product} for \emph{continuous}
product distributions.
\begin{proof}[Proof of \cref{thm:proper-learning-monotone-product} for continuous distributions]
Our proper learner is defined as follows. Given labelled sample access to a monotone function $f
\colon \bbR^n \to \zo$ with samples from continuous product distribution $\cD$:
\begin{enumerate}
\item Run the algorithm in \cref{thm:hadley-improper-learner} with parameter $\epsilon/3$ to obtain
functions $\block$ and $g$, with $N = \Theta(n/\epsilon)$.
\item Compute (by brute force) the monotone function $h \colon [N]^n \to \zo$ that is closest to
$g$ under $\Unif([N]^n)$.
\item Output the function $x \mapsto h(\block(x))$.
\end{enumerate}
Since $h$ and $\block$ are both monotone, the output $h(\block(\cdot))$ is also monotone. Let us
verify that it is close to $f$ under $\cD$. Assume the improper learner succeeds. Then
\begin{enumerate}
\item $\Pru{\bm y \sim \Unif([N]^n)}{ g(\bm y) \neq h(\bm y) } < \epsilon/3$.
\item $\Pru{\bm x \sim \cD}{ g(\block(\bm x)) \neq f(\bm x) } < \epsilon/3$.
\item $\dist_\TV(\block(\cD), \Unif([N]^n)) < \epsilon/3$.
\end{enumerate}
We first claim that
\[
    \Pru{\bm x \sim \cD}{ g(\block(\bm x)) \neq h(\block (\bm x)) } < 2\epsilon/3 .
\]
Let $X \subset [N]^n$ be the set of points $y \in [N]^n$ where $g(y) \neq h(y)$. Then
\begin{align*}
    \dist_\cD(g(\block(\cdot)), h(\block(\cdot)))
        &= \Pru{\bm x \sim \cD}{ \block(\bm x) \in X }  \\
        &\leq \Pru{\bm y \sim \Unif([N]^n)}{ \bm y \in X }
            + \dist_\TV(\block(\cD), \Unif([N]^n)) \\
        &\leq 2\epsilon/3 ,
\end{align*}
which proves the claim. Now, by the triangle inequality,
\[
    \dist_{\cD}(f, h(\block(\cdot)))
    \leq \dist_\cD(f, g(\block(\cdot))) + \dist_\cD(g(\block(\cdot)), h(\block(\cdot)))
    < \epsilon . \qedhere
\]
\end{proof}

Let us now briefly sketch the proof for the more general case of $\Prodreal$, which contains
arbitrary product distributions. The method for doing this is described in Appendix A
of \cite{HarmsYoshida22} and it involves too many technicalities to responsibly include here. It
proceeds essentially as follows:
\begin{enumerate}
\item Extend the domain from $\bbR^n$ to $\bbR^n \times [0,1]^n$ by doing this:
    \begin{enumerate}
        \item Replace sample $\bm x \sim \cD$ with sample $(\bm x, \bm z) \sim \cD \times
            \Unif([0,1]^n)$.
        \item Define input $f' \colon \bbR^n \times [0,1]^n \to \zo$ as
            $f'(x,z) = f(x)$.
    \end{enumerate}
\item The above method for continuous distributions can be adapted to work for the augmented
    domain, and we learn a monotone function $h' \colon \bbR^n \times [0,1]^n \to \zo$
    that is $\epsilon$-close to $f'$.
\item There is a rounding procedure which takes the monotone function $h'$ (whose values on each
restriction $h'(x, \cdot)$ may not be constant) to a monotone function $h \colon \bbR^n \to \zo$.
\end{enumerate}

\subsection{Query Simulators for Conjunctions and Decision Lists}
\label{section:bshouty-appendix}

\paragraph*{Query simulator for conjunctions.}

The paper \cite{Bshouty20} presents an algorithm called \textsc{ApproxC} which satisfies the
following lemma, which can be deduced from Lemma 42 and an inspection of the proof of Lemma 43 in
that paper:
\begin{lemma}[\cite{Bshouty20}, Lemmas 42 \& 43]
Given query access to $f \colon \zo^n \to \zo$ and parameter $\epsilon \in (0,1)$, the algorithm
\textsc{ApproxC} makes $\widetilde O(1/\epsilon)$ queries to $f$
and either outputs $\Reject$, or a set $\bS \subseteq [n]$ and string $\bw \in \zo^n$. Define the
function
\[
    g(x) \define f(x_{\bS} \circ \bw_{\bar{\bS}}) .
    \]
With probability at least $9/10$, the following properties hold:
\begin{enumerate}
\item If the output is not $\Reject$, then $\dist_\cD(f,g) < \epsilon$.
\item If $f$ is a conjunction then the output is not $\Reject$, and $g$ is a conjunction on at
    most $|\bS| = O(\log(1/\epsilon))$ variables.
\end{enumerate}
\end{lemma}

From this lemma, we can obtain \cref{claim:bsh-simulator} by simply outputting the query
simulator that, on input $x$, outputs $f(x_{\bS} \circ \bw_{\bar{\bS}})$, when $\bS$ and $\bw$ are produced by the
algorithm. If the algorithm outputs $\Reject$, we can output a query simulator for the constant 0
function.

\paragraph*{Query simulator for 1-decision lists.}
Let $\bbJ \subseteq \bbP_{\DL}$ be the class of 1-decision-lists of
length at most $K$. The required
subroutine
is now \textsc{ApproxGeneralC} from
\cite[Figure~14 and Lemma~54]{Bshouty20}; the verification of Lemma~54's two hypotheses for decision lists is given in the proof of
\cite[Theorem~55]{Bshouty20}. 
Lemma~54  bounds the number of
relevant variables in the junta decision list by $\lambda k$, for an absolute constant
$\lambda>1$, where $k=\Theta(\log(1/\eta))$; this constant $\lambda$ is absorbed into the hidden constant in the definition of $K$. 
Invoking the subclass and uniform-closeness
parts of Lemma~54 with failure probability at most $\delta/2$ each, the overall failure probability is at most $\delta$.

\end{document}